\documentclass[10pt,a4paper]{article}

\usepackage[margin=2cm]{geometry}
\usepackage{url}%,lineno}
\usepackage{paralist}
\usepackage{sidecap}

\usepackage{tikz,pgf}
\usetikzlibrary{shapes,arrows,shadows} 
\usetikzlibrary{positioning,automata}
\usepackage{multirow}
\usepackage{boxedminipage,epsf}
\usepackage{subcaption}

\usepackage{amsmath,amsthm,amsbsy,amssymb}
\usepackage{graphicx}
\usepackage{cite}
\usepackage{color} 
\usepackage{comment}

\newcommand{\invivo}{{\em in vivo}}

\newcommand{\plus}{$^+$}

\newcommand{\rr}{\raggedright}
\newcommand{\tenpow}{\times10}

\usepackage[british]{babel}

\usepackage{authblk}

\begin{document}

\title{
A mathematical study of CD8\plus\ T~cell responses calibrated with human data
}

\author[1]{John Paul Gosling}
\author[2,3]{Sheeja M.~Krishnan}
\author[2]{Grant Lythe}
\author[4]{Benny Chain}
\author[5]{Cameron Mackay}
\author[2]{Carmen Molina-Par\'is~\thanks{Corresponding author: carmen@maths.leeds.ac.uk}}

\affil[1]{Department of Statistics,
School of Mathematics, University of Leeds, Leeds LS2 9JT, UK}
\affil[2]{Department of Applied Mathematics,
School of Mathematics, University of Leeds, Leeds LS2 9JT, UK}
\affil[3]{Modelling and Economics Unit,
National Infection Service,
Public Health England,
61 Colindale Avenue,
London NW9 5EQ, UK}
\affil[4]{Division of Infection and Immunity, 
UCL, 
Gower St, London WC1E 6BT, UK
}
\affil[5]{Unilever Safety and Environmental Assurance Centre, 
Sharnbrook, 
MK44 1LQ, UK}

\date{20th June 2017 - final version  - \LaTeX}

\maketitle

%%%%%%%%%%%%%%%%%%%%%%%%%%%%

\begin{abstract}

  Complete understanding of the mechanisms regulating the
  proliferation and differentiation that takes place during human
  immune CD8\plus\ T~cell responses is still lacking.  Human clinical
  data is usually limited to blood cell counts, yet the initiation of
  these responses occurs in the draining lymph nodes; 
  antigen-specific effector and memory CD8\plus\ T~cells generated in
  the lymph nodes migrate to those tissues where they are required.
  We use approximate Bayesian computation 
  with deterministic mathematical models of CD8\plus\ T~cell
  populations (naive, central memory, effector memory and effector)
     and yellow fever virus vaccine data to
  infer the dynamics of these CD8\plus\ T~cell populations in three
  spatial compartments: draining lymph nodes, circulation and skin.
  We have made use of the literature to obtain rates of division and
  death for human CD8\plus\ T~cell population subsets and thymic
  export rates.  Under the decreasing potential hypothesis for
  differentiation during an immune response, we find that, as the
  number of T~cell clonotypes driven to an immune response increases,
  there is a reduction in the number of divisions required to
  differentiate from a naive to an effector CD8\plus\ T~cell,
  supporting the ``division of labour'' hypothesis observed in murine
  studies.  We have also considered the reverse differentiation
  scenario, the increasing potential hypothesis.  The decreasing
  potential model is better supported by the yellow fever virus
  vaccine data.

\end{abstract}

%%%%%%%%%%%%%%%%%%%%%%%

\section{Introduction}
\label{intro}

In response to cognate antigen, a single naive T~cell is able to
produce multiple subsets of memory and effector T~cells of different
phenotypes and functional properties~\cite{mahnke2013s}.  Recent
studies suggest that, in humans, this differentiation process follows a
linear progression characterised by the following transitions: from
naive (N) to stem cell memory (SCM), to central memory (CM), to
transitional memory (TM), to effector memory (EM), and to (terminal)
effector (or EMRA) (E)
T~cells~\cite{gattinoni2011human,lugli2013superior}.  Stimulation by
cognate antigen seems to drive less-differentiated cells to generate
more-differentiated progeny.  Yet, the mechanisms that control the
proliferation and differentiation steps during an immune response,
from the activation of naive T~cells to the generation of memory and
effector cells, are not clearly
understood~\cite{kaech2001memory,farber2014human,thome2014spatial,thome2016early}.

In the case of CD8\plus\ T~cells, a few of these mechanisms have
already been identified and have helped decipher some of the rules
that govern CD8\plus\ T~cell differentiation (see
Ref.~\cite{kaech2001memory}).  This study showed that, after antigenic
stimulation, naive CD8\plus\ T~cells become committed to 
multiple rounds of division, then differentiate into effector and memory
cells.  That is, once the parental naive CD8\plus\ T~cell has been
activated, a developmental program is triggered, in which daughter
cells continue to divide and differentiate without further antigenic
stimulation.  Thus, the initial antigen encounter initiates an
instructive developmental programme that leads to effector and memory
CD8\plus\ T~cell formation and confers protective
immunity~\cite{kaech2001memory,farber2014human}.

Mathematical models, together with experimental and clinical data,
have improved our understanding of CD8\plus\ T~cell responses and the
generation of effector and memory cells during infection.  For
example, Ref.~\cite{le2014mathematical} combines mathematical
modelling with experimental data from clinical studies of yellow fever
virus (YFV) to explore kinetic details of the human immune response to
vaccination.  Some important results from this study include: i) the
estimation of the doubling time of effector CD8\plus\ T~cells to be
around two days, ii) that the peak of the CD8\plus\ T~cell-mediated
immune response depends on the rate of T~cell proliferation, and iii)
that the observed expansion of the YFV-specific CD8\plus\ T~cell
population was achieved in fewer than nine cell
divisions~\cite{le2014mathematical}.  The authors made use of a simple
mathematical model, based on
Refs.~\cite{de2001recruitment,althaus2007dynamics}, to describe the
fraction of YFV-specific CD8\plus\ T~cells in the secondary lymphoid
organs (SLOs) and circulation (blood).  The model includes migration
from SLOs to blood and from blood to tissues, as well as proliferation
in the SLOs (before the peak of the response) and death both in the
SLOs and in blood, only after the peak of the response.  They
calibrated their mathematical model with the average (among all
patients) fraction of specific CD8\plus\ T~cells from the YFV vaccine
data in Ref.~\cite{akondy2009yellow} (see Figure~3B in
Ref.~\cite{akondy2009yellow}).

In two other examples of combined experimental and modelling studies,
the authors have followed the clonal expansion of transgenic OT-1
cells in mice.  Their experiments identified a heterogeneous
population of CD8\plus\ T~cells arising from naive cells during
bacterial infection, and showed that the clonal expansion and
differentiation of individual naive T~cells is highly
stochastic~\cite{buchholz2013disparate, gerlach2013heterogeneous}.
These stochastic events, from multiple individual precursors, gave
rise to a robust cellular fate.  Mathematical modelling and parameter
calibration indicated that CD8\plus\ T~cells follow a linear
developmental path, with long-lived slowly-proliferating cells
differentiating to short-lived, highly proliferating
cells~\cite{buchholz2013disparate}.  Finally, Gong {\em et al.} have
recently developed a hybrid approach, that uses an agent-based model
in the lymph nodes (LNs) and an ordinary differential equation (ODE)
model in the blood compartment~\cite{gong2014harnessing}.  This
multi-compartmental model considers the following events: the
interaction of naive T~cells with antigen-bearing dendritic cells
(DCs), as well as cellular proliferation and differentiation in two
spatial locations (LNs and blood)~\cite{gong2014harnessing}.  The
model was calibrated with mice lymphocytic choriomeningitis virus
(LCMV) infection data and the authors concluded that the cellular
heterogeneity observed can be attributed to the number of
antigen-bearing DCs that each naive T~cell is responding
to~\cite{gong2014harnessing}.

A number of questions about the kinetics of CD8\plus\ T~cell-mediated
immune responses remain unanswered.  For example, the sequence of
differentiation steps during a CD8\plus\ T~cell response is unknown:
is it from naive to effector to memory or from naive to memory to
effector? A related second question is more technical. As human
clinical data is usually limited to blood cell counts, how feasible is
it to develop a mathematical model with spatial compartments, such as
the draining lymph nodes, circulation (or blood) and tissues, that can
be parameterised with data from only the blood compartment?  Finally,
flow cytometry and tetramer staining allow us to identify and measure,
within the blood compartment, the fraction of antigen-specific
CD8\plus\ T~cells in the total CD8\plus\ T~cell population.  It
remains a challenge to develop mathematical models of the kinetics of
CD8\plus\ T~cell-mediated immune responses that include different
CD8\plus\ populations (naive, central memory, effector memory and
effector) in the three previously mentioned spatial compartments, that
can be calibrated with human data.

In this paper, we analyse population-average models of CD8\plus\
T~cell dynamics that include four cellular populations: naive, central
memory, effector memory and effector T~cells, and three spatial
compartments: the draining lymph nodes, circulation and skin.
We note that each CD8\plus\ T~cell subpopulation is defined in terms
of its homeostatic ability to proliferate or not, its survival
capability, migration pattern, its division rates in the presence
of antigen, as well as the number of divisions required to
differentiate (see Section~\ref{math})~\cite{thomas2008comprehensive}.
Since our mathematical models will be calibrated with the fraction of
total specific CD8\plus\ T~cells from the YFV vaccine data in
Ref.~\cite{akondy2009yellow} (see Figure~3B in
Ref.~\cite{akondy2009yellow}), our definition of naive, central
memory, effector memory and effector CD8\plus\ T~cells is not based on
their expression levels of CD45RA and CCR7 as provided by flow
cytometry data~\cite{appay2002memory,gerritsen2015memory}. In this sense, our definition of naive, central
memory, effector memory and effector CD8\plus\ T~cells is based on
cell function instead of phenotype.

In order to decipher the sequence of differentiation events during an
immune response, we first consider the decreasing potential (DP)
hypothesis for generating CD8\plus\ T~cell
heterogeneity~\cite{buchholz2013disparate}, with differentiation
events linked to division~\cite{schlub2009division}.  We have made use
of the literature to obtain rates of division and
death for each human CD8\plus\ T~cell population subset, as well as
naive T~cell thymic export rates.  Approximate Bayesian computation
(ABC) has been used together with the mathematical model and YFV
vaccine data from~Ref.~\cite{akondy2009yellow} (see Figure~3B in
Ref.~\cite{akondy2009yellow}) to obtain posterior distributions of
the subset of parameters related to the immune response, such as the
number of divisions in the differentiation programme, the time to
first division, the time to subsequent divisions, the number of
specific clonotypes involved in the response, the duration of the
immune response, and migration rates.  We find that as the number of
clonotypes driven to an immune response increases, there is a
reduction in the number of divisions required to differentiate from a
naive to an effector CD8\plus\ T~cell, thus supporting the ``division
of labour'' already observed in murine
studies~\cite{gerlach2013heterogeneous,harty2008shaping}.
We then consider the
increasing potential (IP) hypothesis for generating CD8\plus\ T~cell
heterogeneity~\cite{kaech2012transcriptional}, and compare it with the
decreasing potential model.  Mathematical modelling, Bayesian
computation and the YFV data provide marginally stronger support for
the DP than the IP hypothesis.  We have also performed a sensitivity
analysis of both models to determine the extent to which the
parameters influence the population dynamics and circulatory kinetics
of CD8\plus\ T~cells through the draining LNs, circulation and skin.

The structure of the paper is as follows: Section~\ref{materials}
provides a brief description of the mathematical models used to
describe CD8\plus\ T~cell dynamics (homeostasis, decreasing potential
and increasing potential hypotheses), the choice of model parameters and the
computational algorithm. We also describe the sensitivity analysis
carried out, as well as the YFV vaccination data used to calibrate the
models. The results of our study are presented in
Section~\ref{results}, namely model calibration for both DPM and IPM
alongside a comparison of the two models. We discuss our results in
Section~\ref{discussion}. In Section~\ref{appendix-methodology}, we
provide full details of the mathematical models (ordinary differential
equations for the homeostasis, decreasing potential and increasing
potential models). We also provide an extensive literature review
highlighting the sources for our parameter estimates for the number
of cells, carrying capacities, division rates, death
rates, thymic export rates, migration rates, times to first and
subsequent divisions, as well as the number of clonotypes recruited to
an immune response and the duration of the immune response.  Finally,
we describe the methods involved in the sensitivity analysis
and Bayesian calibration of the mathematical models.

%%%%%%%%%%

\section{Materials and methods}
\label{materials}

%%%%%%%%%%

\subsection{Mathematical models}
\label{math}

%%%%%%%%%%

\subsubsection{Mathematical model of T~cell homeostasis}
\label{math-homeo}

A subset of the diverse population of CD8\plus\ T~cells generated
during an immune response is preserved once antigen is cleared.  This
is due to the homeostatic mechanisms that are in place to maintain
both the size and T~cell receptor (TCR) diversity of the CD8\plus\
T~cell population.  Thus, we require a mathematical model of CD8\plus\
T~cell homeostasis (in the absence of cognate antigen).  Full details
of the model are provided in Section~\ref{method-model-homeostasis}
below.  We now present a brief summary of the model.

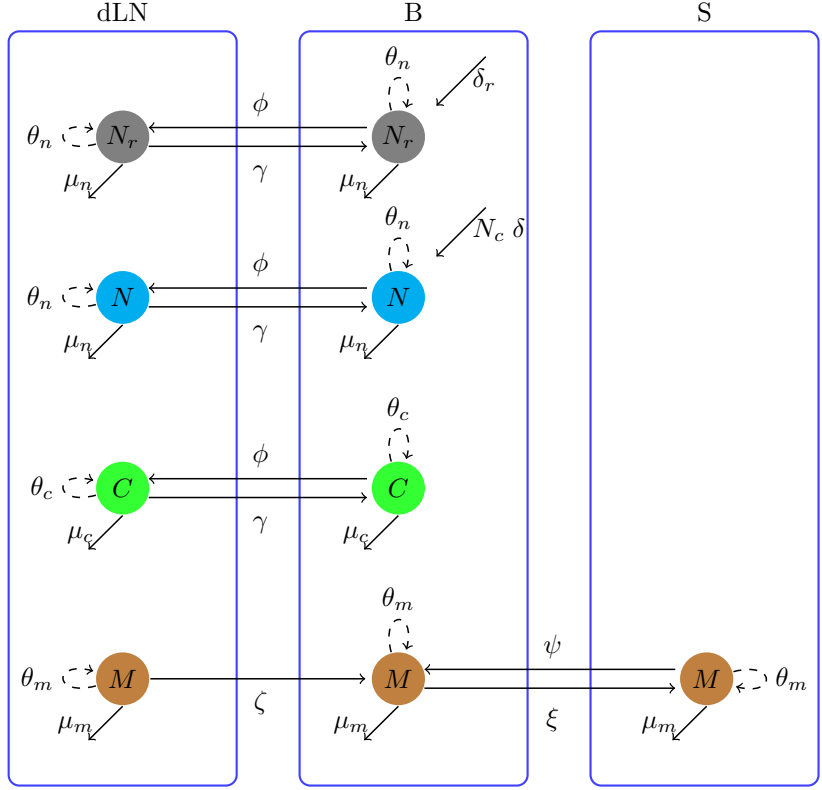
\begin{figure}[h!]
\begin{center}
\begin{tikzpicture} 
[block/.style   ={rectangle, draw=blue, thick,
                    text width=2cm, rounded corners, opacity=0.75,font=\small,
                    minimum height=10cm,minimum width=3cm},
cell/.style={circle,color=black,thick, minimum size=7mm, inner sep= 0pt},
line/.style    ={draw, -latex',shorten >=2pt, line width=0.2mm}]
 \matrix [column sep=8mm, row sep=20mm]
 {
 \node [cell, fill=gray] (naive)[yshift=36mm] {$N_r$};
 \node [cell, fill=cyan] (nai)[below =of naive,yshift=-4mm] {$N$};

 \node [cell, fill=green!80] (CM)[below =of nai,yshift=-8mm] {$C$};
 \node [cell, fill=brown] (EM) [below =of CM,yshift=-8mm]{$M$};
 \node [block] (ln)[label=dLN]{}; &
`
 \node [cell, fill=gray] (naiveb)[right of =naive, xshift=-12mm] {$N_r$};
 \node [cell, fill=cyan] (naib)[right of =nai, xshift=-12mm] {$N$};

 \node [cell, fill=green!80] (CMb)[below =of naib,yshift=-8mm] {$C$};
 \node [cell, fill=brown!100] (EMb) [below =of CMb,,yshift=-8mm]{$M$};
 \node [block] (blood)[label=B]{}; &

 \node [cell, fill=brown!100] (EMs) [right =of EMb,xshift=-15mm]{$M$};
 \node [block] (skin)[label=S]{}; &

\\
  };
 \begin{scope}[every path/.style=line]
	
	\path[dashed,yshift=5mm] (naive) edge  [loop left] node {$\theta_n$} ();
  \path[dashed,yshift=5mm] (nai) edge  [loop left] node {$\theta_n$} ();
	\path[dashed,yshift=5mm] (CM) edge  [loop left] node {$\theta_c$} ();
	\path[dashed,yshift=5mm] (EM) edge  [loop left] node {$\theta_m$} ();
	\path[dashed,yshift=5mm] (naiveb) edge  [loop above] node {$\theta_n$} ();
  \path[dashed,yshift=5mm] (naib) edge  [loop above] node {$\theta_n$} ();
	\path[dashed,yshift=5mm] (CMb) edge  [loop above] node {$\theta_c$} ();
	\path[dashed,yshift=5mm] (EMb) edge  [loop above] node {$\theta_m$} ();
	\path[dashed,yshift=5mm] (EMs) edge  [loop right] node {$\theta_m$} ();

	\path [->] (naive.south)--node[left]{$\mu_n$}++(-0.5,-0.5);
	\path [->] (nai.south)--node[left]{$\mu_n$}++(-0.5,-0.5);
	\path [->] (CM.south)--node[left]{$\mu_c$}++(-0.5,-0.5);
	\path [->] (EM.south)--node[left]{$\mu_m$}++(-0.5,-0.5);
	\path [->] (naiveb.south)--node[left]{$\mu_n$}++(-0.5,-0.5);
	\path [->] (naib.south)--node[left]{$\mu_n$}++(-0.5,-0.5);
	\path [->] (CMb.south)--node[left]{$\mu_c$}++(-0.5,-0.5);
	\path [->] (EMb.south)--node[left]{$\mu_m$}++(-0.5,-0.5);
	\path [->] (EMs.south)--node[left]{$\mu_m$}++(-0.5,-0.5);
   
	\draw[->](naive.340) to node [auto,yshift=-6mm] {$\gamma$} (naiveb.200);
	\draw[<-](naive.20) to node [auto,yshift=0.25mm] {$\phi$} (naiveb.160);	
		\draw[->](nai.340) to node [auto,yshift=-6mm] {$\gamma$} (naib.200);
	\draw[<-](nai.20) to node [auto,yshift=0.25mm] {$\phi$} (naib.160);	
	\draw[->] (CM.340) to node [auto,yshift=-6mm] {$\gamma$} (CMb.200);
	\draw[<-](CM.20) to node [auto,yshift=0.25mm] {$\phi$} (CMb.160);
	\draw[->] (EM.360) to node [auto,yshift=-6mm] {$\zeta$} (EMb.180);
	\draw[->](EMb.340) to node [auto,yshift=-7mm] {$\xi$} (EMs.200);
	\draw[<-](EMb.20) to node [auto,yshift=0.25mm] {$\psi$} (EMs.160);

	\path [<-] ([xshift=3mm,yshift=-10mm]blood.north)--node[right]{$\delta_r$}++(0.7,0.7);
	\path [<-] ([xshift=3mm,yshift=-30mm]blood.north)--node[right]{$N_c \; \delta$}++(0.7,0.7);

\end{scope} 
\end{tikzpicture}
\end{center}
\caption {Mathematical model of CD8\plus\ T~cell homeostasis.  We
  consider non-specific (grey) and specific (blue) naive ($N_r$ and
  $N$), central memory ($C$) (green) and effector memory ($M$) (brown)
  CD8\plus\ T~cells.  Each cell type can be in three different spatial
  compartments: draining LNs (dLNs), blood and resting LNs (B) or skin
  (S).  Each cell type can divide with rates
  $\theta_n$, $\theta_c$ and $\theta_m$, or die with rates $\mu_n$,
  $\mu_c$ and $\mu_m$, respectively. Non-specific and specific naive
  CD8\plus\ T~cells populate the peripheral blood (B) compartment from
  the thymus with rates $\delta_r$ and $N_c \; \delta$,
  respectively. Naive and CM cells migrate between dLNs and B, with
  rates $\gamma$ and $\phi$, respectively.  EM cells migrate from dLNs
  to B with rate $\zeta$, as well as between B and S, with rates $\xi$
  and $\psi$, respectively.  The rates have been labelled with
  subscripts $n$, $c$, and $m$ for naive, central memory, and effector
  memory cells, respectively.}
 \label{CD8homeo} 
 \end{figure}

The homeostasis model includes three subsets of CD8\plus\ T~cells:
naive (N), central memory (CM), and effector memory (EM) cells.  Naive
cells are further divided into antigen-specific and non-specific naive
T~cells. CM and EM cells are assumed to be specific to a fiducial
skin-delivered antigen (or cognate antigen).  The naive and central
memory populations reside in the draining LNs (dLNs) and blood
compartments. Effector memory cells can also reside in the
skin~\cite{gebhardt2012local} (see Figure~\ref{CD8homeo}).

 In what follows we describe, for each cellular subtype, the processes
 that are included in the homeostasis model:

\begin{itemize}

\item Naive cells are released from the thymus into the circulation,
  and we assume a constant thymic export rate per clonotype, denoted
  by $\delta$.  We denote by $\delta_r$, the total thymic non-specific
  export rate (see Table~\ref{table-parameters-fixed}).

\item Division is assumed for N, CM and EM CD8\plus\
  T~cells to be logistic~\cite{antia1998models}, with proliferation
  rate, $\theta$ and carrying capacity per clonotype, $\kappa$.  This
  logistic term encodes competition for resources, and thus avoids
  unlimited exponential growth of the populations.  The rates are
  denoted by the subscripts $n$, $c$, and $m$ for naive, central
  memory, and effector memory cells, respectively (see
  Table~\ref{table-parameters-fixed}).
	
\item Each subset of CD8\plus\ T~cells has a finite lifespan, and we
  assume that the death rates are independent of the spatial location.
  We denote them by $\mu$ and make use of the subscripts $n$, $c$, and
  $m$ for naive, central memory, and effector memory cells,
  respectively (see Table~\ref{table-parameters-fixed}). The death
  term of each subset is proportional to its population size.
  
\item Migration for any cell type is assumed to be proportional to its
  population size.  We assume naive and central memory cells circulate
  between dLNs and the blood compartment.  Effector memory cells
  migrate from the dLNs to the blood compartment, and are assumed to
  circulate between skin and the blood compartment (see
  Table~\ref{table-parameters-bayesian}).
 
\end{itemize}

We have assumed that each population maintains its numbers
independently~\cite{tanchot1997differential, freitas2000population,
  johnson2014population}.  Naive cells are released from the thymus
and divide due to TCR interactions with
self-peptides expressed on antigen-presenting cells, or IL-7 cytokine
signalling~\cite{takada2009naive}.  Effector memory cells maintain
their numbers by homeostatic mechanisms that involve cytokine IL-15,
while central memory cells use both IL-7 and
IL-15~\cite{boyman2009homeostatic,farber2014human}.  We assume that
the death and division rate of each CD8\plus\ T~cell
subtype is independent of its spatial location.  Naive and central
memory cells migrate between dLNs and circulation, while effector
memory cells can also migrate to the skin.  The migration rates of
naive and central memory cells are assumed to be
same~\cite{berard2002qualitative}, as the trafficking molecules CD62L
and CCR7 are expressed on both populations~\cite{nolz2011naive}.
Effector memory cells migrate from dLNs to blood and preferentially
migrate to skin~\cite{masopust2001preferential}.  Effector memory
cells do not migrate from blood to dLNs, as they lack lymph node
homing molecules, but express skin homing chemokine receptors, such as
CCR4 and CCR10~\cite{mueller2013memory, benechet2014visualizing}.  We
do not include effector cells in the homeostasis model, as they are
terminally differentiated cells with a short
lifespan~\cite{gattinoni2011human,lugli2013superior}.

%%%%%%%%%%

\subsubsection{Mathematical model of T~cell differentiation: decreasing potential model}
\label{math-DPM}

During antigen challenge, antigen presenting cells (APCs) migrate to
the dLNs for efficient antigen presentation to naive T~cells.  With
the general observation that upon activation by APCs, antigen-specific
naive cells give rise to long-lived memory cells and effector cells,
several potential mechanisms have been suggested for T~cell
diversification~\cite{ahmed2009precursors,kaech2012transcriptional}.
Experimental studies involving the adoptive transfer and \invivo\ fate
mapping of single CD8\plus\ T~cells~\cite{buchholz2013disparate},
chromatin state transitions~\cite{zhu2013genome}, and metabolism shift
during cell differentiation~\cite{sukumar2013inhibiting}, support a
model of progressive
differentiation~\cite{restifo2013lineage,farber2014human}.  This model
of differentiation is also supported by recent studies of human T~cell
compartmentalisation~\cite{thome2014spatial,thome2016early}.
   
\begin{figure}[h!]
\begin{center}
\begin{tikzpicture} 
[cell/.style={circle,color=black,thick, minimum size=7mm, inner sep= 0pt},
line/.style    ={draw, -latex',shorten >=4pt,line width=0.2mm}]
 \matrix []
 {
 \node [cell, fill=cyan] (naive)[] {$N$};
 \node[star, color=black,fill=black!60!white,star points=4,star point
ratio=0.3] (APC1)[right of =naive, xshift=6mm,yshift=8mm,font=\footnotesize]{$APC$};
 \node [cell, fill=green!100] (CM)[right of =naive, xshift=20mm] {$C$};
 \node[star, color=black,fill=black!60!white,star points=4,star point
ratio=0.3] (APC2)[right of =CM, xshift=6mm,yshift=8mm,font=\footnotesize]{$APC$};

 \node [cell, fill=brown!100] (EM) [right of =CM, xshift=20mm]{$M$};
 \node[star, color=black,fill=black!60!white,star points=4,star point
ratio=0.3] (APC3)[right of =EM, xshift=6mm,yshift=8mm,font=\footnotesize]{$APC$};

 \node [cell, fill=red!80] (E)[right of =EM, xshift=20mm]{$E$};\\
 };
\begin{scope}[every path/.style=line]
\draw [ ->,] (naive.east) -- (CM.west);
\path [ ->] (CM.east) -- (EM.west);
\path [ ->] (EM.east) -- (E.west);
\end{scope}
\end{tikzpicture}
\end{center}
\caption{CD8\plus\ T~cell differentiation model during an immune
  response: decreasing potential model (DPM).  Differentiation is
  mediated by the interaction between specific CD8\plus\ T~cells with
  antigen presenting cells (APCs) and proceeds from naive ($N$) to
  central memory ($C$), to effector memory ($M$), and finally to fully
  differentiated effector ($E$) cells.}
\label{dpm} 
\end{figure}
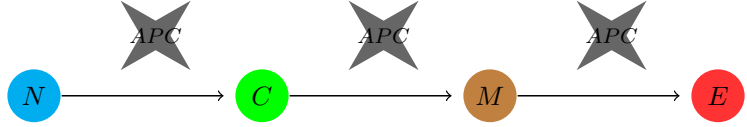

According to this progressive differentiation model, sustained antigenic
stimulation of T~cells drives differentiation towards
effector function.  
Thus, from naive T~cells, central
memory cells are generated, followed by effector memory and effector
cells, with further antigen stimulation at each differentiation stage.
This model, known as the ``decreasing potential model''
(DPM)~\cite{ahmed1996immunological,lanzavecchia2002progressive} is
supported by experimental
observations~\cite{huster2006unidirectional,joshi2007inflammation,schlub2009division,gattinoni2011human,buchholz2012origin,cieri20137}
(see Figure~\ref{dpm}).
 
 %%%%%%%%%%%%

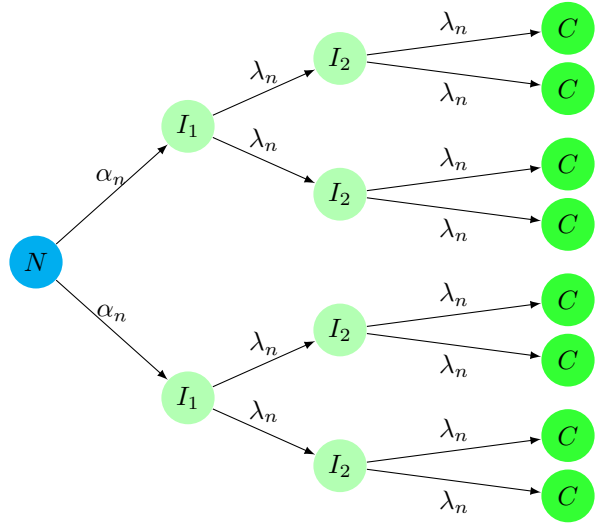
\begin{figure}[h!]
\begin{center}
\begin{tikzpicture}[grow=right, sloped]
\tikzstyle{cell} = [circle,color=black,fill=cyan,minimum size=7mm, inner sep=0pt]
\tikzstyle{mid} = [circle,color=black,fill=green!30, minimum size=7mm,inner sep=0pt]
\tikzstyle{end} = [circle,color=black,fill=green!80, minimum size=7mm,inner sep=0pt]
\tikzstyle{level 1}=[level distance=2cm, sibling distance=3.6cm]
\tikzstyle{level 2}=[level distance=2cm, sibling distance=1.8cm]
\tikzstyle{level 3}=[level distance=3cm, sibling distance=0.8cm]
\node[cell] {$N$}       
        child {node[mid] {$I_1$}
                child {node[mid] {$I_2$}
			 			child{node[end] {$C$}
                     		edge from parent[-latex]
                    		 node[below,rotate=8] {$\lambda_n$}
                		}	
 						child{node[end] {$C$}
                     		edge from parent[-latex]
                    		 node[above,rotate=-8] {$\lambda_n$}
                		}
                    	 edge from parent[-latex]
                    	 node[above, rotate=25] {$\lambda_n$}
               	 }
                child{node[mid] {$I_2$}
					child{node[end] {$C$}
                     		edge from parent[-latex]
                    		 node[below,rotate=8] {$\lambda_n$}
                	}
 					child{node[end] {$C$}
                     		edge from parent[-latex]
                    		 node[above,rotate=-8] {$\lambda_n$}
                	}
                     edge from parent[-latex]
                     node[above,rotate=-25] {$\lambda_n$}
                }
                edge from parent[-latex]
                node[above,rotate=40] {$\alpha_n$}
        }
        child {node[mid] {$I_1$}
                child {node[mid] {$I_2$}
					child{node[end] {$C$}
                     		edge from parent[-latex]
                    		 node[below,rotate=8] {$\lambda_n$}
                	}
 					child{node[end] {$C$}
                     		edge from parent[-latex]
                    		 node[above,rotate=-8] {$\lambda_n$}
                	}
                     edge from parent[-latex]
                     node[above, rotate=25] {$\lambda_n$}
                }
                child {node[mid] {$I_2$}
					child{node[end] {$C$}
                     		edge from parent[-latex]
                    		 node[below,rotate=8] {$\lambda_n$}
                	}
 					child{node[end] {$C$}
                     		edge from parent[-latex]
                    		 node[above,rotate=-8] {$\lambda_n$}
                	}
                     edge from parent[-latex]
                     node[above, rotate=-25] {$\lambda_n$}
                }
                edge from parent[-latex]
                node[above,rotate=-40] {$\alpha_n$}
        }
   ;
\end{tikzpicture}
\end{center}
\caption{Division-linked differentiation hypothesis: a number of
  divisions, $g$, is required for each CD8\plus\ T~cell
  differentiation step.  For example, differentiation from naive ($N$)
  (blue) to central memory ($C$) (green) subtype requires $g_n$
  divisions steps (shown here for $g_n=3$). Each division step gives
  rise to two intermediate daughter cells ($I$) (light green).  The
  first division step requires interaction with an APC and
  has rate $\alpha_n$.  Each subsequent division step
  is part of a programme of $g_n$ divisions and has
  rate $\lambda_n$. The last division step gives rise to central
  memory cells.}
\label{div-linked-diff}
\end{figure}

We will assume that a single APC contact is required to trigger a
programme of multiple rounds of division (or clonal expansion) before
differentiation to the next stage of the differentiation pathway can
take place~\cite{kaech2001memory} (see Fig.~\ref{dpm} and
Fig.~\ref{div-linked-diff}).  By following cellular divisions using
carboxyfluorescein succinimidyl ester (CFSE), experimental studies
have shown in both B~cells and T~cells, that class switching and
differentiation depends on the number of
divisions~\cite{hodgkin1996b,gett1998cell,schlub2009division,hogan2013clonally,marchingo2014antigen}. Each
division enhances the probability of differentiating to a new
identifiable subtype~\cite{hasbold1999quantitative}.

%%%%%%%%%%%%

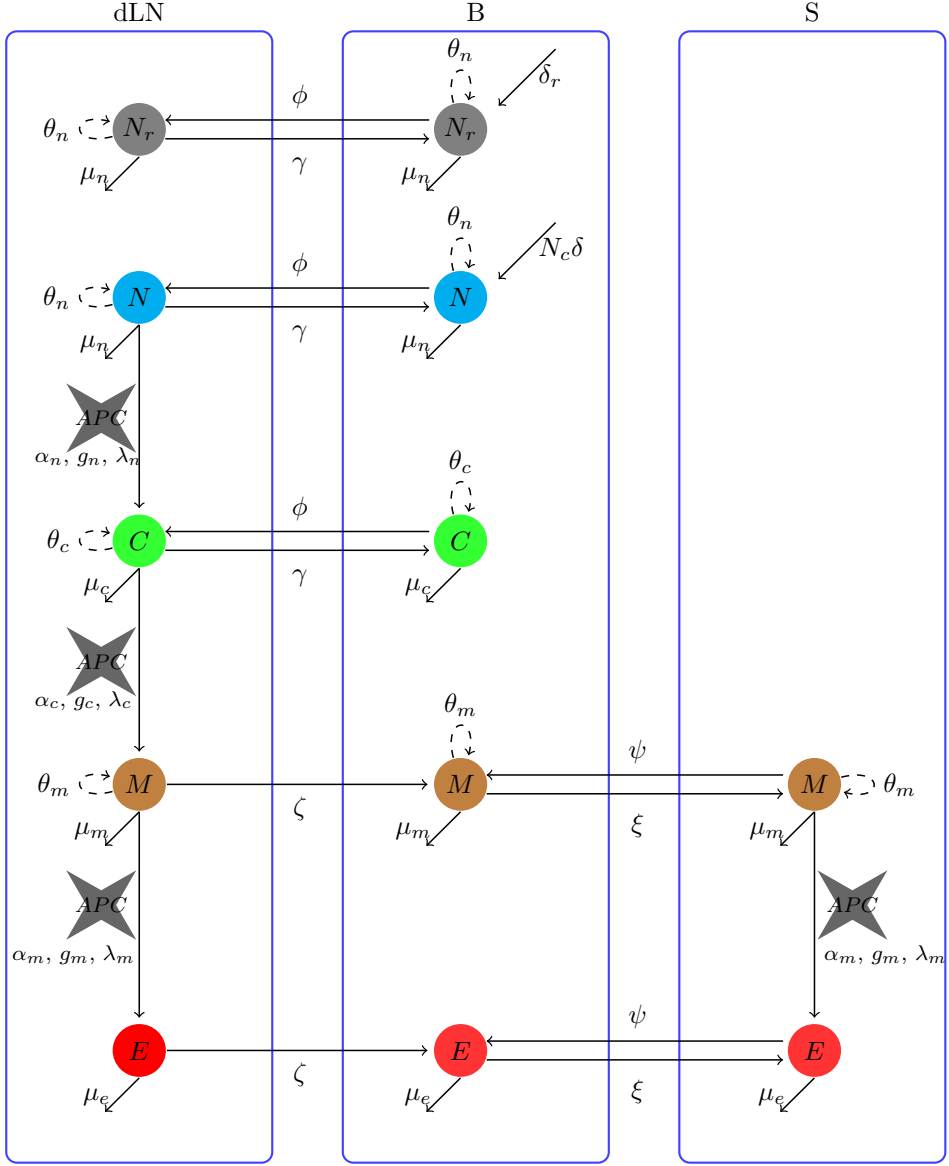
\begin{figure}[h!]
\begin{center}
\begin{tikzpicture} 
[block/.style   ={rectangle, draw=blue, thick,
                    text width=2cm, rounded corners, opacity=0.75,font=\small,
                    minimum height=15cm,minimum width=3.5cm},
cell/.style={circle,color=black,thick, minimum size=7mm, inner sep= 0pt},
line/.style    ={draw, -latex',shorten >=2pt, line width=0.2mm}]
 \matrix [column sep=9mm, row sep=20mm]
 {
 \node [cell, fill=gray] (nr)[yshift=62mm] {$N_r$};
 \node [cell, fill=cyan] (naive)[below =of nr,yshift=-5mm] {$N$};
 \node[star, color=black,fill=black!60!white,star points=4,star point
ratio=0.3] (APC3)[below of =naive,xshift=-5mm, yshift=-6mm,font=\footnotesize]{$APC$};

 \node [cell, fill=green!80] (CM)[below =of naive,yshift=-15mm] {$C$};
 \node[star, color=black,fill=black!60!white,star points=4,star point
ratio=0.3] (APC3)[below of =CM,xshift=-5mm, yshift=-6mm,font=\footnotesize]{$APC$};

 \node [cell, fill=brown!100] (EM) [below =of CM,yshift=-15mm]{$M$};
 \node[star, color=black,fill=black!60!white,star points=4,star point
ratio=0.3] (APC3)[below of =EM,xshift=-5mm, yshift=-6mm,font=\footnotesize]{$APC$};

 \node [cell, fill=red] (E)[below =of EM,yshift=-18mm]{$E$};
 \node [block] (ln)[label=dLN]{}; &

 \node [cell, fill=gray] (nrb)[right of =nr, xshift=-12mm] {$N_r$};
\node [cell, fill=cyan] (naiveb)[right of =naive,xshift=-12mm] {$N$ };
 \node [cell, fill=green!80] (CMb)[below =of naiveb,yshift=-15mm] {$C$};
 \node [cell, fill=brown!100] (EMb) [below =of CMb,,yshift=-15mm]{$M$};
 \node [cell, fill=red!80] (Eb)[below =of EMb,yshift=-18mm]{ $E$};
 \node [block] (blood)[label=B]{}; &

 \node [cell, fill=brown!100] (EMs) [right =of EMb,xshift=-15mm]{$M$};
 \node[star, color=black,fill=black!60!white,star points=4,star point
ratio=0.3] (APC3)[below of =EMs,xshift=5mm, yshift=-6mm,font=\footnotesize]{$APC$};

 \node [cell, fill=red!80] (Es)[below =of EMs,yshift=-18mm]{$E$};
 \node [block] (skin)[label=S]{}; &

\\
  };
 \begin{scope}[every path/.style=line]

	\draw[->](naive.270) to node [auto,xshift=-15mm,yshift=-5mm,font=\footnotesize] {$\alpha_n$, $g_n$, $\lambda_n$} (CM.90);
	
  \draw[->](CM.270) to node [auto,xshift=-15mm,yshift=-5mm,font=\footnotesize] {$\alpha_c$, $g_c$, $\lambda_c$} (EM.90);
	\draw[->](EM.270) to node [auto,xshift=-18mm,yshift=-5mm,font=\footnotesize] {$\alpha_m$, $g_m$, $\lambda_m$} (E.90);
	\draw[->](EMs.270) to node [auto,xshift=0mm,yshift=-5mm,font=\footnotesize] {$\alpha_m$, $g_m$, $\lambda_m$} (Es.90);
	
		\path[dashed,yshift=5mm] (nr) edge  [loop left] node {$\theta_n$} ();
	\path[dashed,yshift=5mm] (naive) edge  [loop left] node {$\theta_n$} ();
	\path[dashed,yshift=5mm] (CM) edge  [loop left] node {$\theta_c$} ();
	\path[dashed,yshift=5mm] (EM) edge  [loop left] node {$\theta_m$} ();
	\path[dashed,yshift=5mm] (naiveb) edge  [loop above] node {$\theta_n$} ();
	\path[dashed,yshift=5mm] (nrb) edge  [loop above] node {$\theta_n$} ();
	\path[dashed,yshift=5mm] (CMb) edge  [loop above] node {$\theta_c$} ();
	\path[dashed,yshift=5mm] (EMb) edge  [loop above] node {$\theta_m$} ();
	\path[dashed,yshift=5mm] (EMs) edge  [loop right] node {$\theta_m$} ();

	\path [->] (nr.south)--node[left]{$\mu_n$}++(-0.5,-0.5);
	\path [->] (naive.south)--node[left]{$\mu_n$}++(-0.5,-0.5);
	\path [->] (CM.south)--node[left]{$\mu_c$}++(-0.5,-0.5);
	\path [->] (EM.south)--node[left]{$\mu_m$}++(-0.5,-0.5);
	\path [->] (E.south)--node[left]{$\mu_e$}++(-0.5,-0.5);
		\path [->] (nrb.south)--node[left]{$\mu_n$}++(-0.5,-0.5);
	\path [->] (naiveb.south)--node[left]{$\mu_n$}++(-0.5,-0.5);
	\path [->] (CMb.south)--node[left]{$\mu_c$}++(-0.5,-0.5);
	\path [->] (EMb.south)--node[left]{$\mu_m$}++(-0.5,-0.5);
	\path [->] (Eb.south)--node[left]{$\mu_e$}++(-0.5,-0.5);
	\path [->] (EMs.south)--node[left]{$\mu_m$}++(-0.5,-0.5);
	\path [->] (Es.south)--node[left]{$\mu_e$}++(-0.5,-0.5); 
	
 \draw[->](nr.340) to node [auto,yshift=-6mm] {$\gamma$} (nrb.200);
	\draw[<-](nr.20) to node [auto,yshift=0.25mm] {$\phi$} (nrb.160);	

  \draw[->](naive.340) to node [auto,yshift=-6mm] {$\gamma$} (naiveb.200);
	\draw[<-](naive.20) to node [auto,yshift=0.25mm] {$\phi$} (naiveb.160);	
	\draw[->] (CM.340) to node [auto,yshift=-6mm] {$\gamma$} (CMb.200);
	\draw[<-](CM.20) to node [auto,yshift=0.25mm] {$\phi$} (CMb.160);
	\draw[->] (EM.360) to node [auto,yshift=-6mm] {$\zeta$} (EMb.180);
	\draw[->](EMb.340) to node [auto,yshift=-7mm] {$\xi$} (EMs.200);
	\draw[<-](EMb.20) to node [auto,yshift=0.25mm] {$\psi$} (EMs.160);
	\draw[->] (E.360) to node [auto,yshift=-6mm] {$\zeta$} (Eb.180);
	\draw[->](Eb.340) to node [auto,yshift=-7mm] {$\xi$} (Es.200);
	\draw[<-](Eb.20) to node [auto,yshift=0.5mm] {$\psi$} (Es.160);
	
	\path [<-] ([xshift=3mm,yshift=-10mm]blood.north)--node[right]{ $\delta_r$}++(0.8,0.8);
	\path [<-] ([xshift=3mm,yshift=-33mm]blood.north)--node[right]{ $ N_c \delta$}++(0.8,0.8);

\end{scope} 
\end{tikzpicture}
\end{center}
\caption{Mathematical model of CD8\plus\ T~cell dynamics during an
  immune response. The model considers naive, central memory, effector
  memory and effector CD8\plus\ T~cells and three spatial
  compartments. T~cell interaction
  with a cognate APC triggers a programme of division steps that leads to
  differentiation (decreasing potential model): from naive to CM, to
  EM and to fully differentiated effector T~cells.  Vertical arrows
  describe the differentiation programme (linked to division) set by
  APC interactions.  As described in the model of homeostasis, N, CM
  and EM divide (oval arrows) ($\theta_n,
  \theta_c, \theta_m$), die ($\mu_n, \mu_c, \mu_m$) and migrate.
  Effector cells have the same migratory behaviour as EM cells.  Cell
  subtype specific rates are labelled by the subscripts $n$, $c$, $m$,
  and $e$ for naive, central memory, effector memory, and effector
  cells, respectively.
  We assume no bystander activation during an immune response and thus,
non-specific  naive CD8\plus\ T~cells are described by the homeostasis model.
  }
\label{cd8-immune-response}
\end{figure}

The mathematical model of CD8\plus\ T~cell dynamics during an immune
response under the decreasing potential hypothesis (see
Fig.~\ref{cd8-immune-response}) includes the following cell
types~\footnote{Full details of the mathematical model have been
  provided in Section~\ref{method-model-immune-response-DPM} and in
  this section we only present a brief summary of the model.}:
specific naive cells ($N$), intermediate cells in the differentiation
pathway between naive and central memory cells ($I$), central memory
cells ($C$), intermediate cells in the differentiation pathway between
central and effector memory cells ($J$), effector memory cells ($M$),
intermediate cells in the differentiation pathway between effector
memory and effector cells ($K$), and effector cells ($E$).
We assume no bystander activation during an immune response and thus,
non-specific  CD8\plus\ T~cell are not considered~\cite{gerlach2013heterogeneous}. 

%%%%%%%%%%%%%%%%

In what follows, we describe, for each cellular subtype, the processes
that are included in the DPM:

\begin{itemize}

\item Antigen-induced proliferation is triggered by a contact with a cognate
  antigen presenting cell (see Figure~\ref{dpm}).  This contact starts
  a programme of division events, characterised by the total number of
  divisions, $g$, that leads to differentiation (see
  Figure~\ref{div-linked-diff}). The rate of contact with an antigen
  presenting cell is denoted by $\alpha$ and is the inverse of the
  time to first division.  The rate of subsequent divisions is denoted
  by $\lambda$ and is the inverse of the time to subsequent divisions.
  The rates are denoted by the subscripts $n$, $c$, and $m$ for naive,
  central memory, and effector memory cells, respectively (see
  Table~\ref{table-parameters-bayesian}).  The division rate of every
  population is assumed to be proportional to its corresponding population size.
  Antigen presentation can take place in the dLNs and skin
  compartment.
		
\item The death rates of naive, central memory and effector memory
  T~cells are the same as in the homeostasis model.  For effector
  cells, we denote their death rate by $\mu_e$.  Intermediate cell
  types have identical death rate to those of their parents, so that
  the death rate of $I$, $J$ and $K$ cells is $\mu_n$, $\mu_c$, and
  $\mu_m$, respectively (see Table~\ref{table-parameters-fixed}).  As
  in the homeostasis model, death terms are proportional to the
  population size of each subset.

\item Migration rules for naive, central and effector memory T~cells
  are as in the homeostasis model.  Effector cells migrate with the
  same rules and rates as effector memory cells (see
  Table~\ref{table-parameters-bayesian}).  Intermediate cells in the
  dLNs and skin compartment have no migratory capability (see
  Fig.~\ref{cd8-immune-response}).

\end{itemize}
    
As the frequency of precursor naive T~cells
varies~\cite{moon2007naive,obar2008endogenous} and there may be
multiple TCR clonotypes responding to a given cognate
antigen~\cite{pewe2004very}, we introduce $N_c$, the number of
different TCR clonotypes that are taking part in the immune response,
as an additional parameter of the model. Together the different TCR
clonotypes constitute the antigen-specific CD8\plus\ naive T~cell
population.  We exclude any non-specific naive CD8\plus\ T~cells in
the immune response model and thus, neglect any potential bystander
activation~\cite{ahmed2011insights,gerlach2013heterogeneous}.
  
%%%%%%%%%%%%%%%%%%%%%%%%%%%%%%%%%%%%%%%%%

\subsubsection{Mathematical model of T~cell differentiation: increasing potential model}
\label{math-IPM}

The mathematical model of CD8\plus\ T~cell dynamics under the
increasing potential hypothesis has been described in
Section~\ref{method-model-immune-response-IPM}, and we only summarise
the model in this section.

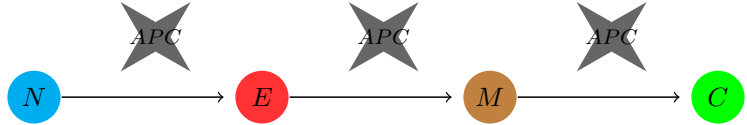
\begin{figure}[h!]
\begin{center}
\begin{tikzpicture} 
[cell/.style={circle,color=black,thick, minimum size=7mm, inner sep= 0pt},
line/.style    ={draw, -latex',shorten >=4pt,line width=0.2mm}]
 \matrix []
 {
 \node [cell, fill=cyan] (naive)[] {$N$};
 \node[star, color=black,fill=black!60!white,star points=4,star point
ratio=0.3] (APC1)[right of =naive, xshift=6mm,yshift=8mm,font=\footnotesize]{$APC$};
 \node [cell, fill=red!80] (CM)[right of =naive, xshift=20mm] {$E$};
 \node[star, color=black,fill=black!60!white,star points=4,star point
ratio=0.3] (APC2)[right of =CM, xshift=6mm,yshift=8mm,font=\footnotesize]{$APC$};

 \node [cell, fill=brown!100] (EM) [right of =CM, xshift=20mm]{$M$};
 \node[star, color=black,fill=black!60!white,star points=4,star point
ratio=0.3] (APC3)[right of =EM, xshift=6mm,yshift=8mm,font=\footnotesize]{$APC$};

 \node [cell, fill=green!100] (E)[right of =EM, xshift=20mm]{$C$};\\
 };
\begin{scope}[every path/.style=line]
\draw [ ->,] (naive.east) -- (CM.west);
\path [ ->] (CM.east) -- (EM.west);
\path [ ->] (EM.east) -- (E.west);
\end{scope}
\end{tikzpicture}
\end{center}
\caption{CD8\plus\ T~cell differentiation model during
an immune response: increasing potential model.  
Differentiation is mediated by the interaction between specific CD8\plus\ T~cells
with antigen presenting cells (APCs) and proceeds from naive ($N$) to effector ($E$), 
to effector memory ($M$), and finally to central memory ($C$) cells.}
\label{ipm} 
\end{figure}

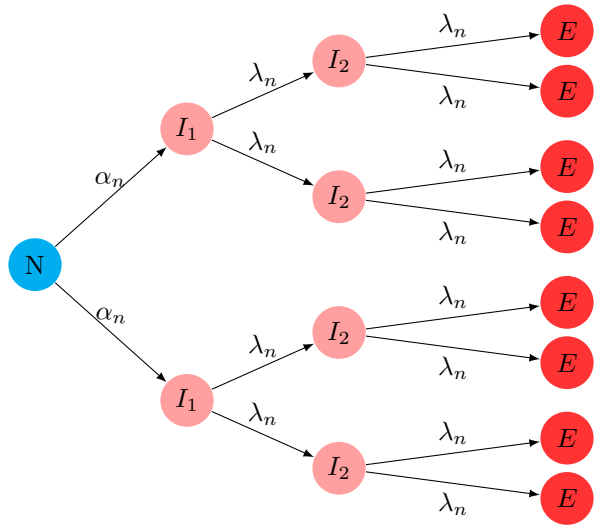
\begin{figure}[h!]
\begin{center}
\begin{tikzpicture}[grow=right, sloped]
\tikzstyle{cell} = [circle,color=black,fill=cyan,minimum size=7mm, inner sep=0pt]
\tikzstyle{mid} = [circle,color=black,fill=pink!150, minimum size=7mm,inner sep=0pt]
\tikzstyle{end} = [circle,color=black,fill=red!80, minimum size=7mm,inner sep=0pt]
\tikzstyle{level 1}=[level distance=2cm, sibling distance=3.6cm]
\tikzstyle{level 2}=[level distance=2cm, sibling distance=1.8cm]
\tikzstyle{level 3}=[level distance=3cm, sibling distance=0.8cm]
\node[cell] {N}       
        child {node[mid] {$I_1$}
                child {node[mid] {$I_2$}
			 			child{node[end] {$E$}
                     		edge from parent[-latex]
                    		 node[below,rotate=8] {$\lambda_n$}
                		}	
 						child{node[end] {$E$}
                     		edge from parent[-latex]
                    		 node[above,rotate=-8] {$\lambda_n$}
                		}
                    	 edge from parent[-latex]
                    	 node[above, rotate=25] {$\lambda_n$}
               	 }
                child{node[mid] {$I_2$}
					child{node[end] {$E$}
                     		edge from parent[-latex]
                    		 node[below,rotate=8] {$\lambda_n$}
                	}
 					child{node[end] {$E$}
                     		edge from parent[-latex]
                    		 node[above,rotate=-8] {$\lambda_n$}
                	}
                     edge from parent[-latex]
                     node[above,rotate=-25] {$\lambda_n$}
                }
                edge from parent[-latex]
                node[above,rotate=40] {$\alpha_n$}
        }
        child {node[mid] {$I_1$}
                child {node[mid] {$I_2$}
					child{node[end] {$E$}
                     		edge from parent[-latex]
                    		 node[below,rotate=8] {$\lambda_n$}
                	}
 					child{node[end] {$E$}
                     		edge from parent[-latex]
                    		 node[above,rotate=-8] {$\lambda_n$}
                	}
                     edge from parent[-latex]
                     node[above, rotate=25] {$\lambda_n$}
                }
                child {node[mid] {$I_2$}
					child{node[end] {$E$}
                     		edge from parent[-latex]
                    		 node[below,rotate=8] {$\lambda_n$}
                	}
 					child{node[end] {$E$}
                     		edge from parent[-latex]
                    		 node[above,rotate=-8] {$\lambda_n$}
                	}
                     edge from parent[-latex]
                     node[above, rotate=-25] {$\lambda_n$}
                }
                edge from parent[-latex]
                node[above,rotate=-40] {$\alpha_n$}
        }
   ;
\end{tikzpicture}
\end{center}
\caption{ Division-linked differentiation hypothesis: a number of
  divisions, $g$, is required for CD8\plus\ T~cell differentiation.
  For example, differentiation from naive ($N$) (blue) to effector
  ($E$) (red) subtype requires $g_n$ divisions steps (shown here for
  $g_n=3$). Each division step gives rise to two intermediate cells
  ($I$) (pink).  The first division step requires interaction with an
  APC and is characterised by the rate $\alpha_n$.  Each subsequent
  division step is part of a programme of $g_n$ divisions and is
  characterised by the rate $\lambda_n$ and the last division step
  gives rise to $E$ cells.}
\label{div-linked-diff-ipm}
\end{figure}

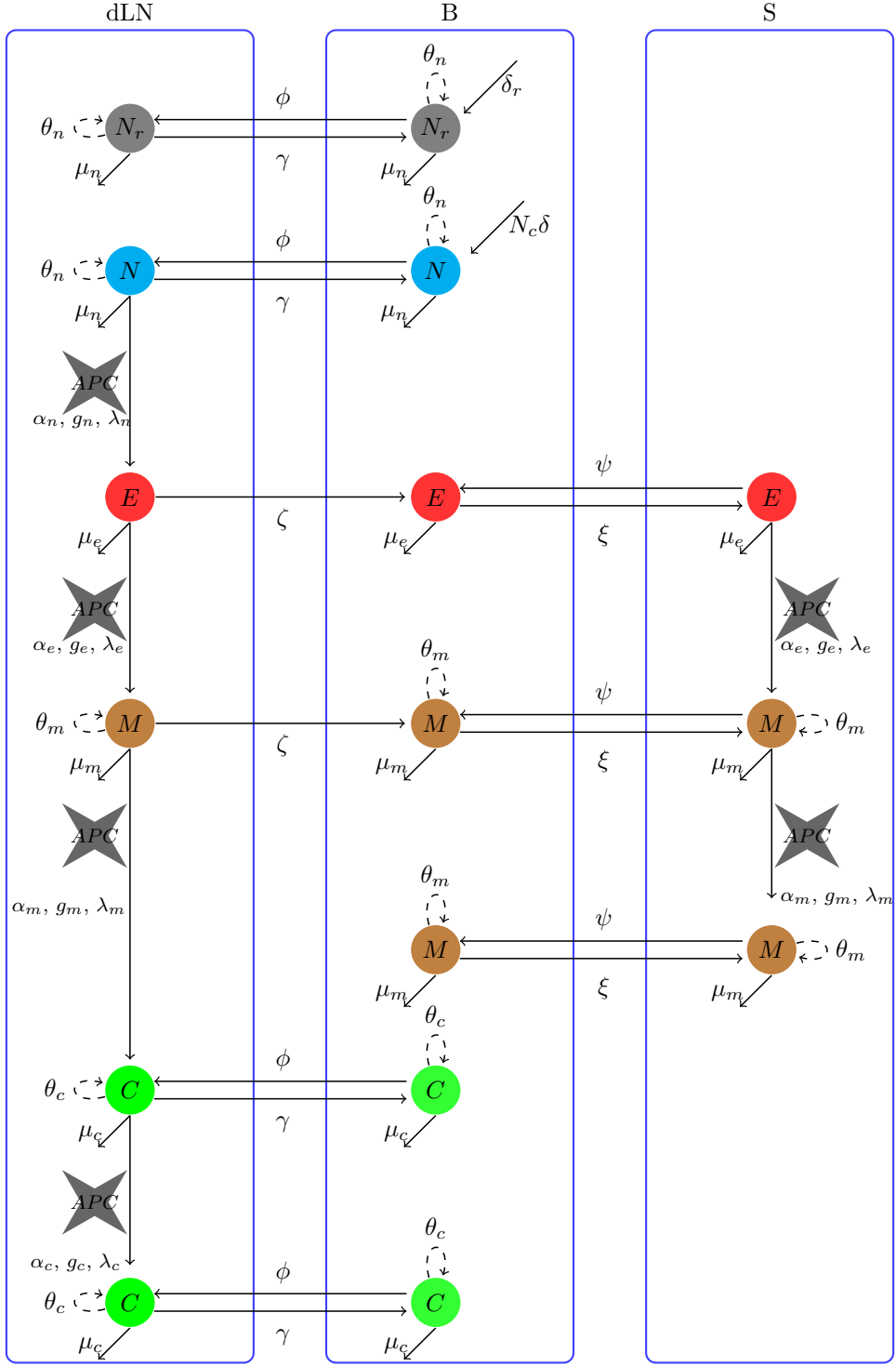
\begin{figure}[htp!]
\begin{center}
\begin{tikzpicture} 
[block/.style   ={rectangle, draw=blue, thick,
                    text width=2cm, rounded corners, opacity=0.75,font=\small,
                    minimum height=19cm,minimum width=3.5cm},
cell/.style={circle,color=black,thick, minimum size=7mm, inner sep= 0pt},
line/.style    ={draw, -latex',shorten >=2pt, line width=0.2mm}]
 \matrix [column sep=10mm, row sep=20mm]
 {
 \node [cell, fill=gray] (nr)[yshift=81mm] {$N_r$};
 \node [cell, fill=cyan] (naive)[below=of nr, yshift=-3mm] {$N$};
 \node[star, color=black,fill=black!60!white,star points=4,star point
ratio=0.3] (APC3)[below of =naive,xshift=-5mm, yshift=-6mm,font=\footnotesize]{$APC$};

 \node [cell, fill=red!80] (E)[below =of naive,yshift=-15mm] {$E$};
 \node[star, color=black,fill=black!60!white,star points=4,star point
ratio=0.3] (APC3)[below of =E,xshift=-5mm, yshift=-6mm,font=\footnotesize]{$APC$};

 \node [cell, fill=brown!100] (EM) [below =of E,yshift=-15mm]{$M$};
 \node[star, color=black,fill=black!60!white,star points=4,star point
ratio=0.3] (APC3)[below of =EM,xshift=-5mm, yshift=-6mm,font=\footnotesize]{$APC$};

 \node [cell, fill=green] (CM)[below =of EM,yshift=-35mm]{$C$};
 \node[star, color=black,fill=black!60!white,star points=4,star point
ratio=0.3] (APC4)[below of =CM,xshift=-5mm, yshift=-6mm,font=\footnotesize]{$APC$};
\node [cell, fill=green] (CMis)[below =of CM,yshift=-13mm]{$C$};
 \node [block] (ln)[label=dLN]{}; &

 \node [cell, fill=gray] (nrb)[right of =nr, xshift=-12mm] {$N_r$};
 \node [cell, fill=cyan] (naiveb)[right of =naive, xshift=-12mm] {$N$};
 \node [cell, fill=red!80] (Eb)[below =of naiveb,yshift=-15mm] {$E$};
 \node [cell, fill=brown!100] (EMb) [below =of Eb,yshift=-15mm]{$M$};
 \node [cell, fill=brown!100] (EMb2) [below =of EMb,yshift=-15mm]{$M$};

 \node [cell, fill=green!80] (CMb)[below =of EMb,yshift=-35mm]{$C$};
 \node [cell, fill=green!80] (CMb2)[below =of CMb,yshift=-13mm]{$C$};

 \node [block] (blood)[label=B]{}; &
 
 \node [cell, fill=red!80] (Es) [right =of Eb,xshift=-15mm]{$E$};
 \node[star, color=black,fill=black!60!white,star points=4,star point
ratio=0.3] (APC3)[below of =Es,xshift=5mm, yshift=-6mm,font=\footnotesize]{$APC$};

 \node [cell, fill=brown!100] (EMs) [right =of EMb,xshift=-15mm]{$M$};
 \node[star, color=black,fill=black!60!white,star points=4,star point
ratio=0.3] (APC3)[below of =EMs,xshift=5mm, yshift=-6mm,font=\footnotesize]{$APC$};
\node [cell, fill=brown!100] (EMis) [below=of EMs,yshift=-15mm]{$M$};
\node [block] (skin)[label=S]{}; &

\\
  };
 \begin{scope}[every path/.style=line]

	\draw[->](naive.270) to node [auto,xshift=-15mm,yshift=-5mm,font=\footnotesize] {$\alpha_n$, $g_n$, $\lambda_n$} (E.90);
 \draw[->](E.270) to node [auto,xshift=-15mm,yshift=-5mm,font=\footnotesize] {$\alpha_e$, $g_e$, $\lambda_e$} (EM.90);
	\draw[->](EM.270) to node [auto,xshift=-18mm,yshift=-0mm,font=\footnotesize] {$\alpha_m$, $g_m$, $\lambda_m$} (CM.90);
	\draw[->](Es.270) to node [auto,xshift=0mm,yshift=-5mm,font=\footnotesize] {$\alpha_e$, $g_e$, $\lambda_e$} (EMs.90);
	
   \path [->] (CM.south)--node[left,yshift=-10mm,font=\footnotesize]{$\alpha_c$, $g_c$, $\lambda_c$}++(0,-2.2);
	
   \path[dashed,yshift=5mm] (naive) edge  [loop left] node {$\theta_n$} ();
	\path[dashed,yshift=5mm] (CM) edge  [loop left] node {$\theta_c$} ();
	\path[dashed,yshift=5mm] (EM) edge  [loop left] node {$\theta_m$} ();
	\path[dashed,yshift=5mm] (naiveb) edge  [loop above] node {$\theta_n$} ();
	\path[dashed,yshift=5mm] (CMb) edge  [loop above] node {$\theta_c$} ();
	\path[dashed,yshift=5mm] (EMb) edge  [loop above] node {$\theta_m$} ();
	\path[dashed,yshift=5mm] (CMb2) edge  [loop above] node {$\theta_c$} ();
	\path[dashed,yshift=5mm] (EMb2) edge  [loop above] node {$\theta_m$} ();
  \path[dashed,yshift=5mm] (CMis) edge  [loop left] node {$\theta_c$} ();
	\path[dashed,yshift=5mm] (EMis) edge  [loop right] node {$\theta_m$} ();

	\path[dashed,yshift=5mm] (EMs) edge  [loop right] node {$\theta_m$} ();

  \path [->] (EMs.south)--node[ auto,xshift=0mm,yshift=-10mm,font=\footnotesize]{$\alpha_m$, $g_m$, $\lambda_m$}++(0,-2.2);
 
	\path [->] (naive.south)--node[left]{$\mu_n$}++(-0.5,-0.5);
	\path [->] (CM.south)--node[left]{$\mu_c$}++(-0.5,-0.5);
	\path [->] (EM.south)--node[left]{$\mu_m$}++(-0.5,-0.5);
	\path [->] (E.south)--node[left]{$\mu_e$}++(-0.5,-0.5);
	\path [->] (naiveb.south)--node[left]{$\mu_n$}++(-0.5,-0.5);
	\path [->] (CMb.south)--node[left]{$\mu_c$}++(-0.5,-0.5);
	\path [->] (CMb2.south)--node[left]{$\mu_c$}++(-0.5,-0.5);
		\path [->] (CMis.south)--node[left]{$\mu_c$}++(-0.5,-0.5);

	\path [->] (EMb.south)--node[left]{$\mu_m$}++(-0.5,-0.5);
		\path [->] (EMb2.south)--node[left]{$\mu_m$}++(-0.5,-0.5);
		\path [->] (EMis.south)--node[left]{$\mu_m$}++(-0.5,-0.5);

	\path [->] (Eb.south)--node[left]{$\mu_e$}++(-0.5,-0.5);
	\path [->] (EMs.south)--node[left]{$\mu_m$}++(-0.5,-0.5);
	\path [->] (Es.south)--node[left]{$\mu_e$}++(-0.5,-0.5); 

  \draw[->](naive.340) to node [auto,yshift=-6mm] {$\gamma$} (naiveb.200);
	\draw[<-](naive.20) to node [auto,yshift=0.25mm] {$\phi$} (naiveb.160);	
	\draw[->] (CM.340) to node [auto,yshift=-6mm] {$\gamma$} (CMb.200);
	\draw[<-](CM.20) to node [auto,yshift=0.25mm] {$\phi$} (CMb.160);
	\draw[->] (EM.360) to node [auto,yshift=-6mm] {$\zeta$} (EMb.180);
	\draw[->](EMb.340) to node [auto,yshift=-7mm] {$\xi$} (EMs.200);
	\draw[<-](EMb.20) to node [auto,yshift=0.25mm] {$\psi$} (EMs.160);
	\draw[->] (E.360) to node [auto,yshift=-6mm] {$\zeta$} (Eb.180);
	\draw[->](Eb.340) to node [auto,yshift=-7mm] {$\xi$} (Es.200);
	\draw[<-](Eb.20) to node [auto,yshift=0.5mm] {$\psi$} (Es.160);

\draw[->] (CMis.340) to node [auto,yshift=-6mm] {$\gamma$} (CMb2.200);
	\draw[<-](CMis.20) to node [auto,yshift=0.25mm] {$\phi$} (CMb2.160);
		\draw[->](EMb2.340) to node [auto,yshift=-7mm] {$\xi$} (EMis.200);
	\draw[<-](EMb2.20) to node [auto,yshift=0.25mm] {$\psi$} (EMis.160);

		\path [->] (nr.south)--node[left]{$\mu_n$}++(-0.5,-0.5);

		\path [->] (nrb.south)--node[left]{$\mu_n$}++(-0.5,-0.5);
   \path[dashed,yshift=5mm] (nr) edge  [loop left] node {$\theta_n$} ();
	\path[dashed,yshift=5mm] (nrb) edge  [loop above] node {$\theta_n$} ();
	
	 \draw[->](nr.340) to node [auto,yshift=-6mm] {$\gamma$} (nrb.200);
	\draw[<-](nr.20) to node [auto,yshift=0.25mm] {$\phi$} (nrb.160);	
		\path [<-] ([xshift=2mm,yshift=-12mm]blood.north)--node[right]{ $\delta_r$}++(0.8,0.8);
	\path [<-] ([xshift=3mm,yshift=-32mm]blood.north)--node[right]{ $ N_c \delta$}++(0.8,0.8);

\end{scope} 
\end{tikzpicture}
\end{center}
\caption{Mathematical model of CD8\plus\ T~cell dynamics during an
  immune response (increasing potential model). The model considers
  naive, central memory, effector memory and effector CD8\plus\
  T~cells and three spatial compartments.  The mathematical model of
  the dynamics during an immune response is based on the assumption
  that a T~cell interaction with an APC triggers a programme of
  division steps that leads to differentiation (increasing potential
  model): from naive to E, to EM and to central memory (CM) T~cells.
  Vertical arrows describe the differentiation programme (linked to
  division) set by APC interactions.  As described in the model of
  homeostasis, (antigen-specific and non-specific) N, CM and EM
  CD8\plus\ T~cells divide (oval dashed arrows)
  ($\theta_n, \theta_c, \theta_m$), die ($\mu_n, \mu_c, \mu_m$) and
  migrate.  Naive cells have the same migratory behaviour as CM cells
  and effector cells have the same migratory behaviour as EM cells.
  Cell subtype specific rates are labelled by the subscripts $n$, $c$,
  $m$, and $e$ for naive, central memory, effector memory, and
  effector cells, respectively.
  We assume no bystander activation during an immune response and thus,
non-specific  naive CD8\plus\ T~cells are described by the homeostasis model.}
\label{cd8-immune-response-ipm}
\end{figure}

The differentiation route in the IPM takes naive T~cells cells to
effector, followed by effector memory and central memory, with further
antigen stimulation at each differentiation stage (see
Fig.~\ref{ipm}).  The mathematical model considers naive
(antigen-specific and non-specific) T~cells, as well as effector,
effector memory and central memory cells in three spatial
compartments.  The events for any given cell type during an immune
response are, as in the DPM case (see Section~\ref{math-DPM}):
encounter with an APC, which starts a programme of division-linked
differentiation steps (see Fig.~\ref{div-linked-diff-ipm}), death, and
migration. Non-specific naive T~cells are assumed to behave according
to the homeostasis model~\cite{ahmed2011insights,gerlach2013heterogeneous} (see Section~\ref{method-model-homeostasis}).
The IPM is depicted in Fig.~\ref{cd8-immune-response-ipm}).  As
discussed above, we will assume that $N_c$ different and independent TCR clonotypes
are taking part in the immune response, and that all of them can be
described with identical rates of activation, proliferation, death and
migration, as well as thymic
export~\cite{antia1998models,antia2003models,gerlach2013heterogeneous}.  In this sense and for
the mathematical models considered in this manuscript, $N_c$ is the
parameter that encodes how broad the CD8\plus\ immune response is, as
it quantifies how many different TCR clonotypes are driven to
proliferate and differentiate in response to skin-delivered antigen.

%%%%%%%%%%%%

%%%%%%%%%%%%%%%%%%%%

%%%%%%%%%%%%%%%%%%%%%%%%%%%%

%%%%%%%%%%

\subsection{Model parameters and computational algorithm}
\label{math-parameters-computational}

We have made use of the published literature to obtain parameter
estimates for the number of cells, carrying capacities, division
rates, death rates and thymic export rates for the
different CD8\plus\ T~cell populations considered in the DPM and IPM.
Full details can be found in Section~\ref{method-model-parameters},
and, in this section, we provide a brief summary of our literature
review.

Model parameters can be classified into two different types, according
to whether they are fixed (see Table~\ref{table-parameters-fixed}) or
included in the Bayesian learning when we carry out model
calibration (see Table~\ref{table-parameters-bayesian}). Fixed
parameters include the carrying capacities, division
rates, death rates and thymic export rates. These parameters are
assumed to have known values and are set at the same value for both
the DPM and IPM.

\begin{table}[h!]
\begin{center}
\begin{tabular}{| c || l | c |}
\hline
{\bf Parameter} & {\bf Definition} & {\bf Value}
\\
\hline
\hline
$\kappa_n$ & Carrying capacity  per clonotype of naive cells &  1,940 cells
\\
\hline
 & Carrying capacity  of specific naive cells &  $N_c \times \kappa_n$
\\
\hline
$\kappa_r$ & Carrying capacity  of non-specific naive cells &   $(N_R-N_c) \times \kappa_n$
\\
\hline
$\kappa_c$ & Carrying capacity  per clonotype of specific central memory cells & 87,750 cells
\\
\hline
$\kappa_m$ & Carrying capacity  per clonotype  of specific effector memory cells & 87,750 cells
\\
\hline
\hline
$\theta_n$ & Division rate of naive cells & 5.63$\times10^{-4}$ per day
\\
\hline
$\theta_c$  & Division rate of central memory cells & 6.49$\times10^{-3}$ per day
\\
\hline
$\theta_m$ & Division rate of effector memory cells & 6.49$\times10^{-3}$ per day
\\
\hline
\hline
$\mu_n$ & Death rate of naive cells & 4.46$\times10^{-4}$ per day
\\
\hline
$\mu_c$ & Death rate of central memory cells & 3.67$\times10^{-3}$ per day
\\
\hline
$\mu_m$ & Death rate of effector memory cells & 3.67$\times10^{-3}$ per day
\\
\hline
$\mu_e$ & Death rate of effector cells & 3.57$\times10^{-2}$ per day
\\
\hline
\hline
$\delta$ & Thymic export rate per clonotype & 0.12 cells per day
\\
\hline
 & Thymic export rate of specific cells & $N_c \times \delta$
\\
\hline
$\delta_r$  & Thymic export rate of non-specific cells & $(N_R-N_c) \times \delta$
\\
\hline
\end{tabular}
\caption{Summary of parameters that are considered fixed in the mathematical models. 
Details of  parameter estimation from published literature has been
 provided in Section~\ref{method-model-parameters}.}
\label{table-parameters-fixed}
\end{center}
\end{table}

The rest of the parameters, which are to be included in the Bayesian
analysis, are those related to the immune response, such as the number
of divisions in the differentiation programme ($g$), the time to first
division ($1/\alpha$), the time to subsequent divisions ($1/\lambda$),
the number of specific clonotypes involved in the response ($N_c$),
the duration of the immune response $\tau_E$, and all migration rates.
These parameters will be part of the model calibration based upon YFV
vaccine data from~Ref.~\cite{akondy2009yellow} (see Figure~3B in
Ref.~\cite{akondy2009yellow}).  As described in
Section~\ref{method-calibration}, given the prior distributions
provided in Table~\ref{table-parameters-bayesian}, we will obtain
posterior distributions for this subset of parameters given a
mathematical model and the data.  We use uniform distributions for
these parameters because we could not say anything more about the
parameters prior to the analysis and we did not want to bias the
analysis.  If we had access to further expertise on the parameters, we
could have employed formal expert knowledge elicitation techniques to
refine the prior distributions by incorporating expert judgements on
the parameter values~\cite{craig1998,ohagan2006}.
  
\begin{table}[h!]
\begin{center}
\begin{tabular}{| c || l | c |}
\hline
{\bf Parameter} (unit) & {\bf Definition} & {\bf Prior distribution}  (unit)
\\
\hline
\hline
$g_n,g_c,g_m,g_e$ & Number of generations & $g_n,g_c,g_m,g_e$ $\sim$ $\mathcal{U}\{1,2,\dots,11\}$
\\ 
\hline
\hline
$\alpha_n,\alpha_c,\alpha_m,\alpha_e$ 
(per day) &  Rate of first division & $\frac{1}{\alpha_n}, \frac{1}{\alpha_c}, \frac{1}{\alpha_m},
\frac{1}{\alpha_e} \sim \mathcal{U} (0.25,5.0)$ 
(day)
\\
\hline
\hline
$\lambda_n,\lambda_c,\lambda_m,\lambda_e$ (per day)
& Rate of subsequent divisions & $\frac{1}{\lambda_n},\frac{1}{\lambda_c},\frac{1}{\lambda_m},
\frac{1}{\lambda_e} \sim \mathcal{U} (0.25,5.0) $ 
(day)
\\ 
\hline
\hline
$\gamma$ (per day)  & Migration rate from dLN to B for N and CM cells
& $\frac{1}{\gamma} \sim \mathcal{U} (6.94 \times 10^{-4},10.0) $ (day)
\\ 
\hline
$\phi$ (per day) & Migration rate from B to  dLN for N and CM cells & 
$\frac{1}{\phi} \sim \mathcal{U} (6.94 \times 10^{-4},10.0) $ (day)
\\ 
\hline
$\zeta$ (per day)& Migration rate from dLN to B for EM and E cells 
& $\frac{1}{\zeta} \sim \mathcal{U} (6.94 \times 10^{-4},10.0) $ (day)
\\
\hline
$ \xi$ (per day) & Migration rate from B to S for EM and E cells
& $\frac{1}{\xi} \sim \mathcal{U} (6.94 \times 10^{-4},10.0)$ (day)
\\
\hline
$\psi$ (per day)&  Migration time from S to B for EM and E cells
& $\frac{1}{\psi} \sim \mathcal{U} (6.94 \times 10^{-4},10.0) $ (day)
\\ 
\hline
\hline
$N_c$ & \rr Number of TCR clonotypes & 
$N_c \sim \mathcal{U} \{1,2,\dots,N_c^{max}\}$\\ 
\hline
\hline
$\tau_E$ (day)& Duration of immune challenge & $\tau_E$ $\sim$ $\mathcal{U} (5,60)$ (day)
\\ 
\hline
\hline
$\sigma^2$ & Variance in the error term & $\sigma^2$ $\sim$ $\mathcal{IG} (3,1)$
\\
\hline
\end{tabular}
\caption{Summary of parameters and their prior distributions  used in the Bayesian analysis for
the calibration of the DP and IP models. 
Details of the estimation of these parameters from the literature 
has been provided in Section~\ref{method-model-parameters}.
We have chosen $N_c^{max}$ = 10$^5$ (see Section~\ref{method-number-clonotypes}).
}
\label{table-parameters-bayesian}
\end{center}
\end{table}

Once a choice of model parameters has been made, the first step is to
provide a solution to the system of equations of the
DPM~\eqref{dpm-naive-dln}-\eqref{dpm-effector-s} for the
antigen-specific cells, as well as to find a solution to
equations~\eqref{non-specific-naive-dLN}
and~\eqref{non-specific-naive-B} for the non-specific naive T~cell
populations.  The second step is to choose initial conditions for all
the cell types in the three spatial compartments. The choice of
initial conditions depends on the immune scenario under consideration
and a comprehensive discussion has been provided in
Section~\ref{method-computational-algorithm}.

Finally, given a choice of parameters and initial conditions, the ODEs
will be solved using a 4th order Runge-Kutta method implemented using
\emph{Python}.  Thus, parameters, initial conditions and the numerical
solver implemented in \emph{Python} constitute the computational
algorithm that will be referred to as the simulator of the
mathematical model. For the case of the DPM model, the simulator will
numerically integrate the equations described in
Section~\ref{method-model-immune-response-DPM}, and, for the IPM, it
will integrate those equations in
Section~\ref{method-model-immune-response-IPM}.
		
%%%%%%%%%%%%%%%%%%%%%%%
				
\subsection{Sensitivity analysis}
\label{maths-sensitivity}		
		
We used global sensitivity analysis methods~\cite{saltelli1} to check
that the model parameters were having the expected effects on the
various model outputs. To achieve this, we have used probabilistic
sensitivity analysis techniques and produced main effect plots using
the methods of Ref.~\cite{oakley1}. These main effect plots show the
average output response as we vary each parameter in turn. We have
also calculated main effect and total effect indices (that provide a
measure of each parameter's importance in terms of contribution to
output uncertainty) in order to establish which inputs were important
for a given model output. This additional analysis has allowed us to
focus our efforts when assessing input parameter uncertainty (using
the methods described in Refs.~\cite{sobol1,oakley1}).  In order to
run these analyses, we needed to specify plausible ranges for each of
the input parameters, and these are given in
Table~\ref{table-parameters-bayesian}.

In our sensitivity analyses, we found that both models had a level of
redundancy in that some parameters had only a limited effect on model
output over the plausible ranges. For the DPM, we found that five of
the model parameters had relatively limited effects on the outputs
corresponding to the data we had available to calibrate the model.
Similarly, for the IPM, there were six parameters that had negligible
effect when varied. This means that we could not hope to reduce our
uncertainty about these parameters in the light of the data and that
carefully modelling of our prior beliefs was not
required. Further details of the sensitivity analyses have been
provided in Section~\ref{method-sensitivity-analysis}.

%%%%%%%%%%

\subsection{Yellow fever virus vaccine human data and model calibration}
\label{model-calibration-data}

%%%%%%%%%%

\subsubsection{Yellow fever virus vaccine human data}
\label{stats-YFV-data}

We make use of clinical data from the kinetics of virus-specific
CD8\plus\ T~cell responses.  Healthy volunteers (21-32 years of age)
were vaccinated with the yellow fever virus (YFV) vaccine 17D (YF-17D)
as described in detail in Ref.~\cite{akondy2009yellow}. Volunteers
were vaccinated with 0.5 mL of the YF-17D vaccine and virus-specific
CD8\plus\ T~cells were selected on days 3, 11, 14, 30 and 90 with
tetramer staining (HLA-A2 restricted CD8\plus\ T~cells specific for
the NS4B epitope of YFV).  We have made use of the fraction of
specific CD8\plus\ T~cells out of the total CD8\plus\ T~cell population
at days 11, 14, 30 and 90 post-vaccination for each volunteer (see
Ref.~\cite{akondy2009yellow}, Figure~3B).

 %%%%%%%%%%

\subsubsection{Model calibration}
\label{maths-calibration}

In order to calibrate the models in the light of the yellow fever data
described in Section~\ref{stats-YFV-data}, we have employed a Bayesian
methodology similar to the ones described in
Refs.~\cite{gelman1,girolami1,robey}.  The general calibration scheme
is as follows: we start with prior distributions for each input
parameter that encode uncertainty about the true values of the inputs,
then we update the prior distributions in the light of the data by
giving more weight to the input parameters that would allow the model
to reproduce the data.  Prior distributions have been described in
Section~\ref{math-parameters-computational} and detailed in
Table~\ref{table-parameters-bayesian}.  Given the updated
distributions (called the posterior distributions), we are able to
investigate relationships between the parameters and the type of model
behaviours that are supported.  There are two substantial technical
challenges to overcome when updating the prior distributions in the
light of data: the first is producing a statistical framework for
linking the data to the model outputs (that is, specifying a
likelihood function), and the second is the normalisation of the
updated probability distributions.  To circumnavigate these problems,
we used an approximate Bayesian computation algorithm that allows us
to use the simulator directly, with minimal formal modelling of the
statistical link between the model and data.  The simulator and the
ABC algorithm we have employed has been fully described in
Section~\ref{method-computational-algorithm} and in
Section~\ref{method-calibration}, respectively.  Again, using a
Bayesian framework, we can also compare models (in our case the DPM
and the IPM) by asking which model has the highest posterior
probability.  The posterior probabilities in this case are found
through a combination of prior beliefs about which model is best, and
considerations of which model the data supports.  For the prior
probabilities, we opted not to favour one model over the other, so we
set the DPM and the IPM to be equally likely.  Relative adjustments to
the prior probabilities were made with respect to how well the data
could be reconstructed using each model (DPM or IPM).  Again, this
adjustment is achieved by making use of an approximate Bayesian
computation algorithm that is presented in
Section~\ref{method-calibration}.  Specific details concerning
parameter uncertainty and our calibration methods have been provided
in Section~\ref{method-calibration}.

%%%%%%%%%%%%%%%%%%%%%%%%%%%%%%%%%%%%%%%%%%%%%%%%

\section{Results}
\label{results}

%%%%%%%%%%%%%%%%%%%%%%%%%%%%%%%%%%%%%%%%%%%%%%%%%%%%

\subsection{Model calibration in the light of the data}
\label{results-calibration}

%%%%%%%%%%

\subsubsection{Model calibration of the decreasing potential model}
\label{calibration-dpm}

Table~\ref{summary-dpm} provides summary statistics of the posterior
parameter distributions obtained after model calibration of the
decreasing potential model with human YFV vaccination data for
$N_c^{max}=10^5$ (see Table~\ref{summary-dpm}).
    
\begin{table}[h!]
\begin{center}
\begin{tabular}{| c ||   c |  c |  c |}
\hline
      $N_c^{max}$     & \multicolumn{3}{l|}{$10^5$ }       \\
       \hline
       \hline
    {\bf Parameter} 
    & 2.5\% & 50\%  (median) & 97.5\%
    %\multicolumn{2}{l|}{95\% CI} 
    \\
    \hline
           \hline
    $g_n$    &  1 & 3 & 10 \\\hline
    $g_c$    &  1 & 3 & 11 \\\hline
    $g_m$    &  1 & 4 & 11 \\\hline
           \hline
    $\alpha_n$ &  0.21 & 0.36 & 2.48 \\\hline
    $\alpha_c$ & 0.20 & 0.35 & 2.20 \\\hline
    $\alpha_m$ &  0.20 & 0.37 & 2.48 \\\hline
           \hline
    $\lambda_n$ &  0.21 & 0.37 & 2.59 \\\hline
    $\lambda_c$ &  0.20 & 0.35 & 2.42 \\\hline
    $\lambda_m$ &  0.20 & 0.34 & 2.46 \\\hline
           \hline
    $\gamma$ & 0.10 & 0.27 & 7.87 \\\hline
    $\phi$   &  0.10 & 0.16 & 1.30 \\\hline
    $\zeta$   & 0.10 & 0.19 & 4.41 \\\hline
    $\psi$  & 0.10 & 0.18 & 2.56 \\\hline
    $\xi$     & 0.10 & 0.22 & 7.47 \\\hline
           \hline
    $N_c$    & 3,378.65 & 46,342.00 & 97,043.23 \\\hline
           \hline
    $\tau_E$  & 8.70 & 21.56 & 56.72 \\\hline
       \hline
    $\sigma^2$ &   0.17 & 0.43 & 1.18 \\\hline
    \end{tabular}
    \caption{Summary statistics of the posterior parameter distributions 
    obtained after model calibration of the decreasing potential model with human YFV vaccination
    data for  $N_c^{max}=10^5$. For all parameters, median values and 95\% credible intervals (CI) are shown.}
\label{summary-dpm}
\end{center}
\end{table}

Given the data, we learn the most about the following parameters of the decreasing potential model:

\begin{itemize}

\item the number of generations for the three cell types (naive,
  central memory and effector memory): the data suggest that the
  number of generations for each cell type, ($g_n, g_c, g_m$), is
  likely to be below six in agreement with a previous mathematical
  study of the same data, that estimated that the observed expansion
  of the CD8\plus\ T~cell response was achieved in fewer than nine
  divisions~\cite{le2014mathematical},

\item the uncertainty in the migration rates $\gamma$ and $\phi$ has been reduced a little in that higher values of $\gamma$ and lower values of $\phi$ are supported by the data, and

\item the duration of the immune challenge: $\tau_E$ is distributed
  around 20 days with only a 5\% chance of $\tau_E<10$ days.
  
\end{itemize}

\begin{figure}[h!]
\begin{subfigure}[b]{0.32\textwidth}
\includegraphics[page=1,width=\textwidth]{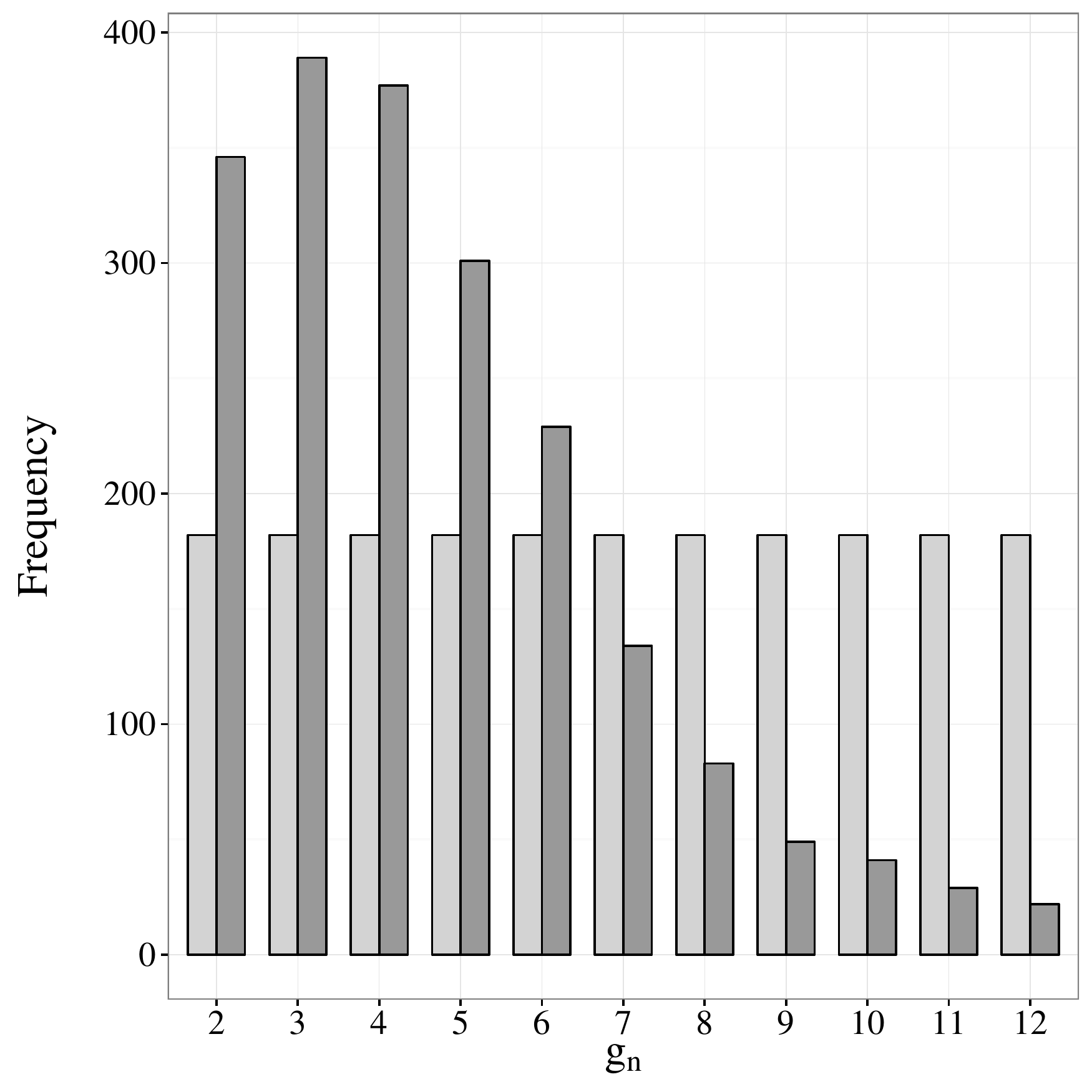}
\end{subfigure}
~
\begin{subfigure}[b]{0.32\textwidth}
\includegraphics[page=2,width=\textwidth]{YFpostgraphsDPM13e5}
\end{subfigure}
~
\begin{subfigure}[b]{0.32\textwidth}
\includegraphics[page=3,width=\textwidth]{YFpostgraphsDPM13e5}
\end{subfigure}

\begin{subfigure}[b]{0.32\textwidth}
\includegraphics[page=10,width=\textwidth]{YFpostgraphsDPM13e5}
\end{subfigure}
~
\begin{subfigure}[b]{0.32\textwidth}
\includegraphics[page=11,width=\textwidth]{YFpostgraphsDPM13e5}
\end{subfigure}
~
\begin{subfigure}[b]{0.32\textwidth}
\includegraphics[page=16,width=\textwidth]{YFpostgraphsDPM13e5}
\end{subfigure}
\caption{Bar charts and histograms of posterior samples (dark grey) 
plotted alongside prior samples (light grey) for six input parameters of the DPM ($g_n, g_c, g_m, 1/\gamma, 1/\phi,
\tau_E$).
}
\label{plot.DPMres}
\end{figure}

The posterior distributions of $g_n, g_c, g_m,1/\gamma, 1/\phi$ and  $\tau_E$
 are given in Figure~\ref{plot.DPMres}.

\begin{figure}[h!]
\centering
\includegraphics[page=18,scale=0.35]{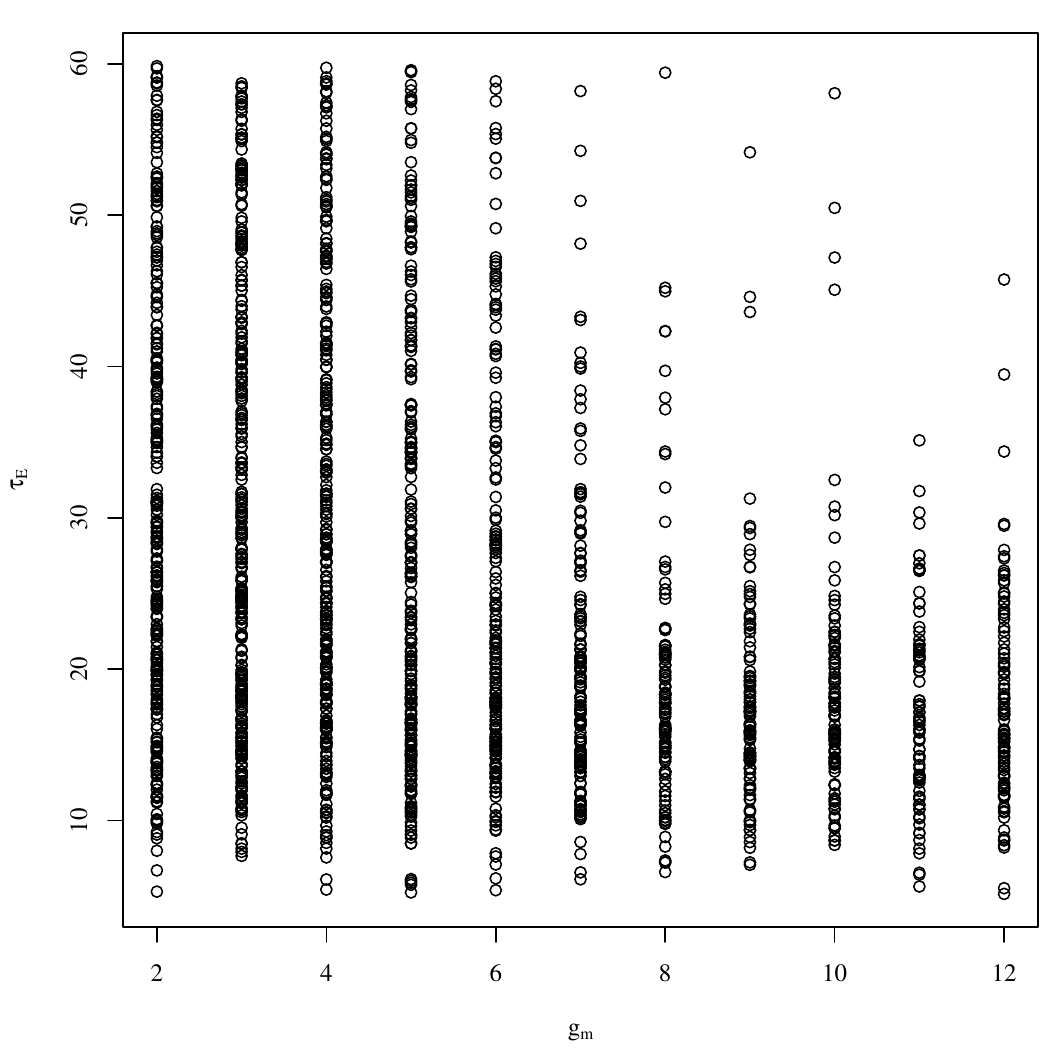}
\caption{Scatter plot of posterior samples for input parameters $g_m$ and $\tau_E$ in the DPM.
}
\label{plot.DPMscat}
\end{figure}

In the posterior distributions, we find a strong relationship between
$g_m$ and $\tau_E$.  It is clear from the posterior distribution that,
if $g_m$ is equal to or greater than seven, then $\tau_E$ is generally
restricted to being at most 30 days, if we want to be able to
reproduce the YFV vaccination data.  This can be seen in the scatter
plot of the posterior samples comparing $g_m$ and $\tau_E$ (see
Figure~\ref{plot.DPMscat}). On the other hand, for lower values of
$g_m$, $\tau_E$ can be any value over seven days and the data can
still be reproduced. This effect is having a relatively small impact
on the results, because there is only a 26\% chance of $g_m$ being
equal to or greater than seven from our posterior distribution.

%%%%%%%%%%%%%%%%%%%%%%%%

\subsubsection{Model calibration of the increasing potential model}
\label{calibration-ipm}

The table below provides a summary statistics of the posterior
parameter distributions obtained after model calibration of the
increasing potential model with human YFV vaccination data for
$N_c^{max}=10^5$ (see Table~\ref{summary-ipm}).
    
\begin{table}[h!]
\begin{center}
\begin{tabular}{| c ||  c |  c |  c |}
\hline
      $N_c^{max}$       & \multicolumn{3}{l|}{$10^5$ }       \\
       \hline
       \hline
    {\bf Parameter} 
    %& median  &\multicolumn{2}{l|}{95\% CI}
     & 2.5\% & 50\%  (median) & 97.5\%
     \\
    \hline
    \hline
	$g_n$     & 1 & 5 & 11 \\\hline
    $g_e$    &  1 & 5 & 11 \\\hline
    $g_m$    & 1 & 6 & 11\\ \hline
    $g_c$    &  1 & 6 & 11 \\\hline
    \hline
    $\alpha_n$ &  0.20 & 0.37 & 2.92 \\\hline
    $\alpha_e$ & 0.20 & 0.34 & 2.50 \\\hline
    $\alpha_m$ &  0.21 & 0.36 & 2.68 \\\hline
        $\alpha_c$ &  0.20 & 0.37 & 2.58 \\\hline
           \hline
    $\lambda_n$ &  0.21 & 0.36 & 2.39 \\\hline
    $\lambda_e$ &  0.20 & 0.33 & 2.429 \\\hline
    $\lambda_m$ &  0.20 & 0.38 & 2.17 \\\hline
        $\lambda_c$ &  0.20 & 0.38 & 2.55 \\\hline
           \hline
    $\gamma$ & 0.10 & 0.21 & 5.89 \\\hline
    $\phi$   &  0.10 & 0.18 & 2.06 \\\hline
    $\zeta$   & 0.10 & 0.20 & 5.48 \\\hline
    $\psi$  & 0.10 & 0.22 & 6.80 \\\hline
    $\xi$     & 0.10 & 0.19 & 2.65 \\\hline
           \hline
    $N_c$    & 948.73 & 43,208.00 & 97,006.47 \\\hline
           \hline
    $\tau_E$  & 7.71 & 15.01 & 43.16 \\\hline
       \hline
    $\sigma^2$ &   0.2 & 0.46 & 1.15 \\\hline
     \end{tabular}
    \caption{Summary statistics of the posterior parameter distributions 
    obtained after model calibration of the increasing potential model with human YFV vaccination
    data for $N_c^{max}=10^5$. For all parameters, median values and 
    95\% credible intervals (CI) are shown.}
\label{summary-ipm}
\end{center}
\end{table}

Given the data, we learn the most about the following parameters of the increasing potential model:

\begin{itemize}

\item 
the model supports lower values of $g_n$ and $g_e$,
 
\item the rate of subsequent divisions per day for effector memory
  cells, $\lambda_m$, is likely to be relatively low (that is, less
  than 0.5 per day), and, for the rate of subsequent divisions per day
  for naive cells, $\lambda_n$, values of 0.35 are favoured by
  the data, and
 
\item
 the duration of the  immune challenge: 
 $\tau_E$ is distributed around 14 days with only  16\% of $\tau_E>20$ days.
 
\end{itemize}

\begin{figure}[h!]
\begin{subfigure}[b]{0.32\textwidth}
\includegraphics[page=1,width=\textwidth]{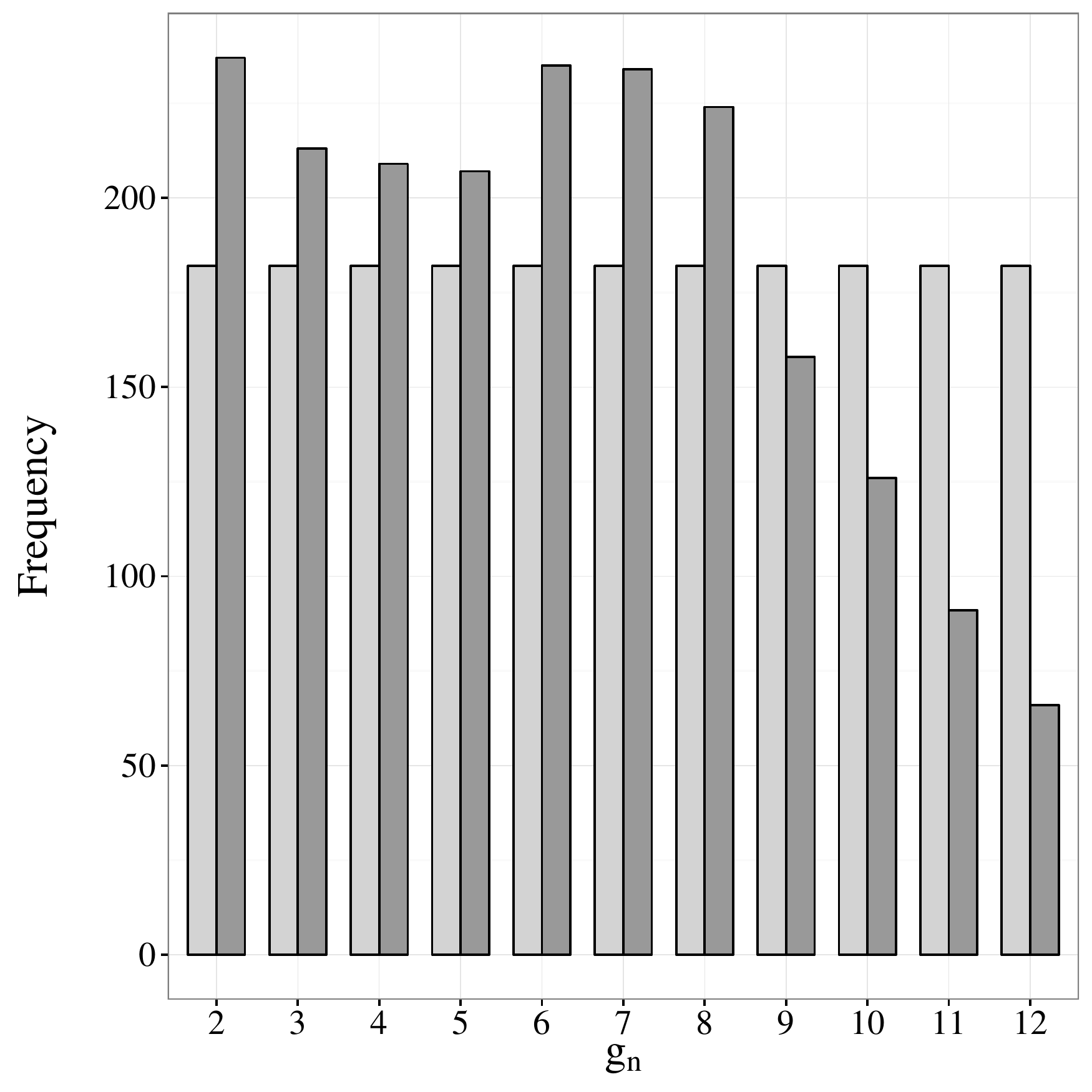}
\end{subfigure}
~
\begin{subfigure}[b]{0.32\textwidth}
\includegraphics[page=2,width=\textwidth]{YFpostgraphsIPM13e5}
\end{subfigure}
~
\begin{subfigure}[b]{0.32\textwidth}
\includegraphics[page=9,width=\textwidth]{YFpostgraphsIPM13e5}
\end{subfigure}

\begin{subfigure}[b]{0.32\textwidth}
\includegraphics[page=11,width=\textwidth]{YFpostgraphsIPM13e5}
\end{subfigure}
~
\begin{subfigure}[b]{0.32\textwidth}
\includegraphics[page=18,width=\textwidth]{YFpostgraphsIPM13e5}
\end{subfigure}
~
\begin{subfigure}[b]{0.32\textwidth}
\includegraphics[page=19,width=\textwidth]{YFpostgraphsIPM13e5}
\end{subfigure}
\caption{Bar charts and histograms of posterior samples (dark grey) plotted alongside prior samples (light grey) for six input parameters of the IPM
($g_n, g_e, 1/\lambda_n, 1/\lambda_m, N_c,
\tau_E$).
}
\label{plot.IPMres}
\end{figure}

Overall, the IPM is more readily able to match the data when $N_c$ is
smaller than for the DPM case.  The posterior samples for six
parameters of the IPM are given in Figure~\ref{plot.IPMres}.

%%%%%%%%%%%%%%%%%%%%%%%%%%%%%%%%%%

\subsubsection{Model comparison}
\label{results-comparison}

We have considered two plausible models of T~cell responses: a
decreasing and an increasing potential model.  In a similar fashion to
the way we have established plausible parameter values given the data,
we can learn which model is more likely given the data.  We do this by
setting the prior probability of each model to 50\% and using an
extended version of the ABC approach to update that probability in the
light of the data.  In short, we randomly choose between the two
models before randomly drawing from the distributions for the
corresponding input parameters.  When we have a big enough posterior
sample, we can count the number of times that each model was accepted.
If one model is more likely than the other, then there will not be an
equal number of acceptances for each model.
  
We use this approach to compare the decreasing potential model with
the increasing potential model.  We find that we need $6.55\times
10^5$ runs to find $2\times 10^3$ acceptable parameter sets for the
IPM, whereas we require just $2.01\times 10^5$ runs to find $2\times
10^3$ acceptable parameter sets for the DPM.  If we assume that, a
priori, the two models are equally likely, then an estimate for the
posterior probability of the DPM over the IPM is given by the
following ratio:
     \begin{equation}
\frac{\frac{2000}{2.01\times 10^5}}{\frac{2000}{2.01\times 10^5}+\frac{2000}{6.55\times 10^5}} = \frac{6.55}{6.55+2.01}=0.76
\; ,
\nonumber
\end{equation}
which leads to a posterior probability of 0.24 for the IPM because we
are only considering two possible models.  Therefore, there is some
evidence to suggest that the DPM is the more appropriate model given
the YFV vaccination data, but this is far from being conclusive: we
note that these probabilities hinge on the fact that we are only
considering two possible models and both models have been found to be
plausible.  The same approach could be used to evaluate a range of
plausible models, and the present comparison is an illustration of the
technique.

Here, because we have not explicitly calculated the model likelihoods,
traditional information-criteria-based approaches to evaluating
relative model performance are not open to us. However, our comparison
approach based upon posterior probabilities automatically accounts for
uncertainty in the model parameters and penalises for model complexity
like many other information-based criteria, such as Akaike information
criterion (AIC) and Bayesian information criterion (BIC).

%%%%%%

 \subsubsection{Division of labour}
\label{labour}

Our results are in line with the concept of
 ``division of labour''~\cite{gerlach2013heterogeneous,harty2008shaping}. That
is, as the naive precursor frequency increases, and thus, a larger
number of antigen-specific naive CD8\plus\ T~cells are recruited into
the immune response, fewer divisions are required in the
differentiation programme of antigen-specific cells, and greater
timescales for the first division are still enough to mount a timely
immune response.  We can provide a more accurate quantification of the
previous statement.  If we let $g = g_n + g_c + g_m$ and $N_c^{max} = 10^5$, we have
estimated the following conditional probabilities:

\begin{eqnarray} 
{\mathbb{P}} (g \ge 9 | N_c < 10^4)       &=&  0.88 \; , \quad \text{with}   \quad (0.82,0.95)  
\quad 99\%  \text{ CI}  
\; ,
\nonumber
\\
{\mathbb{P}}(g \ge 9 | N_c \ge 10^4)     & =&  0.69  \; , \quad \text{with} \quad  (0.65,0.72)
\quad 99\%  \text{ CI}
\; ,  
\nonumber
\\ 
{\mathbb{P}}(g \ge 10 | N_c <  10^4)      & =&  0.79   \; , \quad \text{with} \quad (0.70,0.88)
\quad 99\%  \text{ CI}  
\; ,
\nonumber
\\
{\mathbb{P}}(g\ge10|N_c\ge  10^4)   &=& 0.59   \; , \quad \text{with} \quad (0.56,0.63)
\quad 99\%  \text{ CI}  
\; ,
\nonumber
\\ 
{\mathbb{P}}(g\ge11|N_c<  10^4)      &=& 0.63   \; , \quad \text{with} \quad (0.53,0.74)
\quad 99\%  \text{ CI}  
\; ,
\nonumber
\\
{\mathbb{P}}(g\ge11|N_c\ge  10^4)   &=& 0.52   \; , \quad \text{with} \quad (0.49,0.56)
\quad 99\%  \text{ CI}  
\; .
\nonumber
\end{eqnarray}
From these results, it can be seen that there is a division of labour
in the sense that, if $N_c$ is relatively large, the chance of having
a large number of total divisions in the differentiation programme is
significantly reduced.  We note that all these differences are
statistically significant at an approximate 1\% level, except for when
$g$ is greater than 10.  Finally, for $N_c^{max}$ =
10$^5$ the median total number of generations, $g = g_n + g_c + g_m$,
is 11 (see Table~\ref{summary-dpm}). This value is in line with a
recent study to quantify the kinetics of CD8\plus\ specific T~cells
during YFV vaccination~\cite{le2014mathematical} despite the fact that
this model did not consider the individual kinetics of each CD8\plus\
specific T~cell subset (N, CM, EM, E).
 
%%%%%%%%%%%%%%%%%%%%%%%%%%%%%%%%

\subsubsection{Initial conditions for antigen-specific naive  CD8\plus\
T~cells in the decreasing potential model}
\label{naive-initial-dpm}

\begin{figure}[ht]
\centering
\begin{subfigure}[b]{0.45\textwidth}
\includegraphics[page=1,width=\textwidth]{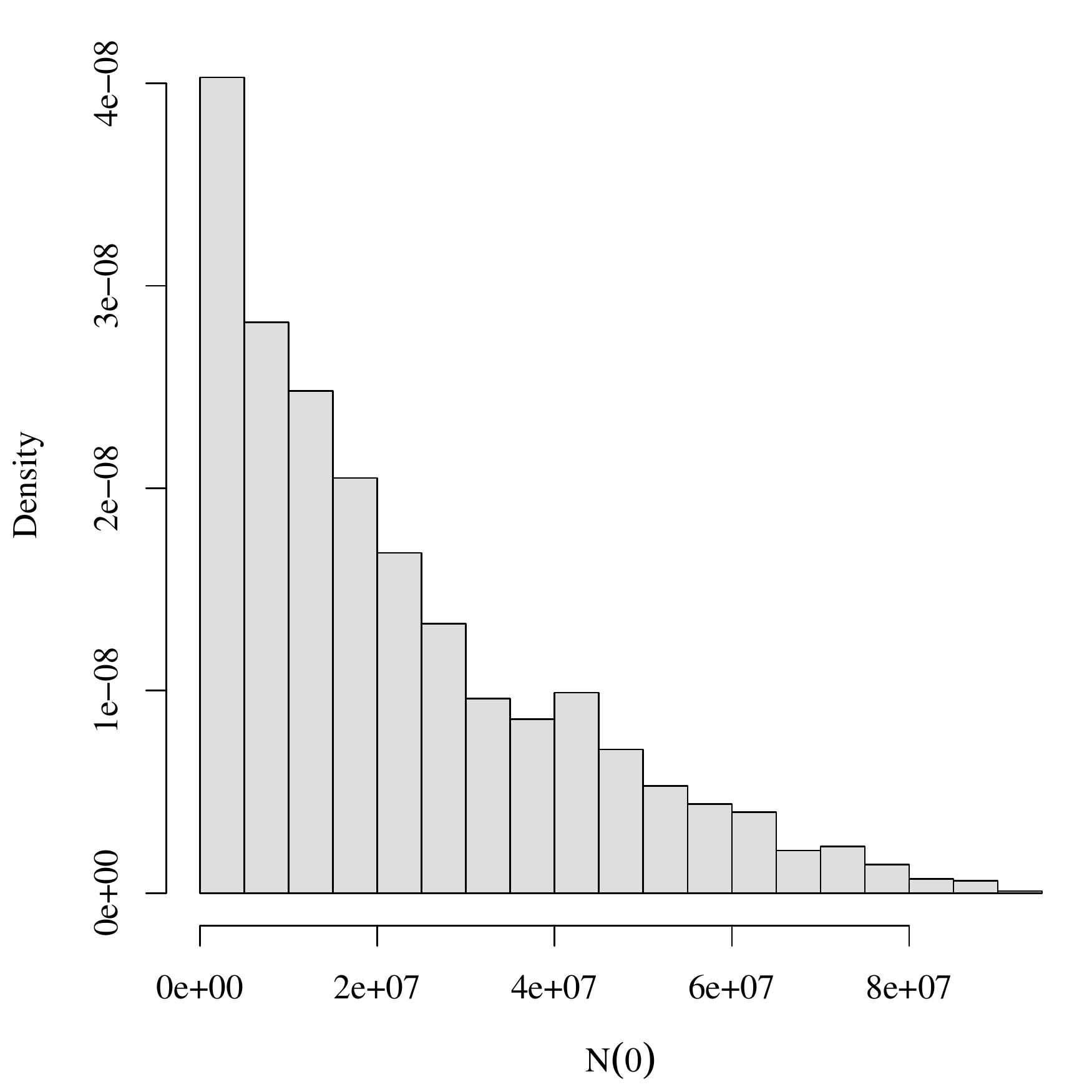}
\end{subfigure}
~
\begin{subfigure}[b]{0.45\textwidth}
\includegraphics[page=2,width=\textwidth]{YFpostgraphsInitialDPM.pdf}
\end{subfigure}
\caption{Histograms of samples
for $N_0$ (left) and $N^{(B)}_0$ (right) based on the posterior distributions for the DPM.
}
\label{plot-initial-naive}
\end{figure}

We have also made use of model calibration to study the posterior
distribution of the initial conditions for antigen-specific naive
CD8\plus\ T~cells in the decreasing potential model.  We have
considered that in the case of YFV vaccination, the initial conditions
correspond to a primary immune response (see
Section~\ref{method-computational-algorithm}), so that, at the time of
the challenge (or initial time, $t=0$), the only CD8\plus\ T~cells
present are naive (non-specific and antigen-specific) and there are no
specific central memory, effector memory or effector T~cells in any
spatial compartment.  A further assumption is the fact that, as
described in detail in Section~\ref{method-computational-algorithm},
prior to the immune challenge (YFV vaccination), naive cells are
assumed to be in homeostatic (or steady-state) conditions (see
Section~\ref{method-model-homeostasis}).  This means that $N_0$ and
$N^{(B)}_0$, the initial conditions for the antigen-specific naive
cell populations in the draining lymph nodes and in the blood
compartments, respectively, are the stable steady-state solutions 
of~\eqref{specific-naive-dLN-homeostasis}
and~\eqref{specific-naive-B-homeostasis}. It is clear from these
equations, that the stable steady-state solutions depend on the fixed
parameters $\theta_n, \kappa_n, \mu_n$ and $\delta$, as well as on the
Bayesian parameters $N_c, \gamma,\phi$.  The posterior distributions
of $N_0$ and $N^{(B)}_0$ are shown in Fig.~\ref{plot-initial-naive}.
We can derive the following probabilities from these distributions:
\begin{eqnarray}
{\mathbb{P}}( N_0 \le 10^{5}) &=& 0.004
\; , \quad {\mathbb{P}}( N^{(B)}_0 \le 10^{5}) = 0.0015
\; ,
\nonumber
\\
{\mathbb{P}}( 10^5 < N_0 \le 10^{6}) &=& 0.0375
\; ,
\quad {\mathbb{P}}( 10^5 < N^{(B)}_0 \le 10^{6}) = 0.017
\; ,
\nonumber
\\
{\mathbb{P}}( 10^6 < N_0 \le 10^{7}) &=& 0.301
\; ,
\quad {\mathbb{P}}( 10^6 < N^{(B)}_0 \le 10^{7}) = 0.1635
\; ,
\nonumber
\\
{\mathbb{P}}( 10^7 < N_0 \le 10^{8}) &=& 0.6575
\; ,
\quad {\mathbb{P}}( 10^7 < N^{(B)}_0 \le 10^{8}) = 0.818
\; ,
\nonumber
\\
{\mathbb{P}}( 10^8 <  N_0 ) &=& 0
\; ,
\quad {\mathbb{P}}( 10^8 < N^{(B)}_0 ) = 0
\; .
\nonumber
\end{eqnarray}
For both $N_0$ and $N^{(B)}_0$, there is the highest probability that
the size is in the interval $[10^7,10^8]$ with probabilities of 0.66
and 0.82 for the dLN and blood, respectively.

%%%%%%%%%%%%%%%%%%%%%%%%%%%%%%%%%%%%

%%%%%%%%%%%%%%%%%%%%%%%%%%%%%%%%%%

\subsection{Model performance in the light of the data}
\label{performance}

In this section, we make use of the results derived from the model
calibration in the light of the YFV data to study the temporal
dynamics of the different CD8\plus\ T~cell populations for both the
DPM and IPM models. We have made use of the mathematical models, the
fixed parameters (see Table~\ref{table-parameters-fixed}), as well as
the summary statistics for each of the models (see
Table~\ref{summary-dpm} and Table~\ref{summary-ipm}, respectively),
and the simulator to generate a time course that corresponds to the
YFV vaccine data, assuming that vaccination takes place at time $t=0$.

\begin{figure}[htp!]
\centering
\includegraphics[scale=0.85]{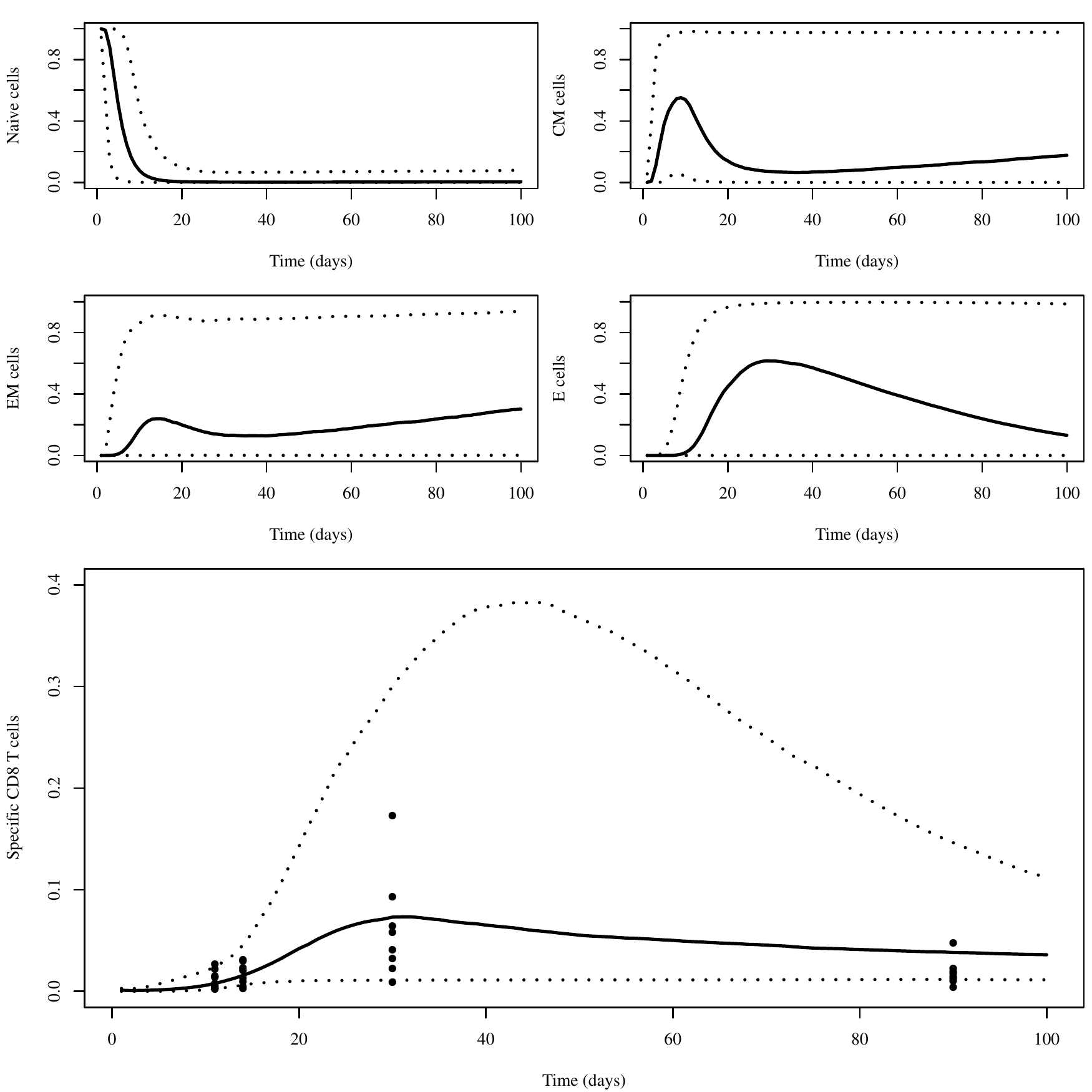}
\caption{Time course of specific CD8\plus\ T~cell subsets in blood generated
  with the simulator of the decreasing potential model. Solid lines
  have been obtained with the median of the posterior parameters and
  dotted lines with the 95\% credible interval (see
  Table~\ref{summary-dpm}).  Top left: fraction of specific naive
  CD8\plus\ T~cells out of total specific CD8\plus\ T~cells.  Top
  right: fraction of specific central memory CD8\plus\ T~cells out of
  total specific CD8\plus\ T~cells.  Bottom left: fraction of specific
  effector memory CD8\plus\ T~cells out of total specific CD8\plus\
  T~cells.  Bottom right: fraction of specific effector CD8\plus\
  T~cells out of total specific CD8\plus\ T~cells.  Lower plot:
  fraction of specific total CD8\plus\ T~cells out of total CD8\plus\
  T~cells.  }
\label{plot-dpm}
\end{figure}

\begin{figure}[htp!]
\centering
\includegraphics[scale=0.85]{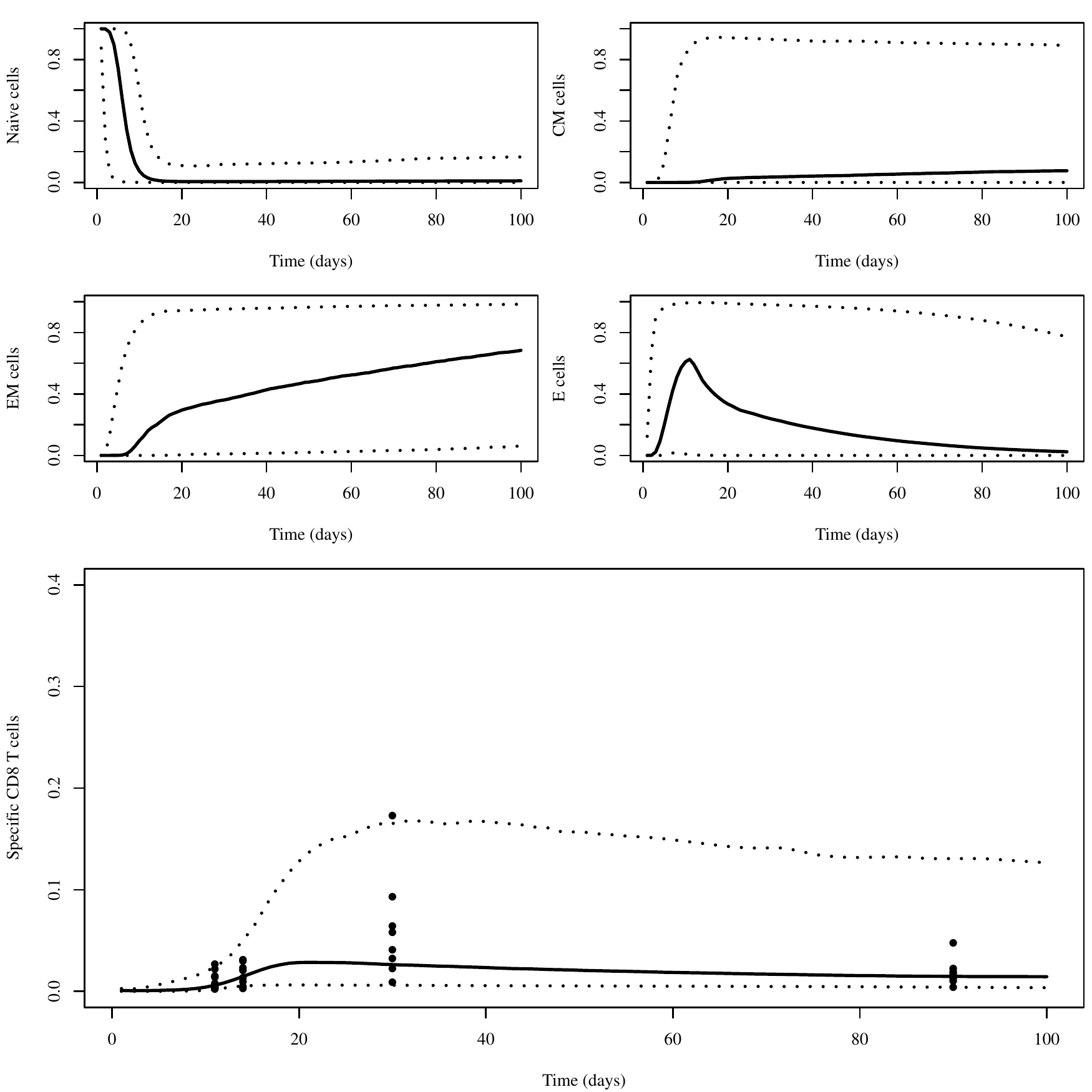}
\caption{Time course of specific CD8\plus\ T~cell subsets  in blood generated
  with the simulator of the increasing potential model. Solid lines
  have been obtained with the median of the posterior parameters and
  dotted lines with the 95\% credible interval (see
  Table~\ref{summary-dpm}).  Top left: fraction of specific naive
  CD8\plus\ T~cells out of total specific CD8\plus\ T~cells.  Top
  right: fraction of specific central memory CD8\plus\ T~cells out of
  total specific CD8\plus\ T~cells.  Bottom left: fraction of specific
  effector memory CD8\plus\ T~cells out of total specific CD8\plus\
  T~cells.  Bottom right: fraction of specific effector CD8\plus\
  T~cells out of total specific CD8\plus\ T~cells.  Lower plot:
  fraction of specific total CD8\plus\ T~cells out of total CD8\plus\
  T~cells.  }
\label{plot-ipm}
\end{figure}

Fig.~\ref{plot-dpm} shows the median time courses for the fraction of
specific CD8\plus\ T~cells and the fractions of the four identified
subtypes in blood. For the DPM model, there is a drop in the fraction of naive
CD8\plus\ T~cells to close to zero 12 days after the vaccination
whereas the fractions of central and effector memory cells peak around
that time. Fig.~\ref{plot-ipm} shows that the IPM exhibits similar
behaviour for the naive CD8\plus\ T~cells, but the fractions of
central and effector memory cells do not have the same peak.

Because the models are currently calibrated on the specific fraction
of CD8\plus\ T~cells (out of total CD8\plus\ T~cells) alone, there is
a great deal of uncertainty shown in the plotted possible time courses
(especially for the four subtypes).  Incorporating reliable phenotype
data into the analysis would help to reduce this uncertainty and would
benefit the model comparison.  Data on the CD8\plus\ T~cell fractions
around days 15 and 20 after vaccination would help to further
differentiate between the models, since we know from the plausible
time courses that this is where the two models have appreciably
different behaviour.  The collection of such data is viable (see
Figure~4B of Ref.~\cite{akondy2009yellow}); however, for the present
analysis, the data on the post-vaccination dynamics of the CD8\plus\
T~cell phenotypes were not made available.

%%%%%%%%%%%%%%%%%%%%%%%%%%%%%%%%%%%%%%%%%%%%%%

\section{Discussion}
\label{discussion}

A population-average mathematical model of CD8\plus\ T~cell dynamics
has been developed
that allows us to study the cellular processes (proliferation,
differentiation, migration and death) that regulate the generation of
a diverse and heterogeneous CD8\plus\ T~cell population during an
immune response.  The model considers four cellular populations,
(naive, central memory, effector memory and effector T~cells) and three
spatial compartments (draining lymph nodes, circulation
and skin). Cell death, division, thymic export, migration, as well as
T~cell activation by antigen, and a programme of
differentiation-linked division are used to functionally characterise
each CD8\plus\ T~cell subtype~\cite{appay2002memory,gerritsen2015memory}. Our mathematical models are
calibrated with the fraction of total specific CD8\plus\ T~cells
from the YFV vaccine data in Ref.~\cite{akondy2009yellow}.

We first consider a differentiation programme based on the ``decreasing
potential'' hypothesis as described recently in
Ref.~\cite{buchholz2013disparate}.  These authors
carried out single-cell kinetic experiments of murine bacterial
infection and concluded that cells differentiate towards phenotypes with
higher proliferation capacity and lower differentiating
capacity~\cite{buchholz2013disparate}.  The reverse differentiation
scenario, called the linear differentiation model or ``increasing
potential
model''~\cite{kaech2012transcriptional,obar2011pathogen,ahmed2009precursors},
where effector cells appear earlier than any other phenotype
(N$\rightarrow $E$\rightarrow$EM$\rightarrow$CM) has also been
considered in our analysis.
 
We have made use of the literature to obtain rates of division, death
and thymic export for each population subset.  Approximate Bayesian
computation (ABC) has been used together with the mathematical model
and YFV vaccination data from~Ref.~\cite{akondy2009yellow} (see
Figure~3B in Ref.~\cite{akondy2009yellow}) to obtain posterior
distributions for the subset of parameters related to the immune
response, such as the number of divisions in the differentiation
programme, the time to first division, the time to subsequent
divisions, the number of specific clonotypes involved in the response,
the duration of the immune response and migration rates.

After model calibration of the DPM (see Table~\ref{summary-dpm}), the
median value of $g=g_n+g_c+g_m$ is 11, which is in line with recent
mathematical modelling results that make use of the YFV vaccination
data~\cite{le2014mathematical,harty2008shaping} and predicted nine cell divisions for
the observed CD8\plus\ T~cell expansion.  We also find that as the
number of naive cells driven to an immune response increases, there is
a reduction in the number of divisions, $g$, required to differentiate
from a naive to an effector CD8\plus\ T~cell, which supports the
``division of labour'' already observed in mice
studies~\cite{gerlach2013heterogeneous}.  Our estimates that median
division rates,
$\alpha_n,\alpha_c,\alpha_m,\lambda_n,\lambda_c,\lambda_m$, are less
than 0.5 per day also agree with the results of
Ref.~\cite{le2014mathematical}, which predicted a doubling time of two
days.  As discussed in Ref.~\cite{wherry2003lineage}, and supported by
our results, we note that memory T~cells (CM and EM), once generated,
are present at higher numbers than naive cells (see
Fig.~\ref{plot-dpm}).  Their long-term maintenance is guaranteed by
homeostatic mechanisms.  Our homeostasis mathematical model encodes
this immunological fact by the choice of parameters (reviewed from the
literature) (values of $\kappa_c > \kappa_n$ and $\theta_c >
\theta_n$).

Model calibration was also carried out for the IPM (see
Table~\ref{summary-ipm}), as well as model comparison between the DPM
and the IPM. Using a Bayesian approach, we have shown that we can use
the data to update parameter distributions and perform model
comparisons.  Despite the data being on the number of specific
CD8\plus\ T~cells, we are able to make inferences about the model
parameters governing differentiation across the different
phenotypes. For the model comparison, our analysis leads to a
posterior probability of the IPM equal to 0.24, so the DPM is favoured
by the data, but we cannot rule out either hypothesis.  The type of
comparison carried out here could be easily extended to cover multiple
models~\cite{chipman2001}, and the associated uncertainty analyses can
be used to help identify where further data could help to aid the
model discrimination (as shown in Section~\ref{performance}).

The model presented here is relatively comprehensive (see some recent
mathematical modelling efforts of CD8\plus\ T~cell
responses~\cite{le2014mathematical,gong2014harnessing}), yet it fails
to include the role of TCR specificity or that of cytokines.  Thus,
improvement of the current model will require the consideration of the
``signal strength'' hypothesis~\cite{gett2003t} and  cross-reactivity~\cite{yan2017modelling}.  This will be
essential to decipher the role of individual T~cell clonotypes, with
different TCR affinities, in the dynamics of a human immune response.
Avenues to explore in the future include the roles of heterogeneity
and
stochastic behaviour at the single cell level. We have limited
ourselves to describing the mean behaviour of the cell populations.
We have also restricted our study to CD8\plus\ T~cells.

We have made use of a deterministic model, yet the expression of
differentiation, or ``binary'', markers by T~cells appears to be
stochastic as not all possible phenotypes can be found in the
population of T~cells that is generated during an immune
response~\cite{mahnke2013s}. We still do not have a full understanding
of the mechanisms that regulate cellular fate
decision~\cite{kaech2012transcriptional}. It is out of the scope of
this paper, but our future effort will require the development of
stochastic mathematical models that can account for the inherent
randomness at the single cell level during immune cell
differentiation~\cite{marchingo2014antigen}.
 
The inclusion of a skin compartment in our model is partly motivated
by a desire to study the dynamics of the CD8\plus\ T~cell response in
humans, where the antigen has been delivered via the dermal route,
such as in
skin sensitisation~\cite{Martin2011}.  The analysis and the results presented suggest that
 the Bayesian methodology
reported will enable us to infer model parameters related to the  T~cell dynamics~\cite{Popple2016} and 
TCR repertoire~\cite{Oakes2016} associated with immune responses via the dermal route.  
The ability to conclusively distinguish between different 
hypotheses (and therefore models) of  CD8\plus\ T~cell differentiation 
will depend critically on the availability of phenotype data from such responses.

%%%%%%%%%%%%%%%%%%%%%%%%%%%%%%%%%%%%%%

\paragraph{Acknowledgement}

The authors wish to thank the members of the ``T~cell Forum'' both
past and present for their creative input and critical review of this
manuscript.  JPG, SMK, GL and CMP acknowledge financial support from
Unilever under contract CH-2011-0828.

\bibliographystyle{vancouver}

%%%%%%%%%%%%%%%%%%%%%%%%%%%%%%%%%%%%%%

\section{Methods}
\label{appendix-methodology}

%%%%%%%%%%%%%%%%%%%%%%%%%%%%%%%%%%%%%%CMP is here!!!

\subsection{Mathematical models of CD8\plus\ T~cell dynamics}
\label{method-DPM-model}

%%%%%%%%%%%%%%%%%%%%%%%%%%%%%%%%%%%%%%

\subsubsection{Mathematical model of CD8\plus\ T~cell homeostasis}
\label{method-model-homeostasis}

We first introduce the mathematical model that describes the dynamics
of CD8\plus\ T~cells during homeostasis.  In homeostasis, we consider
the following CD8\plus\ T~cell subsets: non-specific naive T~cells
($N_r$), antigen-specific naive T~cells ($N$), antigen-specific
central memory T~cells ($C$), and antigen-specific effector memory
T~cells ($M$).  We assume that terminally differentiated
(antigen-specific) effector T~cells ($E$) do not divide
and thus, are not included in the analysis that follows.
We hypothesise that the number of different T~cell clonotypes that are
antigen-specific out of the total TCR diversity, $N_R$, and driven
into the immune response is $N_c$.  In this way, $N$ represents the
total number of antigen-specific naive T~cells that belong to $N_c$
different clonotypes, $C$ represents the total number of
antigen-specific central memory T~cells that belong to $N_c$ different
clonotypes, and $M$ represents the total number of antigen-specific
effector memory T~cells that belong to $N_c$ different clonotypes.
Finally, we assume that all clonotypes (non-specific and
antigen-specific) can be described with the same parameters for
proliferation, death and migration, and thus, are
identical~\cite{antia1998models,antia2003models}.  In this sense and
for the mathematical model considered in this manuscript, $N_c$ is the
parameter that encodes how broad the CD8\plus\ immune response is, as
it quantifies how many different TCR clonotypes are driven to
proliferate and differentiate in response to the specific antigen.

%%%%%%%%%%%%%%%%%%%%%%%%%%%%%%%%%%%%%%%%%%%%%%%%%%

The mathematical model considers CD8\plus\ T~cells in three different
spatial compartments: draining lymph nodes (dLNs), blood and resting
lymph nodes (B) and skin (S).  CD8\plus\ T~cell populations in blood
and resting lymph nodes are labelled with a superscript $(B)$ and
those in skin with a superscript $(S)$.  The CD8\plus\ T~cell
population of the draining lymph nodes is not labelled with a
superscript.  We include the following processes in the homeostasis
model (see Figure~\ref{CD8homeo}):

\begin{itemize}

\item {\bf Thymic output}: naive T~cells that have survived thymic
  selection are incorporated into the peripheral blood compartment,
  $(B)$. We denote by $\delta$ the thymic export rate per T~cell
  clonotype and by $\delta_r$ the total thymic non-specific export
  rate. We note that $N_c \; \delta$ is the total specific naive cell
  thymic exit rate, and $\delta_r = (N_R- N_c) \; \delta$ is the total
  non-specific naive cell thymic exit rate.

\item {\bf Division}: we model cell division with a logistic term,
  with $\theta$ the division rate and $\kappa$, the carrying capacity
  of the population.  The logistic term is a simple and standard way
  to model the competition for limited resources of a population with
  a characteristic equilibrium size of
  $\kappa$~\cite{antia1998models}.  We assume that antigen-specific
  and non-specific naive cells homeostatically proliferate with the
  same rate, but have a different carrying capacity, $\kappa_r$ and
  $\kappa_n$, respectively.  Each T~cell population 
  divides with rate $\theta_n$, $\theta_c$ and $\theta_m$ for
  naive, CM and EM, respectively, and has corresponding carrying
  capacities given by $\kappa_r$, $\kappa_n$, $\kappa_c$ and
  $\kappa_m$.

\item {\bf Cell death}: we assume a constant per cell rate of death
  and we denote it by $\mu$.  We assume that specific and non-specific
  naive cells die with the same rate.  Each cell type can die with
  rates $\mu_n$, $\mu_c$ and $\mu_m$ for naive, CM and EM,
  respectively.

\item {\bf Cell migration}: we assume that naive and central memory
  cells migrate between the dLNs and the resting LNs and blood, but do
  not migrate to the skin. $\gamma$ is their migration rate from the
  dLNs to the resting LNs and blood, and $\phi$ is their migration
  rate from the resting LNs and blood to the dLNs.  We assume that
  specific and non-specific naive cells have the same migration rates.
  Effector memory cells can only migrate from the dLNs to the resting
  LNs and blood with rate $\zeta$.  They also migrate between the
  resting LNs and blood and skin with rates $\xi$ and $\psi$,
  respectively.

\end{itemize}

%%%%%%%%%%%%%%%%%%%%%%%%%%%%%%%%%%%%%%%%%%

The ordinary differential equations (ODEs) that describe the
homeostasis model are presented below and a description of the model
is provided in Figure~\ref{CD8homeo}.  Each proliferation and death
rate is labelled with a subscript that corresponds to the population
subset under consideration, $n$, $c$, and $m$ for naive (N), central
memory (CM), and effector memory (EM) cells, respectively.  In the
dLNs, we write
\begin{eqnarray}
\frac{dN_r}{dt} &=&
\theta_n \; N_r \left( 1-\frac{N_{r}}{\kappa_r} \right)
- \mu_n \; N_r 
-\gamma \; N_r
+ \phi \; N_r^{(B)}
\; ,
\label{non-specific-naive-dLN}
\\ 
 \frac{dN}{dt} &=&
 \theta_n  \; N \left( 1 - \frac{N}{N_c \; \kappa_n} \right) 
 - \mu_n \; N
 -\gamma \; N 
 + \phi \; N^{(B)}
\; ,
\label{specific-naive-dLN-homeostasis}
\\ 
\frac{dC}{dt} &=&
\theta_c \; C \left( 1-\frac{C} {N_c \; \kappa_c} \right)
- \mu_c \; C 
-\gamma \; C
+ \phi \;  C^{(B)}
\; ,
\\ 
\frac{dM}{dt} &=&
 \theta_m \; M \left( 1-\frac{M}{N_c \; \kappa_m} \right)
 - \mu_m \; M 
 -\zeta \; M
 \; .
\end{eqnarray}

In blood and the resting LNs, we have
\begin{eqnarray}
\frac{dN_r^{(B)}}{dt} &=&
\theta_n \; N_r^{(B)} \left( 1-\frac{N_{r}^{(B)}}{\kappa_r} \right)
- \mu_n \; N_r^{(B)}
- \phi \; N_r^{(B)}
+ \gamma \; N_r
+\delta_r
\; ,
\label{non-specific-naive-B}
\\ 
 \frac{dN^{(B)}}{dt} &=&
 \theta_n  \; N^{(B)} \left( 1 - \frac{N^{(B)}}{N_c \; \kappa_n} \right) 
 - \mu_n \; N^{(B)}
 - \phi \; N^{(B)}
  +\gamma \; N 
+ N_c \; \delta  
\; ,
\label{specific-naive-B-homeostasis}
\\ 
\frac{dC^{(B)}}{dt} &=&
\theta_c \; C^{(B)} \left( 1-\frac{C^{(B)}}{N_c \; \kappa_c} \right)
- \mu_c \; C^{(B)}
- \phi \;  C^{(B)}
+\gamma \; C
\; ,
\\ 
\frac{dM^{(B)}}{dt} &=&
 \theta_m \; M^{(B)} \left(1-\frac{M^{(B)}}{N_c \; \kappa_m} \right)
 - \mu_m \; M^{(B)} 
 -\xi \; M^{(B)}
 +\zeta \; M
 + \psi \;  M^{(S)}
 \; .
\end{eqnarray}

Finally, in the skin we have
\begin{eqnarray}
 \frac{dM^{(S)}}{dt} 
 &=&
  \theta_m \; M^{(S)} \left(1-\frac{M^{(S)}}{N_c \; \kappa_m} \right)
 - \mu_m \; M^{(S)} 
 - \psi \;  M^{(S)}
 + \xi \; M^{(B)}
\; .
\end{eqnarray}
In order to solve the previous set of ODEs, we need to provide initial
conditions for the nine different cell types considered in the
homeostasis model.  We introduce the following notation for the
initial conditions of naive T~cells:
\begin{eqnarray}
N_r(0) =N_{r0},
\quad
 N(0) = N_0,
\quad
 N_r^{(B)}(0) =N_{r0}^{(B)},
 \quad
 N^{(B)}(0) = N_0^{(B)}
 \; ,
 \end{eqnarray}
and for central and effector memory T~cells:
 \begin{eqnarray}
 C(0) = C_0,
  \quad
 M(0) = M_0,
  \quad
  C^{(B)}(0) = C_0^{(B)},
   \quad
 M^{(B)}(0) = M_0^{(B)},
  \quad
 M^{(S)}(0) = M_0^{(S)}
 \; .
 \end{eqnarray}
 Initial conditions will be considered and specified in
 Section~\ref{method-computational-algorithm}, where we discuss the
 computational algorithm used to solve the ODEs of the combined
 CD8\plus\ homeostasis, discussed in this section, and immune response
 model~\cite{kaech2002effector} (either the decreasing potential
 model, which is described in
 Section~\ref{method-model-immune-response-DPM}, or the increasing
 potential model, which is introduced in
 Section~\ref{method-model-immune-response-IPM}).

%%%%%%%%%%%%%%%%%%%%%%%%%%%%%%%%%%%%%%

\subsubsection{CD8\plus\ T~cell dynamics during an immune response: decreasing potential model}
\label{method-model-immune-response-DPM}

We now introduce the mathematical model that describes the dynamics of
a CD8\plus\ T~cell immune response. We assume that $N_c$ different TCR
clonotypes are driven into the response and thus, the non-specific
clonotypes, $N_R-N_c$, and their naive cells will follow the dynamics
that were considered in the homeostasis model, and given
by~\eqref{non-specific-naive-dLN} and~\eqref{non-specific-naive-B}
(see Section~\ref{method-model-homeostasis}). We also assume, as
described above, that the different clonotypes driven into the
response can be described with identical rates, and are thus
indistinguishable.
 
We consider the following CD8\plus\ T~cell subsets: antigen-specific
naive T~cells ($N$), antigen-specific central memory T~cells ($C$),
antigen-specific effector memory T~cells ($M$), and antigen-specific
effector T~cells ($E$).  These different T~cell subtypes describe, in
a combined way, the $N_c$ clonotypes driven into the immune
response. In this way, $N$ represents the total number of
antigen-specific naive T~cells that belong to $N_c$ different
clonotypes, $C$ the total number of antigen-specific central memory
T~cells that belong to $N_c$ different clonotypes, $M$ the total
number of antigen-specific effector memory T~cells that belong to
$N_c$ different clonotypes, and $E$ the total number of
antigen-specific effector T~cells that belong to $N_c$ different
clonotypes.  The model considers CD8\plus\ T~cells in three different
spatial compartments and we follow the notation introduced in
Section~\ref{method-model-homeostasis}. Thus, all T~cell populations
in blood and resting lymph nodes are labelled with a superscript
$(B)$, those in skin with a superscript $(S)$, and the CD8\plus\
T~cell population of the draining lymph nodes is not labelled with a
superscript.  We include the following processes in the immune
response model (see Figure~\ref{cd8-immune-response}):

\begin{itemize}

\item {\bf Antigen-driven proliferation and differentiation in the
    dLNs}: we assume that a T~cell contact with an APC starts a
  programme of division-linked-differentiation for antigen-specific
  CD8\plus\ T~cells, according to the decreasing potential model (DPM)
  (see Figure~\ref{dpm}).  Each differentiation step requires an APC
  contact and involves a number of divisions (or generations), $g$,
  that depends on the subtype of the T~cell, whether N, CM or EM (see
  Figure~\ref{div-linked-diff}).  That is, in the dLNs and upon
  antigen stimulation (APC-mediated), specific naive cells divide for
  $g_n$ generations to differentiate to central memory cells at the
  $g_n^{\rm th}$ division.  Similarly, central memory and effector
  memory cells divide for $g_c$ and $g_m$ generations to differentiate
  to effector memory and effector cells, respectively, at the last
  division.  We assume that effector memory cells can also encounter
  APCs in the $S$ spatial compartment, and differentiate to effector
  cells.  We denote by $I$, $J$, and $K$ the precursor (or
  intermediate) cells for $C$, $M$, and $E$, respectively (see
  Figure~\ref{div-linked-diff}).  In this way, the progeny of $N$ is
  denoted by $I_1$, the progeny of $I_1$ by $I_2$, the progeny of
  $I_{g_n-1}$ by $C$, the progeny of $C$ by $J_1$, the progeny of
  $J_1$ by $J_2$, the progeny of $I_{g_c-1}$ by $M$, the progeny of
  $M$ by $K_1$, the progeny of $K_1$ by $K_2$, and finally, the
  progeny of $I_{g_m-1}$ by $E$.  The first division takes place with
  rate $\alpha$ and subsequent divisions occur at rate $\lambda$. Each
  of these rates will include a subscript $n,c,m$ depending on the
  T~cell subtype under consideration, whether N, CM or EM.
 
\item {\bf Thymic output}: naive T~cells are exported from the thymus
  to the peripheral blood compartment, $B$, as described in
  Section~\ref{method-model-homeostasis}.

\item {\bf Division}: we consider homeostatic
  proliferation only for N (antigen-specific and non-specific), CM and
  EM T~cells, as described in Section~\ref{method-model-homeostasis}.
  We neglect division for the populations $I$, $J$
  and $K$, as they are intermediate cells generated during the immune
  response.
 
\item {\bf Cell death}: we assume a constant per cell rate of death
  and we denote it by $\mu$.  Each cell type can die with rates
  $\mu_n$, $\mu_c$, $\mu_m$, and $\mu_e$, respectively.  We assume
  that the intermediate populations $I$, $J$ and $K$ have the same
  death rates as their parent subtypes, N, CM and EM, respectively.

\item {\bf Cell migration}: naive and CM T~cells have the same
  migratory behaviour as described in
  Section~\ref{method-model-homeostasis}. EM and effector T~cells have
  the same migratory behaviour, as described for EM T~cells in
  Section~\ref{method-model-homeostasis}.  We assume that the
  intermediate populations $I$, $J$ and $K$ do not migrate, thus, they
  only need to be considered in the dLN compartment.
 
\end{itemize}

%%%%%%%%%%%%

The ordinary differential equations (ODEs) that describe the immune
response according to the decreasing potential model are presented
below and a description of the model is provided in
Figure~\ref{cd8-immune-response}.  Each proliferation, differentiation
and death rate is labelled with a subscript that corresponds to the
population subset under consideration, $n$, $c$, $m$ and $e$ for
naive, central memory, effector memory, and effector cells,
respectively.  In the dLNs, we can write
\begin{eqnarray}
\frac{dN}{dt} &=&
 \theta_n  \; N \left( 1 - \frac{N}{N_c \; \kappa_n} \right) 
 - \alpha_n \; N
 - \mu_n \; N
 -\gamma \; N 
 + \phi \; N^{(B)}
\; ,
\label{dpm-naive-dln}
\\ 
\frac{dI_1}{dt} &= &
2 \; \alpha_n \; N
- \lambda_n \; I_1
 - \mu_n \; I_1
 \; ,
\\
\frac{dI_i}{dt} &= &
2 \; \lambda_n \; I_{i-1}
 - \lambda_n \; I_{i}
-\mu_n \; I_{i}
\; ,
\quad
\forall i = 2, \ldots, g_{n}-1
\; ,
\\
\frac{dC}{dt} &=&
2 \; \lambda_n \; I_{g_n-1}
+
\theta_c \; C \left( 1-\frac{C}{N_c \; \kappa_c} \right)
- \alpha_c \; C 
- \mu_c \; C 
-\gamma \; C
+ \phi \;  C^{(B)}
\; ,
\\ 
\frac{dJ_1}{dt} &= &
2 \; \alpha_c \; C
- \lambda_c \; J_1
 - \mu_c \; J_1
 \; ,
\\
\frac{dJ_i}{dt} &= &
2 \; \lambda_c \; J_{i-1}
 - \lambda_c \; J_{i}
-\mu_c \; J_{i}
\; ,
\quad
\forall i = 2, \ldots, g_{c}-1
\; ,
\\
\frac{dM}{dt} &=&
2 \; \lambda_c \; J_{g_c-1}
+
 \theta_m \; M \left( 1-\frac{M}{N_c \; \kappa_m} \right)
 - \alpha_m \; M 
 - \mu_m \; M 
 -\zeta \; M
 \; .
 \\
\frac{dK_1}{dt} &= &
2 \; \alpha_m \; M
- \lambda_m \; K_1
 - \mu_m \; K_1
 \; ,
\\
\frac{dK_i}{dt} &= &
2 \; \lambda_m \; K_{i-1}
 - \lambda_m \; K_{i}
-\mu_m \; K_{i}
\; ,
\quad
\forall i = 2, \ldots, g_{m}-1
\; ,
\\
\frac{dE}{dt} &=&
 2 \; \lambda_m \; K_{g_m-1}
  -\mu_e \; E
   -\zeta \; E 
\; .
\label{dpm-effector-dln}
\end{eqnarray}
In blood and resting LNs, we have
\begin{eqnarray}
 \frac{dN^{(B)}}{dt} &=&
 \theta_n  \; N^{(B)} \left( 1 - \frac{N^{(B)}}{N_c \; \kappa_n} \right) 
 - \mu_n \; N^{(B)}
 - \phi \; N^{(B)}
 +\gamma \; N 
 + N_c \; \delta  
\; ,
\label{dpm-naive-b}
\\ 
\frac{dC^{(B)}}{dt} &=&
\theta_c \; C^{(B)} \left( 1-\frac{C^{(B)}}{N_c \; \kappa_c} \right)
- \mu_c \; C^{(B)}
- \phi \;  C^{(B)}
+\gamma \; C
\; ,
\\ 
\frac{dM^{(B)}}{dt} &=&
 \theta_m \; M^{(B)} \left(1-\frac{M^{(B)}}{N_c \; \kappa_m} \right)
 - \mu_m \; M^{(B)} 
 -\xi \; M^{(B)}
 +\zeta \; M
 + \psi \;  M^{(S)}
 \; ,
 \\
  \frac{dE^{(B)}}{dt} &=& 
  - \mu_e \; E^{(B)} 
  -\xi \; E^{(B)}
+  \zeta \; E
  +\psi \; E^{(S)}
  \;  .
  \label{dpm-effector-b}
\end{eqnarray}
Finally, in the skin we have
\begin{eqnarray}
 \frac{dM^{(S)}}{dt} 
 &=&
  \theta_m \; M^{(S)} \left(1-\frac{M^{(S)}}{N_c \; \kappa_m} \right)
 - \alpha_m \; M^{(S)} 
 - \mu_m \; M^{(S)} 
 - \psi \;  M^{(S)}
+\xi \; M^{(B)}
\; ,
\label{dpm-memory-s}
\\
\frac{dK^{(S)}_1}{dt} &= &
2 \; \alpha_m \; M^{(S)} 
- \lambda_m \; K^{(S)}_1
 - \mu_m \; K^{(S)}_1
 \; ,
\\
\frac{dK^{(S)}_i}{dt} &= &
2 \; \lambda_m \; K^{(S)}_{i-1}
 - \lambda_m \; K^{(S)}_{i}
-\mu_m \; K^{(S)}_{i}
\; ,
\quad
\forall i = 2, \ldots, g_{m}-1
\; ,
\\
\frac{dE^{(S)}}{dt} &=&
 2 \; \lambda_m \; K^{(S)}_{g_m-1}
  -\mu_e \; E^{(S)}
   -\psi \; E^{(S)} +
   \xi \;  E^{(B)} 
\; .
\label{dpm-effector-s}
\end{eqnarray}

We note that during an immune response, the dynamics for the
non-specific CD8\plus\ naive T~cells, $N_r$ and $N_r^{(B)}$ are
regulated by division and given by
by~\eqref{non-specific-naive-dLN} and~\eqref{non-specific-naive-B}
(see Section~\ref{method-model-homeostasis}).

The mathematical model allows us to follow in time and to quantify the
number of total specific CD8\plus\ T~cells generated during an immune
response in each spatial compartment. That is, at any time $t \ge 0$,
these populations are given by
\begin{eqnarray}
T_{8} (t) &=& 
N  (t) 
+ \sum_{i=1}^{g_n-1} I_i  (t) 
+ C (t) +\sum_{i=1}^{g_c-1} J_i (t)  
+ M  (t) + \sum_{i=1}^{g_m-1} K_i  (t)
+ E (t)
\; ,
\\ 
T_{8}^{(B)} (t)  &=& N^{(B)}  (t) + C^{(B)} (t)  + M^{(B)} (t)  + E^{(B)} (t)
\; ,
\\ 
T_{8}^{(S)} (t)  &=&  M^{(S)} (t)  + \sum_{i=1}^{g_m-1} K^{(S)}_i  (t) + E^{(S)}  (t)
\; ,
\end{eqnarray}
where $T_{8}(t)$ is the total number of specific CD8\plus\ T~cells in
the dLN compartment, $T_{8}^{(B)} (t) $ is the total number of
specific CD8\plus\ T cells in the blood compartment, and $T_{8}^{(S)}
(t)$ is the total number of specific CD8\plus\ T~cells in the skin
compartment at time $t$.  We also need to account for the non-specific
naive T~cell population and thus, the total number of CD8\plus\
T~cells in each spatial compartment at time $t$, is given by $T_{8}
(t) + N_r (t)$, $T_{8}^{(B)} (t) + N_r^{(B)} (t)$ and $T_{8}^{(S)}
(t)$, respectively.  From cell counts, as defined above, the
mathematical model also allows us to calculate the fraction of
specific CD8\plus\ T~cells out of the total CD8\plus\ T~cell
population, and the fraction of specific CD8\plus\ T~cells of a given
subtype (N, CM, EM, E).  We define the specific fraction of CD8\plus\
T~cells out of the total CD8\plus\ T~cell population in blood, and the
fraction of specific CD8\plus\ naive, central memory, effector memory
and effector T~cells at a given time $t$ in the blood compartment, as
follows
\begin{equation}
f_{8} (t) = \frac{T_{8}^{(B)}(t)}{N_r^{(B)} (t)+ T_{8}^{(B)}(t)}
\; , 
f_{n} (t) = \frac{N^{(B)} (t)}{T_{8}^{(B)}(t)}
\; , 
f_{c} (t)= \frac{C^{(B)}(t)}{T_{8}^{(B)}(t)}
\; , 
f_{m} (t)= \frac{M^{(B)}(t)}{T_{8}^{(B)}(t)}
\; , 
f_{e}(t) = \frac{E^{(B)}(t)}{T_{8}^{(B)}(t)}
\; .
\label{def-fraction}
\end{equation}

In order to solve the previous set of ODEs, we need to provide initial
conditions for the different cell types considered in the immune
response model.  We assume that at the initial time there are no
intermediate cells, that is
\begin{eqnarray}
I_{i}(0) &=& 0 \; , 
\quad \forall i = 1, \ldots, g_{n}-1
\; ,
\\
J_{i}(0) &=& 0 \; , 
\quad \forall i = 1, \ldots, g_{c}-1
\; ,
\\
K_{i}(0) &=& 0 \; , 
\quad \forall i = 1, \ldots, g_{m}-1
\; ,
\\
K^{(S)}_{i}(0) &=& 0 \; , 
\quad \forall i = 1, \ldots, g_{m}-1
\; .
\end{eqnarray}
For a primary immune response, we assume that at the initial time there are no specific central memory, effector
memory or effector T~cells in any spatial compartment, that is
\begin{eqnarray}
C(0)&=& C_0=0,
\quad
M(0)=M_0=0,
\quad
E(0)=E_0=0
\; ,
\\
C^{(B)}(0)&=& C_0^{(B)} = 0,
\quad
M^{(B)}(0)=M_0^{(B)}=0,
\quad
E^{(B)}(0)=E_0^{(B)}=0
\; ,
\\
M^{(S)}(0)&=&M_0^{(S)}=0,
\quad
E^{(S)}(0)=E_0^{(S)}=0
\; .
\end{eqnarray}
In the case of a secondary immune response, the initial conditions for
the specific memory subsets (CM or EM), $C_0, M_0, C_0^{(B)},
M_0^{(B)}$ and $M_0^{(S)}$ will be taken to be the number of cells
that have been generated during a primary immune response and
maintained by homeostasis in each of the spatial compartments.
Initial conditions for effector cells will be zero, that is,
$E_0^{(B)}=E_0^{(S)}=0$.  Initial conditions for naive T~cells
(non-specific and antigen-specific) will be considered and specified
in Section~\ref{method-computational-algorithm}, where we discuss the
computational algorithm used to solve the ODEs of the combined
CD8\plus\ T~cell homeostasis and immune response model.

%%%%%%%%%%%%%%%%%%%%%%%%%%%%%%%%%%%%%%

\subsubsection{Mathematical model of CD8\plus\ T~cell dynamics: increasing potential model}
\label{method-model-immune-response-IPM}

We now introduce the mathematical model that describes the dynamics of
a CD8\plus\ T~cell immune response under the hypothesis of increasing
potential differentiation (IPM).  We assume that $N_c$ different TCR
clonotypes are driven into the response and thus, the dynamics of the
non-specific naive T~cells is described by the homeostasis model (see
Section~\ref{method-model-homeostasis}). We also assume that the
different clonotypes driven into the response can be described with
identical rates, and are thus indistinguishable.
 
We consider the same CD8\plus\ T~cell subsets as in the DPM:
antigen-specific naive T~cells ($N$), antigen-specific central memory
T~cells ($C$), antigen-specific effector memory T~cells ($M$), and
antigen-specific effector T~cells ($E$).  These different T~cell
subtypes describe, in a combined way, the $N_c$ clonotypes driven into
the immune response.  The model considers CD8\plus\ T~cells in three
different spatial compartments and we follow the notation introduced
in Section~\ref{method-model-homeostasis}. Thus, all T~cell
populations in blood and resting lymph nodes are labelled with a
superscript $(B)$, those in skin with a superscript $(S)$, and the
CD8\plus\ T~cell population of the draining lymph nodes is not
labelled with a superscript.  We include the following processes in
the immune response model (see Figure~\ref{cd8-immune-response-ipm}):

\begin{itemize}

\item {\bf Antigen-driven proliferation and differentiation in the
    dLNs}: we assume that a T~cell contact with an APC starts a
  programme of division-linked-differentiation for antigen-specific
  CD8\plus\ T~cells, according to the increasing potential model (see
  Figure~\ref{ipm}).  Each differentiation step requires an APC
  contact and involves a number of divisions (or generations), $g$,
  that depends on the subtype of the T~cell, whether N, E, or EM (see
  Figure~\ref{div-linked-diff-ipm}).  That is, in the dLNs upon
  antigen stimulation (APC-mediated), specific naive cells divide for
  $g_n$ generations to differentiate to effector cells at the
  $g_n^{\rm th}$ division.  Similarly, effector and effector memory
  cells divide for $g_e$ and $g_m$ generations to differentiate to
  effector memory and central memory cells, respectively at the last
  division.  We denote by $I$, $J$, and $K$ the precursors for $E$,
  $M$, and $C$, respectively (see Figure~\ref{div-linked-diff-ipm}).
  We assume that effector cells can also encounter APCs in the $S$
  spatial compartment, which differentiate to effector memory cells.
  In this way, the progeny of $N$ is denoted by $I_1$, the progeny of
  $I_1$ by $I_2$, the progeny of $I_{g_n-1}$ by $E$, the progeny of
  $E$ by $J_1$, the progeny of $J_1$ by $J_2$, the progeny of
  $I_{g_c-1}$ by $M$, the progeny of $M$ by $K_1$, the progeny of
  $K_1$ by $K_2$, and finally, the progeny of $I_{g_m-1}$ by $C$.  The
  first division takes place with rate $\alpha$ and subsequent
  divisions occur at rate $\lambda$. Each of these rates will include
  a subscript $n,e,m$ depending on the T~cell subtype under
  consideration, whether N, E or EM.  In the case of central memory
  cells in the dLNs, we assume that a contact with an APC starts a
  programme of $g_c$ divisions with no further differentiation events.
  That is, each proliferation event gives rise to two central memory
  cells that are identical to the dividing (or mother) cell. The rate
  of first division is $\alpha_c$ and subsequent divisions take place
  with rate $\lambda_c$ (see Figure~\ref{cd8-immune-response-ipm}).
  In a similar way and in the $S$ compartment, a contact between an
  APC and an effector memory cell starts a programme of $g_m$
  divisions with no further differentiation events.  That is, each
  proliferation event gives rise to two effector memory cells that are
  identical to the dividing (or mother) cell. The rate of first
  division is $\alpha_m$ and subsequent divisions take place with rate
  $\lambda_m$ (see Figure~\ref{cd8-immune-response-ipm}).
 
\item {\bf Thymic output}: naive T~cells are exported from the thymus
  to the peripheral blood compartment, $B$, as described in
  Section~\ref{method-model-homeostasis}.

\item {\bf Division}: we consider homeostatic
  proliferation only for N, CM and EM T~cells, as described in
  Section~\ref{method-model-homeostasis}.  We neglect homeostatic
  proliferation for the populations $I$, $J$ and $K$, as they are
  intermediate cells generated during the immune response.
 
\item {\bf Cell death}: we assume a constant per cell rate of death
  and we denote it by $\mu$.  Each cell type can die with rates
  $\mu_n$, $\mu_c$, $\mu_m$, and $\mu_e$, respectively.  We assume
  that the intermediate populations $I$, $J$ and $K$ have the same
  death rates as their parent subtypes, N, E and EM, respectively.

\item {\bf Cell migration}: naive and CM T~cells have the same
  migratory behaviour and described in
  Section~\ref{method-model-homeostasis}. EM and effector T~cells have
  the same migratory behaviour, as described for EM T~cells in
  Section~\ref{method-model-homeostasis}.  We assume that the
  intermediate populations $I$, $J$ and $K$ do not migrate, thus, they
  only need to be considered in the dLN compartment.
 
\end{itemize}

The ordinary differential equations (ODEs) that describe the immune
response for the increasing potential model are presented below and a
description of the model is provided in
Figure~\ref{cd8-immune-response-ipm}.  Each proliferation,
differentiation and death rate is labelled with a subscript that
corresponds to the population subset under consideration, $n$, $c$,
$m$ and $e$ for naive, central memory, effector memory, and effector
cells, respectively.  In the dLNs, we can write
\begin{eqnarray}
\frac{dN}{dt} &=&
 \theta_n  \; N \left( 1 - \frac{N}{N_c \; \kappa_n} \right) 
 - \alpha_n \; N
 - \mu_n \; N
 -\gamma \; N 
 + \phi \; N^{(B)}
\; ,
\\ 
\frac{dI_1}{dt} &= &
2 \; \alpha_n \; N
- \lambda_n \; I_1
 - \mu_n \; I_1
 \; ,
\\
\frac{dI_i}{dt} &= &
2 \; \lambda_n \; I_{i-1}
 - \lambda_n \; I_{i}
-\mu_n \; I_{i}
\; ,
\quad
\forall i = 2, \ldots, g_{n}-1
\; ,
\\
\frac{dE}{dt} &=&
2 \; \lambda_n \; I_{g_n-1}
- \alpha_e \; E 
- \mu_e \; E 
-\zeta \; E
\; ,
\\ 
\frac{dJ_1}{dt} &= &
2 \; \alpha_e \; E
- \lambda_e \; J_1
 - \mu_e \; J_1
 \; ,
\\
\frac{dJ_i}{dt} &= &
2 \; \lambda_e \; J_{i-1}
 - \lambda_e \; J_{i}
-\mu_e \; J_{i}
\; ,
\quad
\forall i = 2, \ldots, g_{e}-1
\; ,
\\
\frac{dM}{dt} &=&
2 \; \lambda_e \; J_{g_e-1}
+
 \theta_m \; M \left( 1-\frac{M}{N_c \; \kappa_m} \right)
 - \alpha_m \; M 
 - \mu_m \; M 
 -\zeta \; M
 \; .
 \\
\frac{dK_1}{dt} &= &
2 \; \alpha_m \; M
- \lambda_m \; K_1
 - \mu_m \; K_1
 \; ,
\\
\frac{dK_i}{dt} &= &
2 \; \lambda_m \; K_{i-1}
 - \lambda_m \; K_{i}
-\mu_m \; K_{i}
\; ,
\quad
\forall i = 2, \ldots, g_{m}-1
\; ,
\\
\frac{dC_0}{dt} &=&
 2 \; \lambda_m \; K_{g_m-1}
+
\theta_c \; C_0 \left( 1-\frac{C}{N_c \; \kappa_c} \right)
- \alpha_c \; C_0 
- \mu_c \; C_0 
-\gamma \; C_0
+ \phi \;  C_0^{(B)}
\; ,
\\ 
\frac{dC_1}{dt} &= &
2 \; \alpha_c \; C_0
+
\theta_c \; C_1 \left( 1-\frac{C}{N_c \; \kappa_c} \right)
- \lambda_c \; C_1
- \mu_c \; C_1
-\gamma \; C_1
+ \phi \;  C_1^{(B)}
 \; ,
\\
\frac{dC_i}{dt} &= &
2 \; \lambda_c \; C_{i-1}
+
\theta_c \; C_i \left( 1-\frac{C}{N_c \; \kappa_c} \right)
 - \lambda_c \; C_{i}
-\mu_c \; C_{i}
-\gamma \; C_i
+ \phi \;  C_i^{(B)}
\; ,
\quad
\forall i = 2, \ldots, g_{c}-1
\; ,
\\
\frac{dC_{g_c}}{dt} &= &
2 \; \lambda_c \; C_{g_c-1} 
+
\theta_c \; C_{g_c} \left( 1-\frac{C}{N_c \; \kappa_c} \right)
-\mu_c \; C_{g_c}
-\gamma \; C_{g_c}
+ \phi \;  C_{g_c}^{(B)}
 \; ,
\end{eqnarray}
where the population of central memory cells in the draining lymph nodes is given by
$C = \sum_{i=0}^{{g_c}} C_i$.

In blood and resting LNs, we have
\begin{eqnarray}
 \frac{dN^{(B)}}{dt} &=&
 \theta_n  \; N^{(B)} \left( 1 - \frac{N^{(B)}}{N_c \; \kappa_n} \right) 
 - \mu_n \; N^{(B)}
 - \phi \; N^{(B)}
 +\gamma \; N 
 + N_c \; \delta  
\; ,
\label{ipm-naive}
\\ 
 \frac{dE^{(B)}}{dt} &=& 
  - \mu_e \; E^{(B)} 
  -\xi \; E^{(B)}
+  \zeta \; E
  +\psi \; E^{(S)}
 \; ,
\label{ipm-effector}
 \\
\frac{dM_0^{(B)}}{dt} &=&
 \theta_m \; M_0^{(B)} \left(1-\frac{M^{(B)}}{N_c \; \kappa_m} \right)
 - \mu_m \; M_0^{(B)} 
 -\xi \; M_0^{(B)}
 +\zeta \; M
 + \psi \;  M_0^{(S)}
\; ,
\label{ipm-m0}
\\ 
\frac{dM_j^{(B)}}{dt} &=&
 \theta_m \; M_j^{(B)} \left(1-\frac{M^{(B)}}{N_c \; \kappa_m} \right)
 - \mu_m \; M_j^{(B)} 
 -\xi \; M_j^{(B)}
 + \psi \;  M_j^{(S)}
\; ,
\quad
\forall j = 1, \ldots, g_{m}
\label{ipm-mj}
\\ 
\frac{dC_i^{(B)}}{dt} &=&
\theta_c \; C_i^{(B)} \left( 1-\frac{C^{(B)}}{N_c \; \kappa_c} \right)
- \mu_c \; C_i^{(B)}
- \phi \;  C_i^{(B)}
+\gamma \; C_i
\; ,
\quad
\forall i = 0, \ldots, g_{c}
\; ,
\label{ipm-ci}
\end{eqnarray}
where the population of effector memory cells in the blood compartment is given by
$M^{(B)} = \sum_{i=0}^{{g_m}} M_i^{(B)}$ and
 the population of central memory cells in the blood compartment is given by
$C^{(B)} = \sum_{i=0}^{{g_c}} C_i^{(B)}$.

Finally, in the skin we have
\begin{eqnarray}
\frac{dE^{(S)}}{dt} &=&
- \alpha_e \; E^{(S)}
  -\mu_e \; E^{(S)}
   -\psi \; E^{(S)} 
   + \xi \;  E^{(B)} 
\; ,
\\    
\frac{dJ_1^{(S)}}{dt} &= &
2 \; \alpha_e \; E^{(S)}
- \lambda_e \; J_1^{(S)}
 - \mu_e \; J_1^{(S)}
 \; ,
\\
\frac{dJ_i^{(S)}}{dt} &= &
2 \; \lambda_e \; J_{i-1}^{(S)}
 - \lambda_e \; J_{i}^{(S)}
-\mu_e \; J_{i}^{(S)}
\; ,
\quad
\forall i = 2, \ldots, g_{e}-1
\; ,
\\
\frac{dM_0^{(S)}}{dt} &=&
2 \; \lambda_e \; J_{g_e-1}^{(S)}
+
 \theta_m \; M_0^{(S)} \left( 1-\frac{M^{(S)}}{N_c \; \kappa_m} \right)
 - \alpha_m \; M_0^{(S)}
 - \mu_m \; M_0^{(S)} 
 - \psi \;  M_0^{(S)}
  +\xi \; M_0^{(B)}
  \; ,
 \\
\frac{dM_1^{(S)}}{dt} &=&
2 \; \alpha_m \; M_0^{(S)}
+
 \theta_m \; M_1^{(S)} \left( 1-\frac{M^{(S)}}{N_c \; \kappa_m} \right)
 - \lambda_m \; M_1^{(S)}
 - \mu_m \; M_1^{(S)} 
 - \psi \;  M_1^{(S)}
  +\xi \; M_1^{(B)}
  \; ,
 \\
\frac{dM_i^{(S)}}{dt} &=&
2 \; \lambda_m \; M_{i-1}^{(S)}
+
 \theta_m \; M_i^{(S)} \left( 1-\frac{M^{(S)}}{N_c \; \kappa_m} \right)
 - \lambda_m \; M_i^{(S)}
 - \mu_m \; M_i^{(S)} 
 - \psi \;  M_i^{(S)}
  +\xi \; M_i^{(B)}
  \; ,
  \quad
\forall i = 2, \ldots, g_{m}-1
 \\
\frac{dM_{g_m}^{(S)}}{dt} &=&
2 \; \lambda_m \; M_{g_m-1}^{(S)}
+
 \theta_m \; M_{g_m}^{(S)} \left( 1-\frac{M^{(S)}}{N_c \; \kappa_m} \right)
 - \mu_m \; M_{g_m}^{(S)} 
 - \psi \;  M_{g_m}^{(S)}
  +\xi \; M_{g_m}^{(B)}
\; ,
\end{eqnarray}
where the population of effector memory cells in the skin  compartment is given by
$M^{(S)} = \sum_{i=0}^{{g_m}} M_i^{(S)}$.

We note that, during an immune response, the dynamics of the
non-specific CD8\plus\ naive T~cells, $N_r$ and $N_r^{(B)}$ is
regulated by cell division and given by
by~\eqref{non-specific-naive-dLN} and~\eqref{non-specific-naive-B}
(see Section~\ref{method-model-homeostasis}).

%%%%%%%%%%%%%%%%%%%%%%%%%%%%%%%%%%%%

The IP mathematical model allows us to follow in time and to quantify
the number of total specific CD8\plus\ T~cells generated during an
immune response in each spatial compartment. That is, at any time $t
\ge 0$, these populations (in the increasing potential model) are
given by
\begin{eqnarray}
T_{8} (t) &=& 
N (t)  + \sum_{i=1}^{g_n-1} I_i (t) + E (t) +\sum_{i=1}^{g_e-1} J_i(t) + M (t) + \sum_{i=1}^{g_m-1} K_i (t) +  \sum_{i=0}^{g_c} C_i(t) 
\; ,
\\ 
T_{8}^{(B)} (t) &=& N^{(B)} (t) + E^{(B)}(t)  +  \sum_{i=0}^{g_m} M_i^{(B)}(t)  +  \sum_{i=0}^{g_c} C_i^{(B)}(t) 
\; ,
\label{t8b-ipm}
\\ 
T_{8}^{(S)}  (t) &=&   E^{(S)} (t)  + \sum_{i=1}^{g_e-1}J_i^{(S)} (t) +  \sum_{i=0}^{g_m} M_i^{(S)}(t) 
\; ,
\end{eqnarray}
where $T_{8} (t)$ is the total number of specific CD8\plus\ T~cells in
the dLN compartment, $T_{8}^{(B)} (t)$ is the total number of specific
CD8\plus\ T~cells in the blood compartment, and $T_{8}^{(S)} (t)$ is
the total number of specific CD8\plus\ T~cells in the skin
compartment.  As mentioned in
Section~\ref{method-model-immune-response-DPM}, the total number of
CD8\plus\ T~cells in each spatial compartment at time $t$, is given by
$T_{8} (t) + N_r (t)$, $T_{8}^{(B)} (t) + N_r^{(B)} (t)$ and
$T_{8}^{(S)} (t)$, respectively.  The IPM will also allows us to
calculate the total fraction of specific CD8\plus\ T~cells out of the
total CD8\plus\ T~cell count in blood and the fraction of specific
CD8\plus\ T~cells of a given subtype in blood. These fractions have
already been defined in~\eqref{def-fraction}.  For the increasing
potential model, we note that ${T_{8}^{(B)}}(t)$ is given
by~\eqref{t8b-ipm}, ${N^{(B)}}(t)$ solves~\eqref{ipm-naive},
${E^{(B)}}(t)$ solves~\eqref{ipm-effector}, ${M^{(B)}}(t)$ is given by
$\sum_{i=0}^{{g_m}} \; M_i^{(B)} (t)$, where ${M^{(B)}_0 }(t)$
solves~\eqref{ipm-m0} and ${M^{(B)}_j}(t)$ solves~\eqref{ipm-mj} for
$1 \le j \le g_m$, and ${C^{(B)}}(t)$ is given by $\sum_{i=0}^{{g_c}}
\; C_i^{(B)} (t)$, where ${C^{(B)}_j }(t)$ solves~\eqref{ipm-ci}.

In order to solve the previous set of ODEs, we need to provide initial
conditions for the different cell types considered in the IPM immune
response model.  We assume that at the initial time there are no
intermediate cells, that is
\begin{eqnarray}
I_{i}(0) = 0 \; , 
\quad &\forall& i = 1, \ldots, g_{n}-1
\; ,
\\
J_{i}(0) = J^{(S)}_{i}(0) =0 \; , 
\quad &\forall& i = 1, \ldots, g_{e}-1
\; ,
\\
K_{i}(0) = M_{i}^{(B)}(0)= M_{i}^{(S)}(0)  =0 \; , 
\quad &\forall&l i = 1, \ldots, g_{m}-1
\; ,
\\
C_{i}(0) = C_{i}^{(B)}(0) =0 \; , 
\quad &\forall& i = 1, \ldots, g_{c}
\; .
\end{eqnarray}
For a primary immune response, and as discussed in
Section~\ref{method-model-immune-response-DPM}, we assume that at the
initial time there are no specific central memory, effector memory or
effector T~cells in any spatial compartment.  In the case of a
secondary immune response, as described in
Section~\ref{method-model-immune-response-DPM}, there are no effector
cells at the initial time.  Initial conditions for the memory subsets
(CM or EM), $C_0, M_0, C_0^{(B)}, M_0^{(B)}$ and $M_0^{(S)}$, will be
taken to be the number of cells that have been generated during a
primary immune response and maintained by homeostasis in each of the
spatial compartments [see
Section~\ref{method-computational-algorithm}].  Initial conditions for
naive cells (non-specific and antigen-specific) will be considered and
specified in Section~\ref{method-computational-algorithm}, where we
discuss the computational algorithm used to solve the ODEs of the
combined CD8\plus\ T~cell homeostasis and immune response model.

%%%%%%%%%%%%%%%%%%%%%%%%%%%%%%%%%%%%%%

\subsection{Parameter estimates from published literature}
\label{method-model-parameters}

%%%%%%%%%%%%%%%%%%%%%%%%%%%%%%%%%%%%%%

In this Section, we make use of the published literature to obtain
parameter estimates for the number of cells, carrying capacities,
division rates, death rates and thymic export rates.

%%%%%%%%%%%%%%%%%%%%%%%%%%%%%%%%%%%%%%

\subsubsection{Cell numbers}
\label{method-cell-numbers}

%%%%%%%%%%%%%%%%%%%%%%%%%%%%%%%%%%%%%%

\begin{enumerate}[(i)]

\item
{\bf Naive TCR diversity.}

The human naive TCR diversity  (or number of different TCR clonotypes) has been estimated to be
$2.5 \times 10^7$ in Ref.~\cite{arstila1999direct}.
We denote this parameter by $N_R$.

\item
{\bf Total number of lymphocytes (in lymph nodes and blood).}

The total number of lymphocytes in humans has been estimated to be
$4.60 \times 10^{11}$ from a quantitative assessment of lymphocyte
numbers in mucosal and lymphoid tissues~\cite{ganusov2007most}.  These
authors also report that the total number of lymphocytes in the lymph
nodes is $1.90 \times 10^{11}$ and the total number of lymphocytes in
the blood is $1.0 \times 10^{10}$ (see Table~1 of
Ref.~\cite{ganusov2007most}).

\item 
{\bf Total number of CD8\plus\ T~cells (in lymph nodes and blood).}

In Ref.~\cite{ganusov2007most} the authors make use of the
distribution of CD8\plus\ T~cells in different organs and tissues
(provided in Ref.~\cite{westermann1992distribution}), to estimate that
the total number of CD8\plus\ T~cells in humans is $1.09 \times
10^{11}$.  They are also able to estimate that the total number of
CD8\plus\ T~cells in the LNs is $3.8 \times 10^{10}$ and $2.5 \times
10^9$ in blood (see Table~3 of Ref.~\cite{ganusov2007most}).  We also
note that making use of the estimates provided in
Ref.~\cite{vrisekoop2008sparse} (Table~3 and Table~S1), the mean
number of CD8\plus\ T~cells (for five healthy individuals aged between
20 and 25) in 5 litres (L) of blood is $2.59 \times 10^9$. These
authors have assumed that the total blood volume in humans is 5L and
that $2\%$ of all T~cells reside in
blood~\cite{westermann1992distribution}.  Given these two different
estimates for the total number of CD8\plus\ T~cells in blood, we take
their average to obtain $2.55 \times 10^9$.

\item 
{\bf Total number of naive CD8\plus\ T~cells (in lymph nodes and blood).}

Recent experimental observations allow us to calculate the number of
naive CD8\plus\ T~cells in the lymph nodes and blood in
humans~\cite{sathaliyawala2013distribution}.  The frequency of naive
CD8\plus\ T~cells is found to be 48\% in the inguinal lymph nodes,
39\% in the lung lymph nodes and 40\% in the mesenteric lymph nodes
(see Figure~3B of Ref.~\cite{sathaliyawala2013distribution}).  If we
assume that 42.3\% of all CD8\plus\ T~cells in the LNs are naive
(average of the above percentages), we then obtain that $1.61
\tenpow^{10} = 0.423 \times 3.8 \tenpow^{10}$ is the number of naive
CD8\plus\ T~cells in the LNs.  In blood the frequency of naive
CD8\plus\ T~cells is found to be $39\%$ (see Figure~3B of
Ref.~\cite{sathaliyawala2013distribution}), which implies that $9.95
\times 10^8 = 0.39 \times 2.55 \times 10^9$ is the number of naive
CD8\plus\ T~cells in blood.  This last number can be also estimated
from Table~S1 of Ref.~\cite{vrisekoop2008sparse} .  These authors
estimate that 56.6\% of all CD8\plus\ T~cells in blood are naive.  If
there are $2.55 \times 10^9$ CD8\plus\ T~cells in blood, we conclude
that there are $1.44 \times 10^9$ naive CD8\plus\ T~cells in blood.
Given these two different estimates for the number of naive CD8\plus\
T~cells in blood, we take their average to obtain $1.22 \times 10^9$.
A recent estimate from Ref.~\cite{westera2015lymphocyte} (Table~2)
indicates that the (averaged over young and aged individuals) number
of naive CD8\plus\ T~cells in blood is 142.5 cells per $\mu$L. Thus,
these authors conclude that the total number of naive CD8\plus\
T~cells in blood is $7.13 \times 10^8$.  From the two different
estimates, $1.22 \times 10^9$ and $7.13 \times 10^8$, we assume that
their average, $9.7 \times 10^8$ is the total number of naive
CD8\plus\ T~cells in blood for humans.  Finally, if we assume, as done
by the authors of Ref.~\cite{vrisekoop2008sparse}, that $2\%$ of all
T~cells reside in blood~\cite{westermann1992distribution}, we obtain
$4.85 \times 10^{10}$ to be the total number of naive CD8\plus\
T~cells in humans.  We denote this parameter by $N_8$.

\item 
{\bf Number of naive CD8\plus\ T~cells per clonotype (in lymph nodes and blood).}

Given the TCR diversity, $N_R$, we can now estimate the number of  naive CD8\plus\ T~cells per clonotype
 in the LNs and in blood.
For the LNs,  we obtain $644=1.61 \tenpow^{10}/N_R$ and for blood, we have 
$39 = 9.7 \times 10^8/N_R$.

\item 
{\bf Number of lymph nodes (LNs) in the human body.}

The mathematical models developed in this manuscript consider three
different spatial compartments: the draining lymph nodes (dLNs), the
resting lymph nodes and blood (B) and skin (S).  Naive cells are found
in the first two compartments, and we need to calculate the number of
naive CD8\plus\ T~cells per clonotype in each of these two spatial
locations (dLNs and B).  Ref.~\cite{goroll2012primary} estimates the
number of individual LNs in humans to be 600, as well as
Ref.~\cite{susan2008greys}. Ref.~\cite{pabst2007plasticity} estimates
this number to be in the range $[600,700]$, with a mean of 650.  If we
take the average of 600 and 650, we obtain 625 to be the the number of
individual LNs in humans.  Our interest in skin sensitisation
scenarios (or cutaneous vaccination) leads us to consider the axilla
site to be the dLN compartment~\cite{o1983basic}.  In the axilla site,
it is reported that the number of individual LNs is in the range of
$[20,30]$~\cite{susan2008greys}.  We will consider then that there are
25 individual LNs in the axilla site, which correspond to the draining
LNs. Thus, for our purposes we assume there are 25 dLNs and 600
resting LNs.

\item 
{\bf Number of naive CD8\plus\ T~cells per clonotype in dLN compartment.}

Our previous estimates allow us to conclude that the number of naive
CD8\plus\ T~cells per clonotype in an individual LN is 644/625 (number
of naive CD8\plus\ T~cells per clonotype in LNs divided by the number
of individual LNs in humans).  We approximate 644/625 by 1, that is,
there is one naive CD8\plus\ T~cell per clonotype in an individual
LN. As the dLNs have 25 individual LNs, we conclude that
$n_c^{(dLN)}$, the number of naive CD8\plus\ T~cells per clonotype in
the dLN compartment is 25. Finally, we note that the total number of
naive CD8\plus\ T~cells in the dLN compartment is $N_R \times
n_c^{(dLN)}$.

\item 
{\bf Number of naive CD8\plus\ T~cells per clonotype in B compartment.}

We now compute the number of naive CD8\plus\ T~cells per clonotype in
the blood compartment, which is composed of the resting lymph nodes
and blood. Our previous estimates lead to 600 resting LNs.  Thus,
there are 600 naive CD8\plus\ T~cells per clonotype in the resting
LNs.  We had previously obtained 39 naive CD8\plus\ T~cells per
clonotype in blood.  Together these results imply that the spatial B
compartment consists of 639 naive CD8\plus\ T~cells per clonotype. We
denote this number by $n_c^{(B)}$. If we multiply $n_c^{(B)}$ by
$N_R$, we obtain the total number of naive CD8\plus\ T~cells in the
blood compartment.
 
\item
{\bf Number of specific and non-specific naive CD8\plus\ T~cells in dLN and in B compartments.}

Let us define $N_c$ to be the number of different TCR clonotypes
driven into an immune response.  This implies that $N_R - N_c$ is the
number of non-specific TCR clonotypes.  As $n_c^{(dLN)}$ is the number
of naive CD8\plus\ T~cells per clonotype in the dLN compartment, this
means that $N_c \times n_c^{(dLN)}$ and $(N_R-N_c) \times n_c^{(dLN)}$
is the total number of antigen-specific and non-specific,
respectively, naive CD8\plus\ T~cells in the dLN compartment.  For the
B compartment, the previous results generalise to $N_c \times
n_c^{(B)}$ and $(N_R-N_c) \times n_c^{(B)}$, as the total number of
specific and non-specific, respectively, naive CD8\plus\ T~cells in
the B compartment.

\end{enumerate}

%%%%%%%%%%%%%%%%%%%%%%%%%%%%%%%%%%%%%%

\subsubsection{Carrying capacities}
\label{method-carrying-capacities}

%%%%%%%%%%%%%%%%%%%%%%%%%%%%%%%%%%%%%%

\begin{enumerate}[(i)]

\item
{\bf Carrying capacity per clonotype
 of the naive CD8\plus\ T~cell population.}
 
We first estimate the carrying capacity per clonotype of the naive
CD8\plus\ T~cell population, $\kappa_n$. We assume that $\kappa_n$ is
given by $\frac{N_8}{N_R}$, as $N_8$ is the total number of naive
CD8\plus\ T~cells in humans and $N_R$ its TCR diversity. This means
that if there are $N_c$ clones driven into an immune response, the
carrying capacity of the specific naive CD8\plus\ T~cell population is
$N_c \times \kappa_n$.

\item
{\bf Carrying capacity of the non-specific
 naive CD8\plus\ T~cell population.}
 
For the non-specific naive CD8\plus\ T~cell population its carrying
capacity is denoted by $\kappa_r$ and it is given by $(N_R-N_c) \times
\kappa_n$.

\item
{\bf Carrying capacity per clonotype
 of the central and effector memory CD8\plus\ T~cell population.}

We now estimate the carrying capacity per clonotype of the central,
$\kappa_c$, and effector, $\kappa_m$ memory CD8\plus\ T~cell
populations. To this end, we make use of
Table~\ref{table-memory-percentage}, that provides measurements of the
fraction of both central and effector memory CD8\plus\ T~cell
populations in blood.  Thus, the number of central and effector memory
CD8\plus\ T~cells in blood is given by $4.09 \times 10^8= 16.02 \%
\times 2.55 \times 10^9$ and $4.85 \times 10^8=19.03 \% \times 2.55
\times 10^9$, and the total number of memory CD8\plus\ T~cells in
blood is $8.94 \times 10^8$.  A recent estimate from
Ref.~\cite{westera2015lymphocyte} (Table~2) indicates that the
(averaged over young and aged individuals) number of memory CD8\plus\
T~cells in blood is 101.5 cells per $\mu$L. This implies that the
total number of memory CD8\plus\ T~cells in blood (assuming 5L of
blood in humans) is $5.08 \times 10^8$. We average over these two
different estimates to conclude that there are $7.01 \times 10^8$
memory CD8\plus\ T~cells in blood.  Assuming that $2\%$ of all T~cells
reside in blood~\cite{westermann1992distribution}, we obtain $3.51
\times 10^{10}$ to be the total number of memory CD8\plus\ T~cells in
humans.  In order to obtain the carrying capacity per clonotype, we
require an estimate of memory TCR diversity.
Ref.~\cite{qi2014diversity} reports that the number of clonotypes
observed in the human memory population is about 2$\times 10^5$ (see
Figure~1D of Ref.~\cite{qi2014diversity}). A similar diversity was
estimated in Ref.~\cite{arstila1999direct}, 1-2$\times 10^5$ different
$\beta$ chains, which on average only paired with a single $\alpha$
chain each.  If we assume that $\kappa_c=\kappa_m$, we conclude that
$\kappa_c=\kappa_m=87,750=\frac{3.51 \times 10^{10}}{2 \times 2 \times
  10^5}$.

\end{enumerate}

   %%%%%%%%%%%%%%%%%

\begin{table}[h!]
\begin{center}
\begin{tabular}{| c || c | c | c | c | c | c | }
 \hline
Reference  & Markers  & Measure (\%)& Naive  & Central memory    & Effector memory  & Effector 
\\
\hline
\hline
   Table~2, Ref.~\cite{chamuleau2009immune} & CD45RA and CD27 & median & 44.00 & 28.00 & 7.00  & 7.00 
   \\
   \hline
   Figure~1, Ref.~\cite{riddell2015multifunctional}  & CD45RA and CD27 & mean  & 31.12 & 22.05 & 19.68 & 30.53 
   \\
   \hline
    Table~2, Ref.~\cite{roos2000changes}  & CD45RA and CD27 & median & 52.00 & 33.00 & 3.00  & 11.50 
    \\
    \hline
    Figure~1, Ref.~\cite{hong2004age}  & CD45RA and CCR7 & mean  & 27.82 & 6.65  & 40.06 & 23.09 
    \\
    \hline
    Figure~3B, Ref.~\cite{sathaliyawala2013distribution}  & CD45RA and CCR7 & mean  & 39.00 & 8.03  & 23.00 & 30.19 
    \\
    \hline
    Table~1, Ref.~\cite{saule2006accumulation}  & CD45RA and CCR7 & mean   & 40.40 & 8.55  & 18.45 & 26.80 
     \\
     \hline
    Table S3, Ref.~\cite{thome2014spatial} & CD45RA and CCR7 &  mean     & 44.55 & 5.85  & 22    & 27.61 
    \\
    \hline
    \hline
          &       & {\bf average} & {39.84} & {16.02} & {19.03} & {22.39} 
          \\
          \hline
\end{tabular}
\caption{Reported (percentage) distribution of CD8\plus\ T~cell subsets in blood.}
\label{table-memory-percentage}
\end{center}
\end{table}          

%%%%%%%%%%%%%%%%%%%%%%

\subsubsection{Division rates}
\label{method-homeostatic-proliferation}

\begin{enumerate}[(i)]

\item {\bf Total production rate of naive CD8\plus\ T~cells.}  In
  Ref.~\cite{vrisekoop2008sparse} (Table~3), the median naive
  CD8\plus\ T~cell production per day is estimated to be $2.39 \times
  10^7$ cells.  The authors of Ref.~\cite{westera2015lymphocyte}
  (Table~2) report a production rate (averaged over young and aged
  individuals) of $1.55 \times 10 ^7$ cells per day.  We consider the
  average of these two different estimates to yield $1.97 \times 10
  ^7$ naive CD8\plus\ T~cells per day.

\item {\bf Thymic rate of naive CD8\plus\ T~cells.}  For a human
  adult, 20\% of the production of naive CD8\plus\ T~cells is
  contributed from the thymus~\cite{murray2003naive} and the remaining
  80\% of production is from peripheral proliferation.  In
  Ref.~\cite{den2012maintenance} the authors report that thymic
  production is 11\% of the total. We take the average of these two
  estimates, $15.5\%$, to be the thymic contribution to naive
  CD8\plus\ T~cell production.

\item
{\bf Division rate of naive CD8\plus\ T~cells.}

The above estimate leads to $84.5\%$ to be the peripheral contribution
to naive CD8\plus\ T~cell production.  Given the total production
rate, $1.97 \times 10 ^7$ naive CD8\plus\ T~cells per day, this means
that $1.66 \times 10 ^7$ naive CD8\plus\ T~cells per day are generated
by peripheral proliferation. If $N_8$ is the total number of naive
CD8\plus\ T~cells, then $3.42 \times 10^{-4}=\frac{1.66 \times 10
  ^7}{N_8}$ per day is the homeostatic naive division rate.  The
authors of Ref.~\cite{mclean1995vivo} estimated that naive T~cells
divide once every 3.5 years in the periphery, which leads to $7.83
\times 10^{-4}$ per day as the homeostatic naive division
rate. We take the average of these two different estimates to obtain
$\theta_n=5.63 \times 10^{-4}$ per day.

\item
{\bf Division rate of central and effector memory CD8\plus\ T~cells.}

In order to estimate $\theta_c$ and $\theta_m$, the division rate of
central and effector memory CD8\plus\ T~cells, respectively, we make
use of the estimated proliferation rates provided in
Ref.~\cite{mclean1995vivo}. The authors estimated that memory T~cells
divide once every 22 weeks, which leads to $6.49 \times 10^{-3}$ per
day. We assume $\theta_m=\theta_c$, since the authors do not
distinguish between CM and EM T~cells.

\end{enumerate}

%%%%%%%%%%%%%%%%%%%%%%%%%%%%%%%%%%%%%%

\subsubsection{Death rates}
\label{method-death-rates}

\begin{enumerate}[(i)]

\item
{\bf Death rate of naive CD8\plus\ T~cells.}

Our first estimate for the death rate of CD8\plus\ naive T~cells,
$\mu_n$, has been obtained making use of
Ref.~\cite{vrisekoop2008sparse} (Table~2).  In this reference, the
authors provide the median half-life of naive CD8\plus\ T~cells to be
2,374 days. We make use of the fact that the death rate is given by
$\frac{\log 2}{\text{half-life}}$ to obtain $2.92 \times10^{-4}$ per
day.  A second estimate provides an average turnover rate of 0.06\%
per day (averaged over young and aged individuals).  This is
equivalent to a lifespan of 4.5 years for naive CD8\plus\ naive
T~cells or to a death rate of $6.09 \times10^{-4}$ per day (see
Table~2 of Ref.~\cite{westera2015lymphocyte}).  We take the average of
the above estimates to obtain $\mu_n=4.51\times10^{-4}$ per day.
		
\item
{\bf Death rate of central and effector memory CD8\plus\ T~cells.}

The authors of Ref.~\cite{vrisekoop2008sparse} (Table~2) also provide
the median half-life of memory CD8\plus\ T~cells to be 244 days. Thus,
we estimate $2.84 \times10^{-3}$ per day.  A second estimate provides
an average turnover rate of 0.45\% per day (averaged over young and
aged individuals).  This is equivalent to a lifespan of approximately
222 days for memory CD8\plus\ T~cells or to a death rate of $4.05
\times10^{-3}$ per day (see Table~2 of
Ref.~\cite{westera2015lymphocyte}).  We take the average of the above
estimates to obtain $\mu_c=3.67\times10^{-3}$ per day.  We assume
$\mu_m=\mu_c$, since the authors do not distinguish between CM and EM
T~cells.

\item
{\bf Death rate of effector CD8\plus\ T~cells.}

Finally, the death rate of effector CD8\plus\ T~cells, $\mu_e$, has
been estimated from the observation that effector CD8\plus\ T~cells
have a lifespan of about four weeks~\cite{ahmed2011insights}. As the
death rate = $\frac{1}{\text{lifespan}}$, we obtain $\mu_e = 3.57
\times10^{-2}$ per day.

\end{enumerate}

%%%%%%%%%%%%%%%%%%%%%%%%%%%%%%%%%%%%%%

\subsubsection{Thymic export rates}
\label{method-thymic-export}

\begin{enumerate}[(i)]

\item
{\bf Thymic rate per CD8\plus\ T~cell clonotype.}

We have estimated above that $15.5\%$ is the thymic contribution to
naive CD8\plus\ T~cell production and that the total production rate
of naive CD8\plus\ T~cells is $1.97 \times 10 ^7$ cells per day. This
implies that $0.12 = 15.5\% \times 1.97 \times 10 ^7/N_R$ is the
thymic output per clonotype and per day.  We denote this parameter by
$\delta$. If $N_c$ TCR clonotypes are driven into an immune response,
the total rate of thymic export for antigen-specific naive CD8\plus\
T~cells is $N_c \times \delta$.

\item
{\bf Thymic rate for non-specific CD8\plus\ naive T~cells.}

Given the estimate for $\delta$ above, and that there are $N_R-N_c$ non-specific 
naive CD8\plus\ T~cell clonotypes, the total rate of thymic export for non-specific 
 naive CD8\plus\ T~cells is $\delta_R= (N_R-N_c) \times \delta$.

\end{enumerate}

%%%%%%%%%%%%%%%%%%%%%%%%%%%%%%%%%%%%%%

\subsubsection{Migration rates}
\label{method-migration-rates}

Recent estimates from mice of the timescales of T~cell migration
provide a range between 0.5 minutes to 3
days~\cite{ganusov2014mathematical}.  Naive T~cells in mice have been
estimated to reside in the lymph nodes for timescales that range
between 0.5 and three days~\cite{mandl2012quantification}. Finally,
the timescale of effector T~cells to exit from the lymph nodes have
been estimated to be of the order of minutes to hours (see Figure~2 in
Ref.~\cite{schwab2007finding}).  Given these estimates, we choose to
assume that the timescales of migration in our model, $\gamma^{-1},
\phi^{-1},\zeta^{-1},\psi^{-1}$, and $\xi^{-1}$ are not fixed
parameters and will be sampled from a uniform distribution with
minimum one minute and maximum ten days, as shown in
Table~\ref{table-parameters-bayesian}.

%%%%%%%%%

\subsubsection{Changes in lymph node influx and efflux}
\label{method-shut-down}

\begin{enumerate}[(i)]

\item
{\bf Change in influx rate.}

Bovine data indicate that during an initial period of 3.5 days, lymph
node influx increases from $4.67 \times 10^6$ to $15.04 \times 10^6$
T~cells per hour.  This means a 3.22 fold increase during the first
3.5 days post-infection.  This will be implemented in our model as
follows: during the first 3.5 days post-infection (post-challenge or
post-vaccination) the migration rate $\phi$ will be increased by a
factor of 3.22 to become $3.22 \times \phi$.

\item
{\bf Change in efflux rate.}

The output of lymphocytes in the efferent lymph (from a lymph node
draining a PPD-induced delayed type hypersensitivity reaction in
sheep) has been shown to decrease significantly over the first 24
hours~\cite{seabrook2005novel}. We have taken the average of the
fold-reduction in output observed at 9, 12, 15, 21 and 24 hours, which
was 0.26, 0.19, 0.16, 0.23, 0.29, and 0.37 (see Figure~1 in
Ref.~\cite{seabrook2005novel}), to obtain a value of 3.4
fold-reduction during a period of twenty four hours.  A different set
of bovine data indicates a three fold-reduction in the efflux rate at
12 hours post-exposure to orf virus, which reaches resting levels by
24 hours. We take this set of data to obtain a 3 fold-reduction during
a period of twenty four hours.  The average of these two estimates
leads to a 3.2 fold-reduction in efflux during a period of twenty four
hours.  This will be implemented in our model as follows: during the
first 2 days post-infection (post-challenge or post-vaccination) the
migration rates $\gamma$ (for naive and central memory T~cells) and
$\zeta$ (for effector and effector memory T~cells) will be decreased
by a factor of 3.2 to become $\gamma/3.2$ or $\zeta/3.2$,
respectively.

\end{enumerate}

%%%%%%%%%

\subsubsection{Programme of proliferation: number of generations}
\label{method-generations}

A recent mathematical modelling effort, in combination with human
CD8\plus\ YFV kinetic data~\cite{miller2008human,akondy2009yellow},
has estimated that during the differentiation process CD8\plus\
T~cells undergo fewer than nine divisions~\cite{le2014mathematical}.
Given this estimate and our division-linked differentiation
hypothesis, we choose to assume that the number of divisions, encoded
in the parameters $g_n$, $g_c$, $g_m$ and $g_e$, to undergo
differentiation are not fixed parameters and will be sampled from a
uniform distribution with minimum one division and maximum eleven
divisions (see Table~\ref{table-parameters-bayesian}).

%%%%%%%%%%%%%%%%%%%%%%%%%%%%%%%%%%%%%%

\subsubsection{Time to first division}
\label{method-first-division}

The time to first division for CD8\plus\ T~cells has been measured in
the mouse OT-I model~\cite{hommel2007tcr} and estimated to be
approximately two days, with a range of 29 to 55 hours. In our model
the parameter $\alpha$ is the inverse of the time to first division,
and thus, we will assume that the parameters $\alpha_n^{-1}$,
$\alpha_c^{-1}$, $\alpha_m^{-1}$ and $\alpha_e^{-1}$ for each
CD8\plus\ T~cell subtype (N, CM, EM or E), are not fixed parameters
and will be sampled from a uniform distribution with minimum 0.25 days
and maximum 5 days (see Table~\ref{table-parameters-bayesian}).

%%%%%%%%%%%%%%%%%%%%%%%%%%%%%%%%%%%%%%

\subsubsection{Time to subsequent divisions}
\label{method-later-division}

In Ref.~\cite{buchholz2013disparate}, the authors have made use of
experimental data from the OT-I transgenic mouse model, with CD8\plus\
T~cells that are specific for the SIINFEKL peptide from chicken
ovalbumin (OVA), and mathematical modelling to estimate the doubling
time of these cells. Their estimates range between 0.8 and 1.64 days
(see Figure S17 in Ref.~\cite{buchholz2013disparate}).  The estimate
for the doubling time of CD8\plus\ T~cells provided by the authors of
Ref.~\cite{le2014mathematical} is 1.8 days. These authors have also
made use of a mathematical model, in combination with human CD8\plus\
YFV kinetic data~\cite{miller2008human,akondy2009yellow}, to derive
this estimate from the range 1.4 to 2.66 days for the doubling time of
human CD8\plus\ T~cells.  In our model the parameter $\lambda$ is the
inverse of the time to subsequent divisions (or doubling time), and
thus, we will assume that the parameters $\lambda_n^{-1}$,
$\lambda_c^{-1}$, $\lambda_m^{-1}$ and $\lambda_e^{-1}$ for each
CD8\plus\ T~cell subtype (N, CM, EM or E), are not fixed parameters
and will be sampled from a uniform distribution with minimum 0.25 days
and maximum 5 days (see Table~\ref{table-parameters-bayesian}).

%%%%%%%%%%%%%%%%%%%%%%%%%%%%%%%%%%%%%%

\subsubsection{Maximum number of clonotypes recruited to the immune response}
\label{method-number-clonotypes}

Recent human CD4\plus\ estimates of the number of clonotypes recruited
to the immune response range between 100 and 5,000 (see Figure~1 in
Ref.~\cite{becattini2015functional}).  CD8\plus\ mice data from viral
infections show that more than 1,000 clonotypes have responded to a
given immuno-dominant epitope, with a range between 10$^2$ to 10$^5$
(see the Abstract in Ref.~\cite{pewe2004very}).  Given these
estimates, we choose to assume that the number of clonotypes recruited
to the immune response, $N_c$, will not be a fixed parameter and will
be sampled from a uniform distribution with minimum one and maximum
$N_c^{max}=10^5$, given the above estimates (see
Table~\ref{table-parameters-bayesian}).

%%%%%%%%%%%%%%%%%%%%%%%%%%%%%%%%%%%%%%

\subsubsection{Duration of immune challenge}
\label{method-duration of immune challenge}

The data used to carry out model calibration indicate that viral
titers become undetectable by day 16
post-vaccination~\cite{akondy2009yellow}.  Yet, the fraction of
specific CD8\plus\ T~cells increases until day 30 post-vaccination and
declines by day 90 post-vaccination (see Figure~3 in
Ref.~\cite{akondy2009yellow}).  Given these estimates, we choose to
assume that the the duration of the immune challenge, $\tau_E$, will
not be a fixed parameter and will be sampled from a uniform
distribution with minimum 5 days and maximum 60 days (see
Table~\ref{table-parameters-bayesian}).

%%%%%%%%%%%%%%%%%%%%%%%%%%%%%%%%%%%%%%

\subsection{Computational algorithm}
\label{method-computational-algorithm}

Solving the ODEs that describe the dynamics of antigen-specific and
non-specific CD8\plus\ T~cells during an immune response (according to
the DPM hypothesis), requires finding a solution to the set of
equations~\eqref{dpm-naive-dln}-\eqref{dpm-effector-s} for the
antigen-specific cells, as well as finding a solution to
equations~\eqref{non-specific-naive-dLN}
and~\eqref{non-specific-naive-B}, for the non-specific naive T~cell
populations.  In order to do so, the first thing that we require is to
choose a set of parameters, as described in
Section~\ref{math-parameters-computational} and
Section~\ref{method-model-immune-response-DPM}.  We note that a subset
of parameters are fixed, and given in
Table~\ref{table-parameters-fixed}, and the rest of the parameters
will be sampled from a number of distributions, as described in
Table~\ref{table-parameters-fixed}. Once a set of parameters has been
chosen, and before we can solve the system of ODEs at hand, the next
step is to choose initial conditions for all the cell types in the
three spatial compartments. The choice of initial conditions depends
on the immune scenario under consideration.  For a primary immune
response, we will assume that at the initial time ($t=0$), the only
CD8\plus\ T~cells present are naive (non-specific and
antigen-specific) and thus, there are no specific central memory,
effector memory or effector T~cells in any spatial compartment, that
is $C_0 = M_0 =E_0 =C_0^{(B)} =M_0^{(B)} =E_0^{(B)} =M_0^{(S)}
=E_0^{(S)} =0$.  Naive cells are assumed to be in homeostatic (or
steady-state) conditions prior to the immune challenge.  This means
that $N_{r0}, N_0, N^{(B)}_{r0}$ and $N^{(B)}_0$, the initial
conditions for the naive cell populations in the draining lymph nodes
and in the blood compartments, respectively, are the stable
steady-state solutions of equations~\eqref{non-specific-naive-dLN},
\eqref{specific-naive-dLN-homeostasis}, \eqref{non-specific-naive-B}
and~\eqref{specific-naive-B-homeostasis}.  If we consider the limit in
which the migration terms tend to zero in these equations (without
loss of generality, but in order to simplify the expressions), the
steady-state solutions, labelled with a star, are given by
\begin{eqnarray}
N_r^{\star} &=& \frac{\kappa_r}{\theta_n} \; (\theta_n - \mu_n)
\; , 
\\
N^{\star} &=& \frac{N_c \; \kappa_n}{\theta_n} \; (\theta_n - \mu_n)
\; , 
\\
N_r^{(B) \star} &=& \frac{1}{2 \; \theta_n} \; \left[ \kappa_r \; (\theta_n -\mu_n) + \sqrt{\kappa_r^2 \; (\theta_n - \mu_n)^2 + 4 \; \kappa_r \; 
\delta_r 
\; \theta_n} \right]
\; , 
\\
N^{(B) \star} &=& \frac{1}{2 \; \theta_n} \; \left[ N_c \; \kappa_n\; (\theta_n -\mu_n) + \sqrt{N_c^2 \; \kappa_n^2 \; 
(\theta_n - \mu_n)^2 + 4 \; N_c^2 \; \kappa_n \; 
\delta
\; \theta_n} \right]
\; .
\end{eqnarray}
Given a choice of parameter values, we set $N_{r0}=N_r^{\star},
N_0=N^{\star}, N^{(B)}_{r0} =N_r^{(B) \star}, N^{(B)}_0=N^{(B)
  \star}$, as these steady-state solutions can be shown (in the limit
in which the migration terms tend to zero) to be the unique stable
steady-state solutions of the system of equations given by
~\eqref{non-specific-naive-dLN},
\eqref{specific-naive-dLN-homeostasis}, \eqref{non-specific-naive-B}
and~\eqref{specific-naive-B-homeostasis}. In fact, it is easy to prove
that in this limit, the eigenvalue of the Jacobian matrix is given by
$\mu_n-\theta_n$, which is negative for the parameters described in
Table~\ref{table-parameters-fixed}.

For a secondary immune response, we will assume that at the initial
time ($t=0$), the only CD8\plus\ T~cells present are naive
(non-specific and antigen-specific), central memory and effector
memory cells, but no effector cells in any spatial compartment.  N, CM
and EM CD8\plus\ T~cells are assumed to be in homeostatic (or
steady-state) conditions prior to a secondary immune challenge.  This
means that the initial conditions for these populations in the
draining lymph nodes, in the blood and in the skin compartments, are
the stable steady-state solutions of the equations described in
Section~\ref{method-model-homeostasis} for the DPM.  If we consider
the limit in which the migration terms tend to zero in these equations
(without loss of generality, but in order to simplify the
expressions), the steady-state solutions for these populations,
labelled with a star, are given by
\begin{eqnarray}
N_r^{\star} &=& \frac{\kappa_r}{\theta_n} \; (\theta_n - \mu_n)
\; , 
\\
N^{\star} &=& \frac{N_c \; \kappa_n}{\theta_n} \; (\theta_n - \mu_n)
\; , 
\\
C^{\star} &=& \frac{N_c \; \kappa_c}{\theta_c} \; (\theta_c - \mu_c)
\; , 
\\
M^{\star} &=& 0
\; ,
\\
N_r^{(B) \star} &=& \frac{1}{2 \; \theta_n} \; \left[ \kappa_r \; (\theta_n -\mu_n) + \sqrt{\kappa_r^2 \; (\theta_n - \mu_n)^2 + 4 \; \kappa_r \; 
\delta_r 
\; \theta_n} \right]
\; , 
\\
N^{(B) \star} &=& \frac{1}{2 \; \theta_n} \; \left[ N_c \; \kappa_n\; (\theta_n -\mu_n) + \sqrt{N_c^2 \; \kappa_n^2 \; 
(\theta_n - \mu_n)^2 + 4 \; N_c^2 \; \kappa_n \; 
\delta
\; \theta_n} \right]
\; ,
\\
C^{(B) \star} &=& \frac{N_c \; \kappa_c}{\theta_c} \; (\theta_c - \mu_c)
\; , 
\\
M^{(B) \star} &=& \frac{N_c \; \kappa_m}{\theta_m} \; (\theta_m - \mu_m)
\; ,
\\
M^{(S) \star} &=& \frac{N_c \; \kappa_m}{\theta_m} \; (\theta_m - \mu_m)
\; .
\end{eqnarray}
The previous expressions are for reference only, as we have solved the
equations in their full generality with migration terms included. If
migration terms are included, the analytical expressions of the
steady-states become more complicated, and thus, have not been
included. Yet, we have been able to show that for our choice of
parameters, the equations of the homeostasis model for the naive,
central and effector memory T~cell populations described in
Section~\ref{method-model-homeostasis}, have a unique stable
steady-state that reduces to the solution presented above, if
migratory terms are neglected.

Finally, given a choice of parameters and initial conditions, the ODEs
were solved using a 4th order Runge-Kutta method implemented using
\emph{Python}.  Thus, parameters, initial conditions and the numerical
solver implemented in \emph{Python} constitute the computational
algorithm that will be referred to as the simulator of the
mathematical model. For the case of the DPM model, the simulator will
numerically integrate the equations described in
Section~\ref{method-model-immune-response-DPM}, and for the IPM, it
will integrate the equations presented in
Section~\ref{method-model-immune-response-IPM}.  The initial
conditions for the IPM are chosen in a similar way as done for the
DPM, and the details are not included here.

%%%%%%%%%%%%%%%%%%%%%%%%%%%%%%%%%%%%%%

\subsection{Sensitivity analysis for the decreasing potential model}
\label{method-sensitivity-analysis}

When implementing a mathematical model within some computational
algorithm (or simulator), it is important to verify that the resulting
simulator is behaving in the way that is meant.  It can also be
beneficial to identify the inputs to that simulator that have an
impact on the simulator outputs: if we can identify such important
inputs, we can determine which inputs need to be determined more
carefully and which inputs we have the best chance of learning about
in a parameter estimation scheme.  For the present simulator
(described in Section~\ref{method-computational-algorithm}), we have
performed a global sensitivity analysis as described in
Ref.~\cite{saltelli1}.  In this analysis, we propagate the uncertainty
in the inputs (as discussed earlier) through the simulator to obtain
uncertain outputs.  The uncertainty in the outputs can be apportioned
to each input, in terms of their direct and indirect impacts, and the
underlying principle is that the inputs that are responsible for
causing the most uncertainty in the outputs are the most important.

The direct (that is, the input acting alone) and indirect (that is,
the input acting in conjunction with other inputs) effects that inputs
have on an output can be quantified using main and total effect
indices~\cite{sobol1}. The main effect index for an input gives the
proportion of variance in the output that is directly accounted for by
that input alone. The total effect index for an input gives the
proportion of variance in the output that is accounted for by that
input alone and through interactions that input has with other
inputs. As such, it is not typical for the main effect indices or the
total effect indices to sum to one when considering the indices for
all inputs, because the main effect indices only account for inputs
acting alone and the total effect indices double count interaction
effects.  In addition to these indices, we are able to visualise the
main effect of an input through a plot of the expected value of the
output conditional on fixed values of that input.

Of course, for our simulator, we do not have a single output: we have
multiple time series.  For our sensitivity analysis, we focus on the
model outputs that relate to the data that we will use to calibrate
the model.  In our case, model outputs are $f_8(t), f_n(t), f_c(t),
f_m(t)$ and $f_e(t)$ for $t=11, 14, 30, 90$ days post-vaccine.  This
gives us twenty potential outputs to consider (we, in fact, consider
the logit transformed outputs).  We have calculated main effect
indices and total effect indices considering each of the outputs in
turn.  To get an overall picture of the importance of the inputs
across all outputs simultaneously, we have employed a principal
component analysis (PCA) on the outputs and calculated the indices for
the first four principal components (which account for approximately
90\% of the variability in the sampled outputs).

Using the main and total effect indices from the analysis of the first
four principal components, we find that the most important parameters
are (in order of importance): $\tau_E$, $\lambda_c$, $g_n$, $g_c$,
$g_m$, $\lambda_n$, $\lambda_m$, $\phi$ and $\gamma$.  Although these
parameters have an impact across all of the outputs, it is difficult
to interpret the estimated main effects because the corresponding
principal components are not on the same scale as the simulator's
outputs.

\begin{figure}[h!]
\centering
\includegraphics[page=16,width=0.5\textwidth]{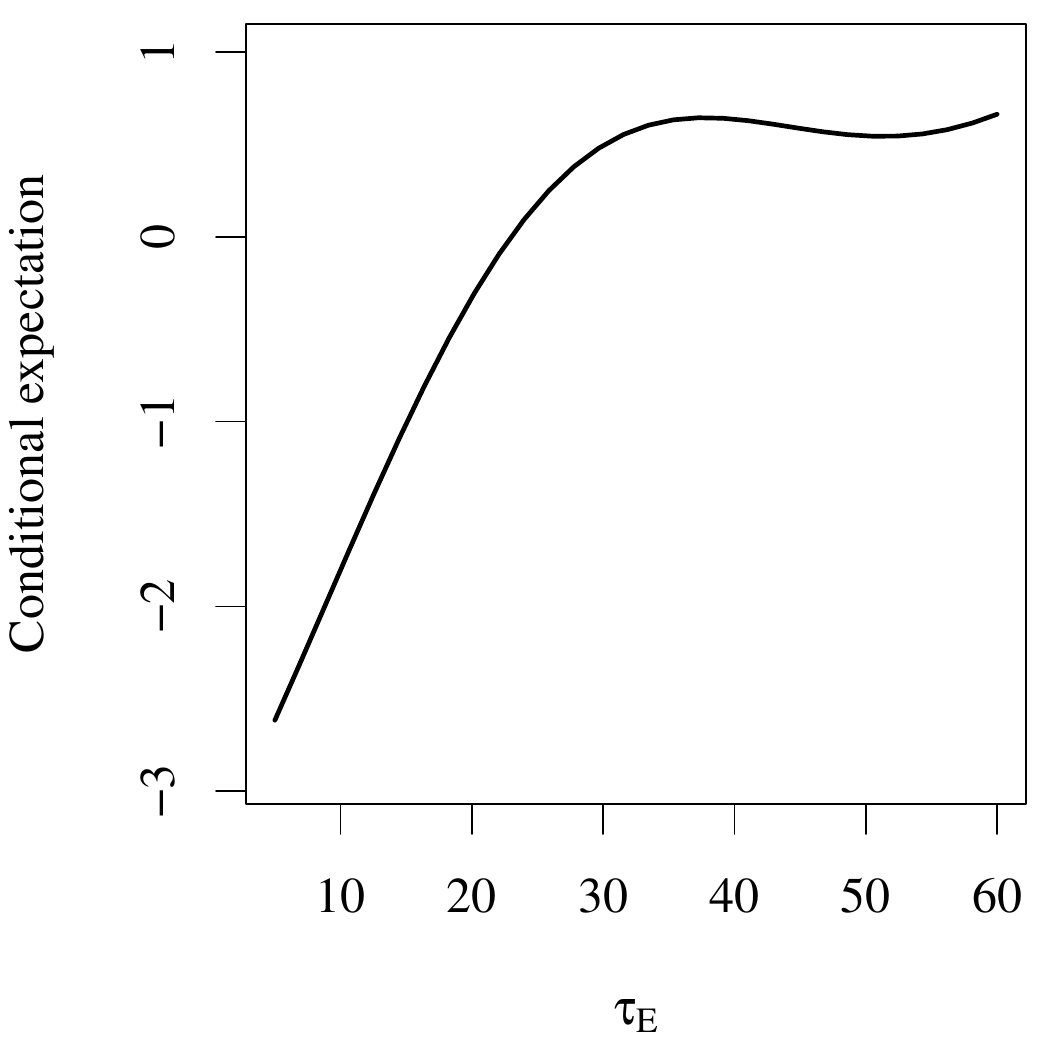}
\caption{Plot of the conditional expectation of $\log[T^{(B)}_8 (t=30)]$ given fixed values of $\tau_E$ for the DPM where we average over the uncertainty in the remaining model parameters.}
\label{plot.ME3te}
\end{figure}

We can therefore investigate the role of the most important parameters
by considering their impact on individual inputs.
Figure~\ref{plot.ME3te} is a plot of $\tau_E$ against the conditional
expectation of $\log\left[T^{(B)}_8 (t=30)\right]$, where this
expectation is calculated by fixing the value of $\tau_E$ and finding
the average value of $\log\left[T_8^{(B)}(t=30)\right]$ with respect
to the uncertainty in the other input parameters.  For all of the
input parameters, we consider their influence on the model outputs
over the ranges specified in Table~\ref{table-parameters-bayesian}.
We can see that as $\tau_E$ increases, the total number of CD8\plus\
antigen-specific T~cells at 30 days increases until $\tau_E=30$ days,
when the number of cells stabilises.

\begin{figure}[h!]
\centering
\begin{subfigure}[b]{0.41\textwidth}
\includegraphics[page=7,width=\textwidth]{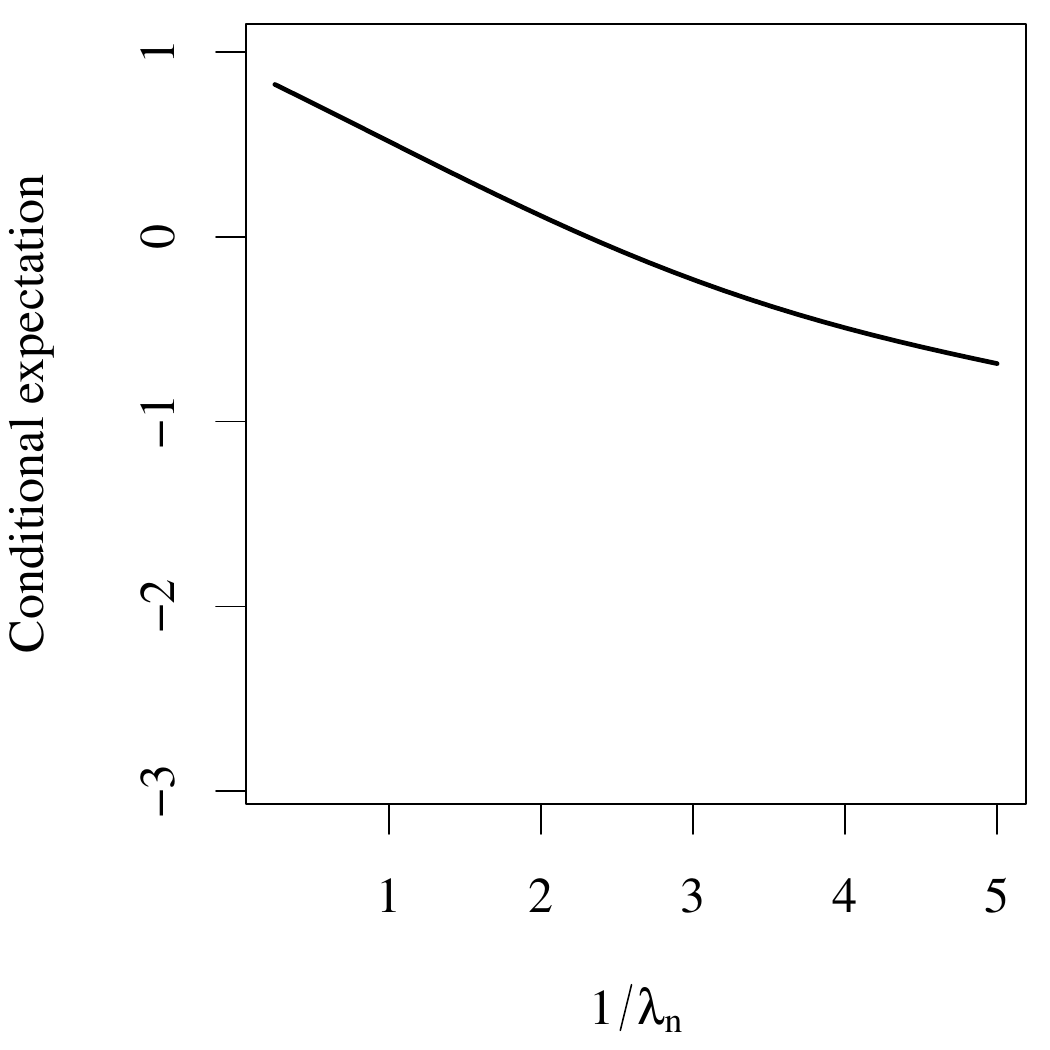}
\end{subfigure}
~
\begin{subfigure}[b]{0.41\textwidth}
\includegraphics[page=8,width=\textwidth]{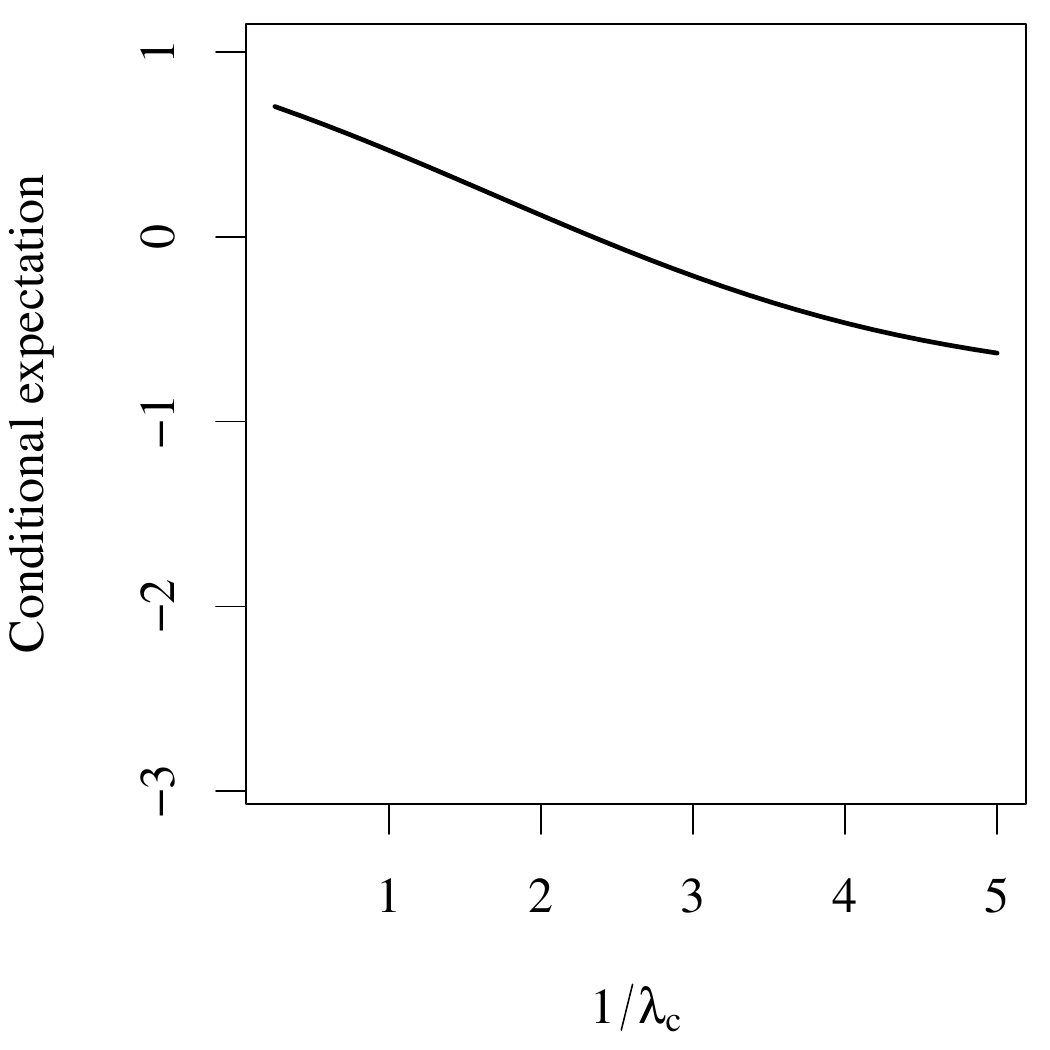}
\end{subfigure}

\begin{subfigure}[b]{0.41\textwidth}
\includegraphics[page=11,width=\textwidth]{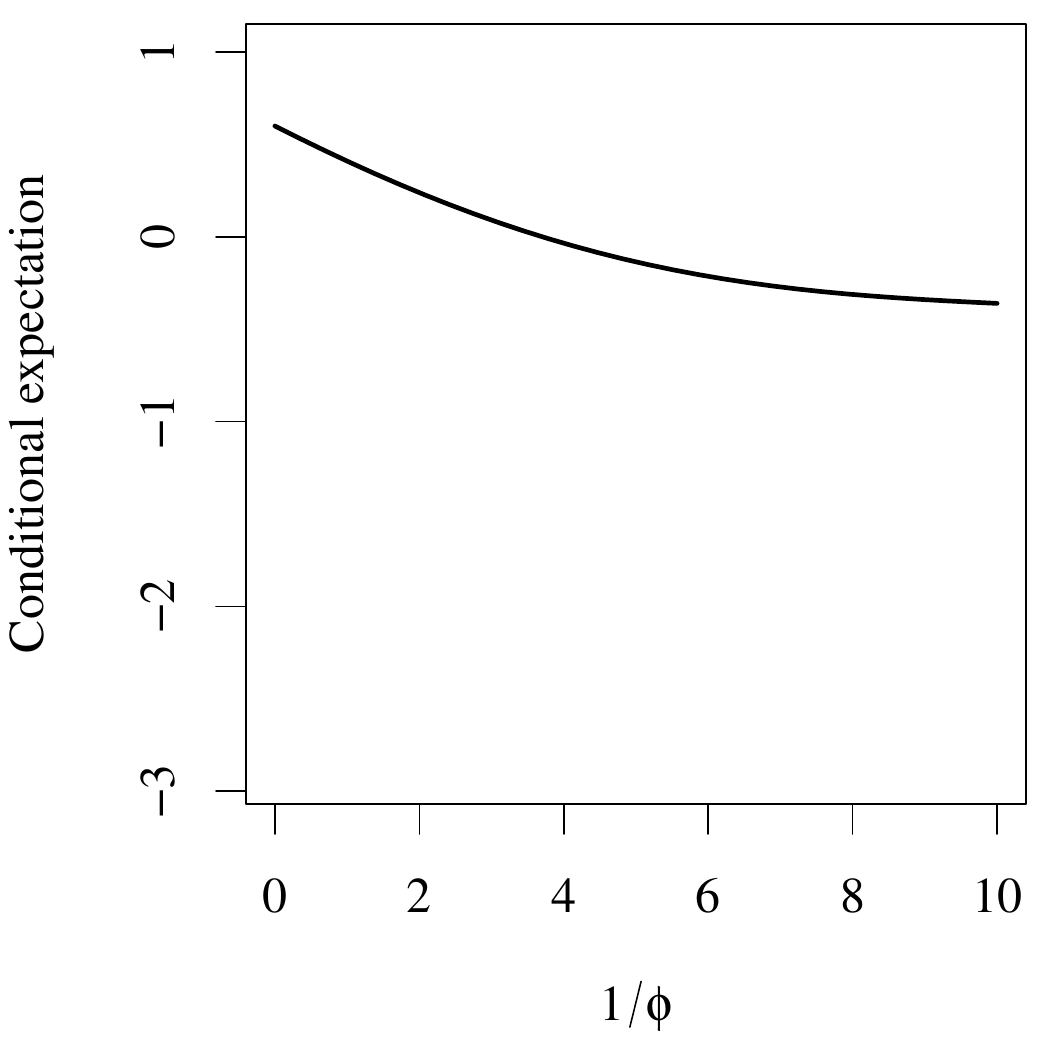}
\end{subfigure}
~
\begin{subfigure}[b]{0.41\textwidth}
\includegraphics[page=10,width=\textwidth]{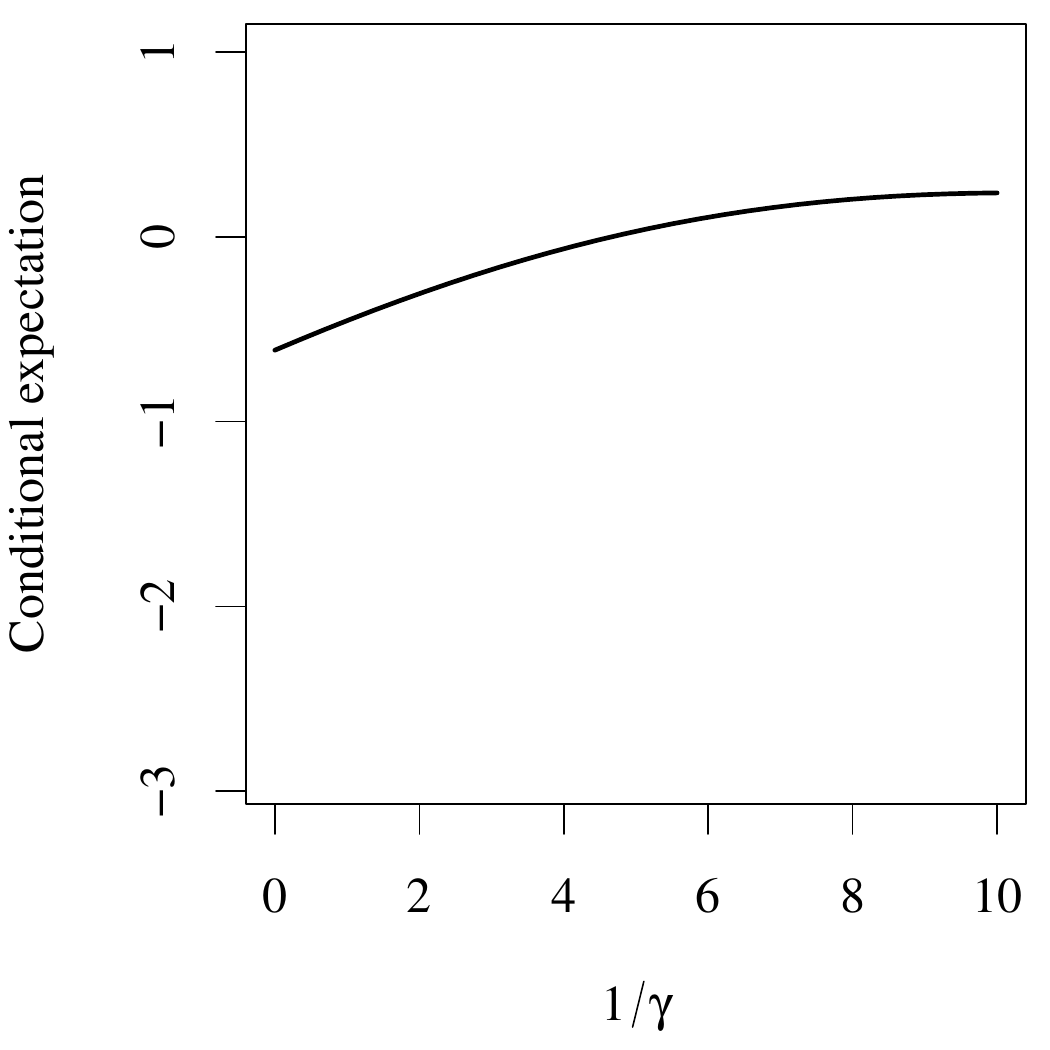}
\end{subfigure}
\caption{Plot of the conditional expectation of $\log[T^{(B)}_8 (t=30)]$ given fixed values of four different input parameters in the DPM ($1/\lambda_n$, $1/\lambda_c$, $1/\phi$ and $1/\gamma$).}
\label{plot.ME3for4}
\end{figure}

Other parameters, such as the number of generations for the different
subtypes, have a much more mundane effect, where the total number of
antigen-specific cells increases as $g_n$, $g_c$ and $g_m$ increase.
For $\lambda_c$, as this parameter increases, the total number of
antigen-specific cells also increases.  The same effect is observed
for $\lambda_n$, $\lambda_m$ and $\phi$, whereas an opposite (but
weaker effect) is seen for $\gamma$ (as shown in
Figure~\ref{plot.ME3for4}).

We also note that we may be able to rule out lower values of $N_c$
when we learn the parameters from data (see
Section~\ref{results-calibration}).  This is because $N_c$ has an
impact on the model outputs for low values alone.  Essentially, the
total number of CD8\plus\ T~cells is reduced for relatively small
values of $N_c$, and this effect disappears for $N_c \gtrsim 25,000$,
where the outputs are seemingly unaffected by $N_c$.

\begin{figure}[h!]
\centering
\begin{subfigure}[b]{0.41\textwidth}
\includegraphics[page=1,width=\textwidth]{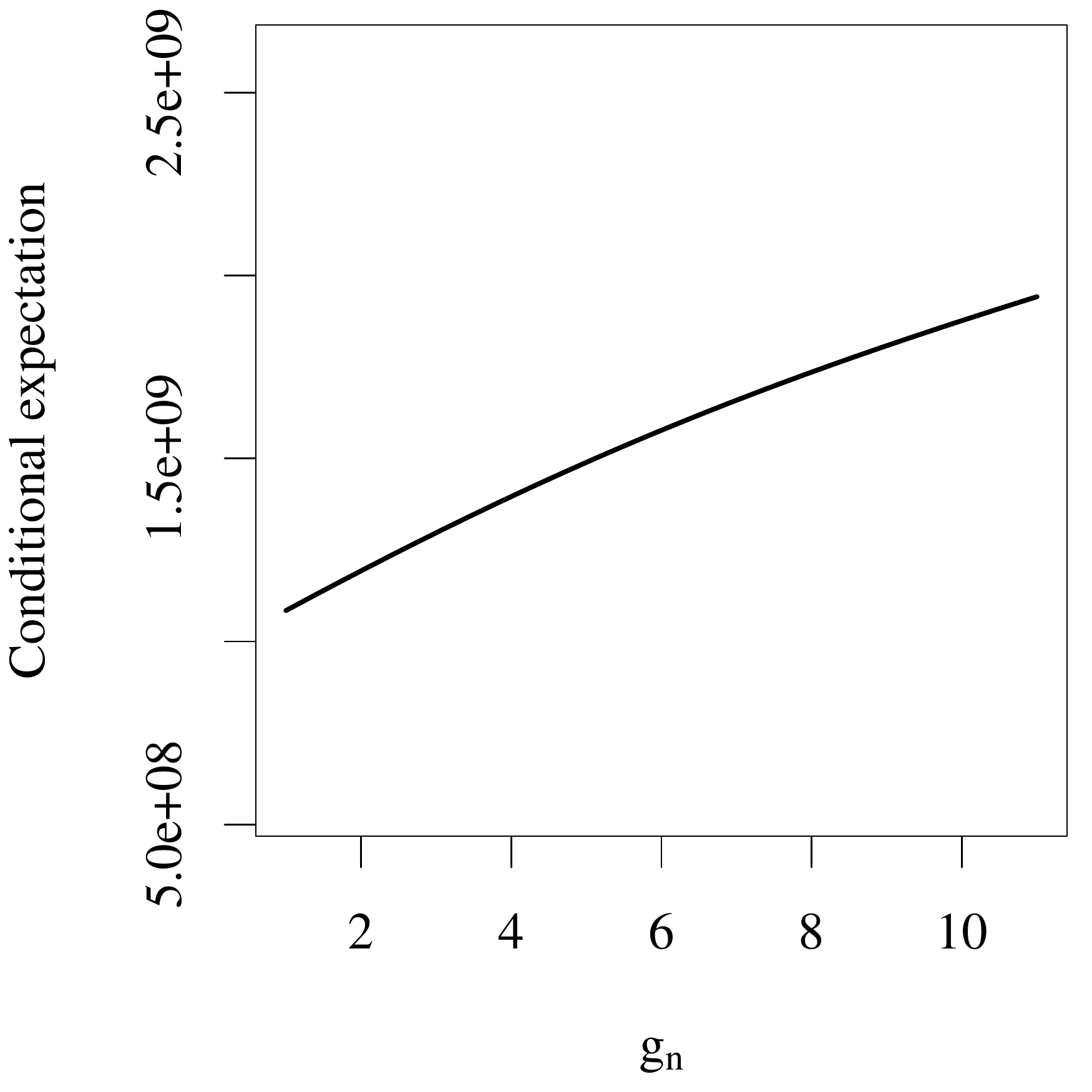}
\end{subfigure}
~
\begin{subfigure}[b]{0.41\textwidth}
\includegraphics[page=2,width=\textwidth]{MainEffectsOutMax}
\end{subfigure}

\begin{subfigure}[b]{0.41\textwidth}
\includegraphics[page=3,width=\textwidth]{MainEffectsOutMax}
\end{subfigure}
~
\begin{subfigure}[b]{0.41\textwidth}
\includegraphics[page=16,width=\textwidth]{MainEffectsOutMax}
\end{subfigure}
\caption{Plot of the conditional expectation for the maximum number
of antigen-specific CD8\plus\ T~cells in blood
 given fixed values of four different input parameters in the DPM ($g_n, g_c, g_m$ and $\tau_E$ fixed one-at-a-time and averaging over the uncertainty in the remaining 15 parameters).}
 \label{plot.ME3max}
\end{figure}

In addition to performing sensitivity analyses on the outputs that
correspond to observed data, we can consider the influence of the
input parameters on other model outputs, such as the maximum number of
antigen-specific CD8\plus\ T~cells in blood and the time at which this
maximum is realised.  For the output $\max_t[ T^{(B)}_8 (t)]$, we
found that $\tau_E$ and the number of generations had a leading role,
alongside $\alpha_m$ and $\phi$. In fact, in order of relevance, the
parameters that influence the output $\max_t[ T^{(B)}_8 (t)]$ are
$\tau_E, g_c, g_n, g_m$ and $\alpha_m$.  It is clear from
Figure~\ref{plot.ME3max} that, as the number of generations and
$\tau_E$ increase, $\max_t[ T^{(B)}_8 (t)]$ increases.

\begin{figure}[h!]
\centering
\includegraphics[page=16,width=0.5\textwidth]{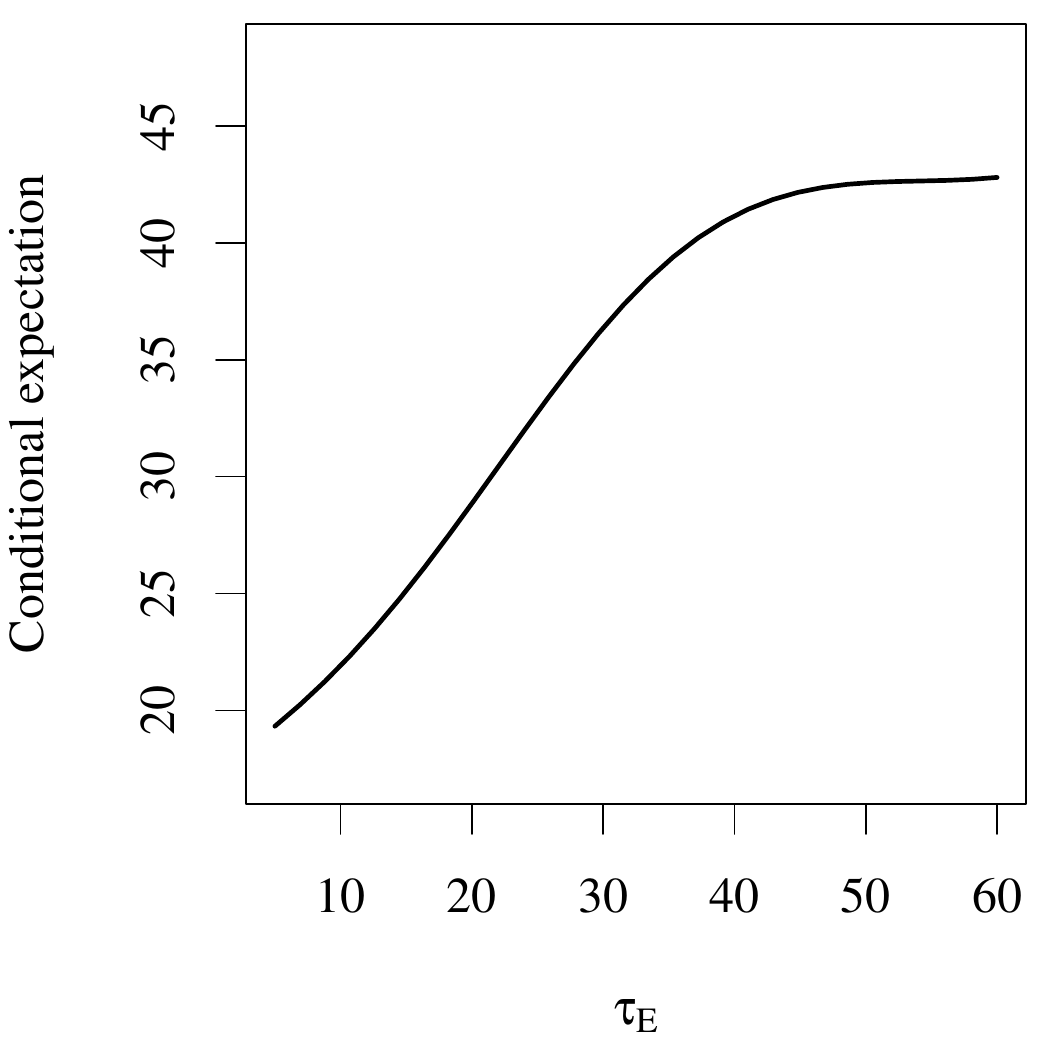}
\caption{Plot of the conditional expectation for the time taken (in days) to reach the maximum 
 number
of antigen-specific CD8\plus\ T~cells in blood,   $\max_t[ T^{(B)}_8 (t)]$, given fixed values of $\tau_E$ in the DPM.}
\label{plot.ME3time}
\end{figure}

\begin{figure}[h!]
\centering
\includegraphics[scale=0.35]{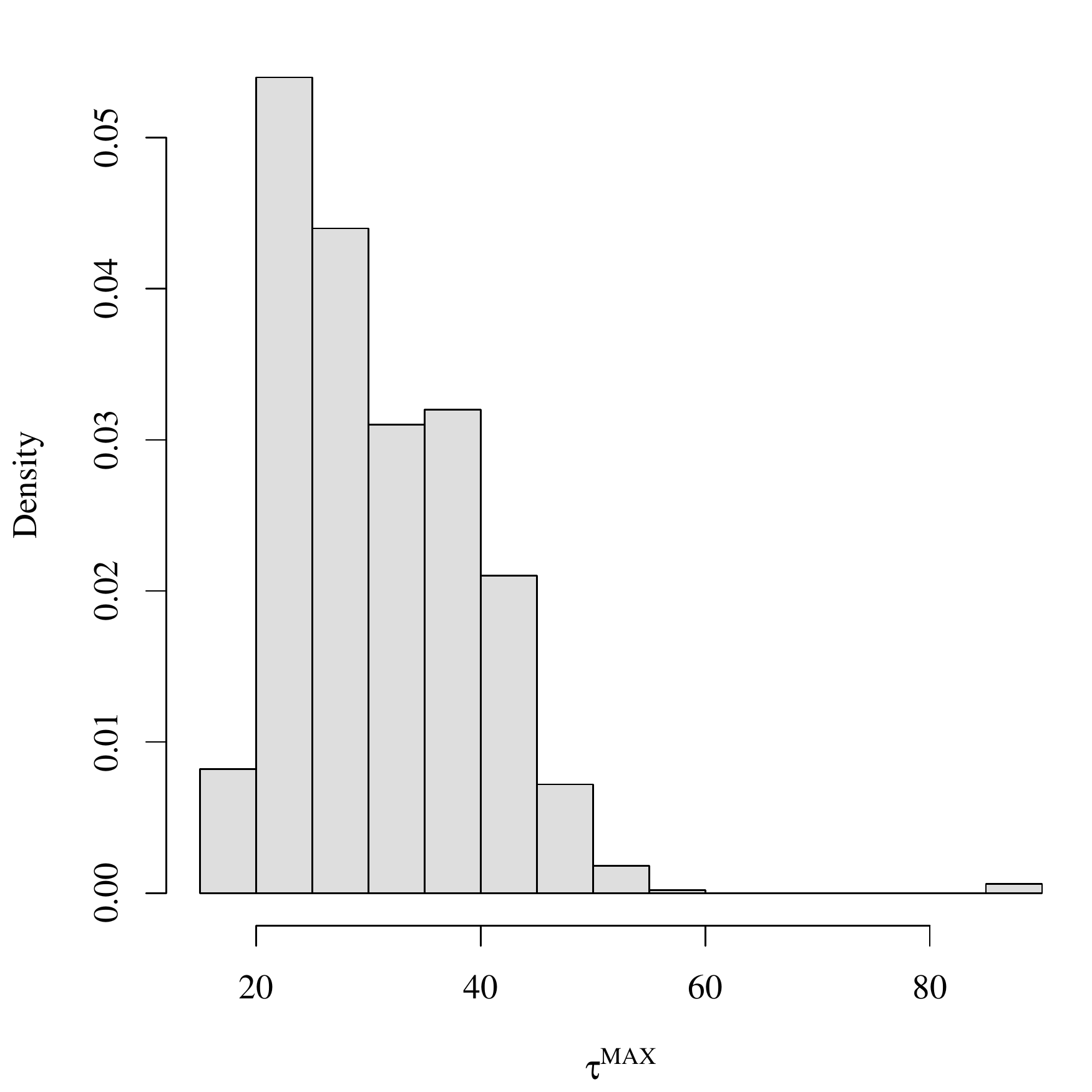}
\caption{Histogram of the
time taken to reach the maximum  
 number
of antigen-specific CD8\plus\ T~cells in blood,
 $\max_t[ T^{(B)}_8 (t)]$,
based upon the posterior distributions of the DPM input parameters.} 
\label{time-max-hist}
\end{figure}

For the time at which $\max_t[T^{(B)}_8 (t)]$ is realised, the most
influential parameter by far is $\tau_E$ (the total effect of $\tau_E$
accounts for 85\% of the variance in the time to the maximum).  As
expected, as $\tau_E$ increases, the time taken to reach the maximum
increases (see Figure~\ref{plot.ME3time}).  For completeness, we also
include a histogram for the time taken to reach the maximum in the DPM
for $N_c^{max}=10^5$ (see Figure~\ref{time-max-hist})

%%%%%%%%%%%%%%%%%%%%

\subsubsection{Sensitivity analysis for the increasing potential model: a brief comment}

For completeness, we conducted the same probabilistic sensitivity
analysis for the increasing potential model (both for individual
outputs and for PCA transformed outputs).  Four parameters are more
influential than any others in that they regularly have relatively
high total effect indices for all the outputs of interest: $\tau_E$,
$g_n$, $\lambda_n$ and $g_e$.  As in the case of the decreasing
potential model, $\tau_E$ has a great impact on the number of
antigen-specific cells in the blood compartment.  There are several
other input parameters that cause some of the variance for some of the
outputs: $g_c$, $\alpha_c$, $\lambda_e$, $\lambda_m$, $\phi$ and
$\xi$.

%%%%%%%%%%%%%%%%%%%%%%%%%%%%%%%%%%%%%%

\subsection{Calibration method}
\label{method-calibration}

We take a Bayesian approach to the calibration of the simulator's
input parameters: we set prior distributions for each of the
parameters and update them in the light of the data and the
mathematical model.  The updated distributions are called posterior
distributions.  Given the complexity of the model and the approximate
nature of the likelihood, we employ an approximate Bayesian
computation (ABC) approach~\cite{marin1}.  For such an approach, we do
not need to specify a likelihood function for the data given the
simulator: we just need to be able to simulate data in a similar form
to the data that we have.  The basic ABC approach that we employ is as
follows:

\begin{enumerate}

\item[0.] Decide on the number of required samples $N$, and set $i=0$.

\item[1.] Make a single draw from the prior distributions of the unknown input parameters.

\item[2.] Run the simulator  in order to use the drawn input parameters to find a set of outputs that correspond to the data.

\item[3.] If the simulated data are sufficiently close to the observed data, add the drawn input parameters to your sample and increase $i$ by one.

\item[4.] If $i<N$, go to step 1.

\end{enumerate}

In step~2, and to simulate data, we can run the simulator for any
given set of input values to get a set of outputs that correspond to
the data that we will observe.  If we make the judgement that the
simulator is providing outputs that correspond to the CD8\plus\ 
specific T~cell
fraction $f_8 (t)$ for the average human
post-vaccination, then we need to also specify a mechanism that can
generate the variability in the human population.  For the 
specific T~cell
fraction at time $t$, $f_8 (t)$, we use a log normal distribution with its mean provided
by the simulator and a variance to be learnt from the data.  These
choices result in a need to learn the input parameters and the
variability parameter value from the data.  Given the data,
plausible distributions are as follows:
\begin{align*}
\log\left(f_{8,\mbox{obs}}(t)\right)|\Theta,\sigma^2 \sim \mbox{N}\left(f_{8,\mbox{model}}(t),\sigma^2\right)~~~~\mbox{for }t\in\{11,14,30,90\},
\end{align*}
where $\Theta$ is the complete set of input parameters and $\sigma^2$ is the variance in the observed data.

In step~3, we compare the simulated data with the observed data. Here
we must choose what distance measure to use and what tolerance is
permitted. Both of these choices have an impact on the accuracy on the
approximation. For instance, if we consider a Euclidean distance
between the simulated and observed data sets and set the tolerance to
be zero, then the algorithm provides an exact posterior result. In our
case, because the data are continuous, there is zero chance of
replicating the data exactly so we must have a non-zero tolerance and
the size of the tolerance will determine the length of time it takes
for the algorithm to produce the sample. We calculate the Euclidean
distance between $\log\left[f_{8,\mbox{obs}}(t)\right]$ and
$\log\left[f_{8,\mbox{model}}(t)\right]$ for each of the time
points and for each of the individuals and compare this with tolerance
$\epsilon_T$. The final value of $\epsilon_T$ we used was 13.5 for
both the DPM and IPM models. Using the same value for the tolerance
was important because the ABC results were used in the model
comparison of Section~\ref{results-comparison}.

%%%%%

\end{document}